\documentclass[10pt,A4paper]{article}

\usepackage{tikz} 
\usepackage{amsmath}
\usepackage{mathrsfs} 
\usepackage{amsfonts} 
\usepackage{pgfplots}

 \usepackage{amssymb}

 \usepackage[colorlinks, linkcolor=blue, anchorcolor=red, citecolor=blue]{hyperref}

 \usepackage{appendix}

\def\ar{\!\!\!&} 

\def\proof{\noindent{\it Proof.~~}} 
\def\qed{\hfill$\Box$\medskip} 
 \usepackage[top=2cm,bottom=2cm, outer=2cm, inner=2cm]{geometry}

\newtheorem{theorem}{Theorem}[section]

\newtheorem{lemma}[theorem]{Lemma}
\newtheorem{corollary}[theorem]{Corollary}
\newtheorem{proposition}[theorem]{Proposition}
\newtheorem{remark}[theorem]{Remark}
\newtheorem{condition}[theorem]{Condition}
\newtheorem{example}[theorem]{Example}

 \newtheorem{assumption}[theorem]{Assumption}

\def\beqlb{\begin{eqnarray}}\def\eeqlb{\end{eqnarray}} 
\def\beqnn{\begin{eqnarray*}}\def\eeqnn{\end{eqnarray*}} 

\def\ar{\!\!\!&}

\renewcommand{\theequation}{\arabic{section}.\arabic{equation}} 

\def\proof{\noindent{\it Proof.~~}}\def\qed{\hfill$\Box$\medskip}

\begin{document}
 \title{\bf\Large The Microstructure of Stochastic Volatility Models with Self-Exciting Jump Dynamics\thanks{Financial support from the Alexander-von-Humboldt-Foundation is gratefully acknowledged.}}

 \author{Ulrich Horst\footnote{Department of Mathematics and School of Business and Economics,  Humboldt-Universit\"at zu Berlin, Unter den Linden 6, 10099 Berlin; email: horst@math.hu-berlin.de}\quad\  and\quad Wei Xu\footnote{Department of Mathematics, Humboldt-Universit\"at zu Berlin, Unter den Linden 6, 10099 Berlin; email: xuwei@math.hu-berlin.de}}
 \maketitle

  \begin{abstract}
We provide a general probabilistic framework within which we establish scaling limits for a class of continuous-time stochastic volatility models with self-exciting jump dynamics. In the scaling limit, the joint dynamics of asset returns and volatility is driven by independent Gaussian white noises and two independent Poisson random measures that capture the arrival of exogenous shocks and the arrival of self-excited shocks, respectively. Various well-studied stochastic volatility models with and without self-exciting price/volatility co-jumps are obtained as special cases under different scaling regimes. We analyze the impact of external shocks on the market dynamics, especially their impact on jump cascades and show in a mathematically rigorous manner that many small external shocks may tigger endogenous jump cascades in asset returns and stock price volatility.  

  \end{abstract}



\renewcommand{\baselinestretch}{1.15}

  \section{Introduction}

 Affine stochastic volatility models have been extensively investigated in the mathematical finance and financial economics literature in the last decades. 
 In the classical Heston \cite{Heston1993} model the volatility process follows square-root mean-reverting Cox-Ingerson-Ross \cite{CIR1985} process. 
 The Heston model introduces a dynamics for the underlying asset that can take into account the asymmetry and excess kurtosis that are typically observed in financial assets returns and provides analytically tractable option pricing formulas. However, it is unable to capture large volatility movements. To account for large volatility movements, the model has been extended to jump-diffusion models by numerous authors. Bates \cite{Bates1996} adds a jump component in the asset price process. Barndorff-Nielsen and Shephard \cite{Barndorff-NielsenShephard2001} consider volatility processes of Ornstein-Uhlenbeck type driven by L\'evy processes. Affine models allowing for jumps in prices and volatilities are considered in Bakshi et al.~\cite{BakshiCaoChen1997}, Bates \cite{Bates1996}, Duffie et al. \cite{DuffiePanSingleton2000}, Pan \cite{Pan2002} and Sepp \cite{Sepp2008}, among many others. Empirical evidence for the presence of (negatively correlated) co-jumps in returns and volatility is given in, e.g.~Eraker \cite{Eraker2004}, Eraker et al. \cite{ErakerJohannesPolson2003}, and Jacod and Todorov \cite{JacodTodorov2010}.

In a standard jump model with arrival rates calibrated to historical data, jumps are inherently rare. Even more unlikely are patterns of multiple jumps in close succession over hours or days. Large moves, however, tend to appear in clusters. For example, as reported in  A\"{\i}t-Sahalia et al.~\cite{SahaliaCacho-DiazLaeven2015} ``from mid-September to mid-November 2008, the US stock market jumped by more than 5\% on 16 separate days. Intraday fluctuations were even more pronounced: during the same two months, the range of intraday returns exceeded 10\% during 14 days.''
\begin{figure}
\begin{center}
	\includegraphics[width=16cm]{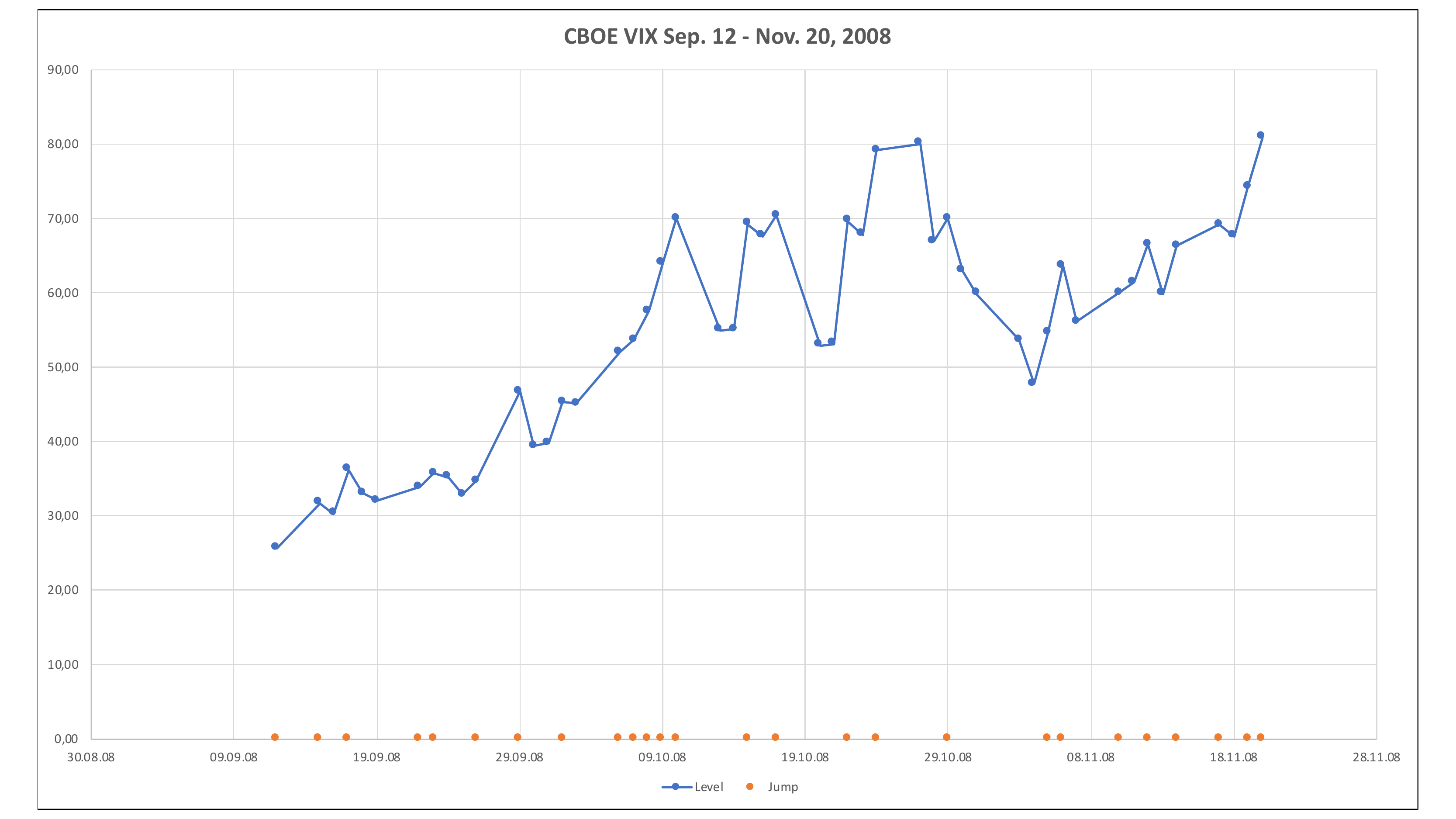}
	\caption{CBOE VIX index based on daily closing values from Sep. 12 to Nov. 20, 2008. Orange dots indicate a jump of 2.5\% or more from previous day closing value.}
\end{center}
\end{figure}
{Jump clusters can also be observed in the volatility. Figure 1 displays the evolution of the Chicago Board of Exchange VIX index and indicates up movements of more than 2.5\% in daily closing values for the above mentioned period; on 26 out of 49 days, the index jump up by 2.5\% or more.} Bates \cite{Bates2019} argues that the dramatic decline in futures prices on Monday, October 19, 1987, from the previous Friday’s closing value ``was the result of an estimated 34 jumps'' in volatility.

 Jump clusters over time have been discussed in the financial econometrics literature by many authors including A\"{\i}t-Sahalia et al. \cite{SahaliaCacho-DiazLaeven2015}, Fulop et al. \cite{FulopLiYu2015}, Lee and Mykland \cite{LeeMykland2008},  Maheu and McCurdy \cite{MaheuMcCurdy2004}, and Yu \cite{Yu2004}. Among the most relevant papers for our work are the ones by Andersen et al. \cite{AndersenFusariTodorov2015} and Bates \cite{Bates2019}. They consider continuous-time models of self-exciting price/volatility co-jumps in intradaily stock returns and volatility. Every small intradaily jump substantially increases the probability of more intradaily cojumps in volatility and returns, and these multiple price jumps can accumulate into the major outliers in daily returns. Bates \cite{Bates2019} finds {``that multifactor models with both exogenous and self-exciting but short-lived volatility spikes'' substantially improve model fits both in-sample and out-of-sample}. He also shows that such models provide more accurate predictions of implied volatility. A similar conclusion on implied volatilities was reached in the recent work by Jiao et al. \cite{JiaoMaScottiZhou2018}.

In order to account for self-exiting jump dynamics one needs to leave the widely applied class of Lévy jump processes. Lévy processes have independent increments and hence do not allow for any type of serial dependence. Hawkes processes are capable of displaying mutually exciting jumps. Originally introduced by Hawkes \cite{Hawkes1971a, Hawkes1971b} to model the occurrence seismic events, Hawkes processes have received considerable attention in the financial mathematics and economics literature as a powerful tool to model financial time series in recent years; we refer to Bacry et al.~\cite{BacryMastromatteoMuzy2015} and references therein for reviews on Hawkes processes and their applications to science and finance. On the more mathematical side, a series of functional limit theorem and large deviation principles for Hawkes processes and marked Hawkes processes has recently been established by, e.g. Bacry et al.~\cite{BacryDelattreHoffmannMuzy2013}, Gao and Zhu \cite{GaoZhu2018b,GaoZhu2018}, and Karabash and Zhu \cite{KarabashZhu2015}. Horst and Xu \cite{HorstXu2019a} introduced Hawkes random measures in order to study limit theorems for limit order book models with self-exciting cross-dependent order flow. In \cite{HorstXu2019b} they established functional limit theorems for marked Hawkes point measures with homogeneous immigration under a light-tailed condition on the arrival dependencies of different events. Under a light-tailed condition on the arrival dependencies of different events, Jaisson and Rosenbaum \cite{JaissonRosenbaum2015} proved that the rescaled intensity process of a Hawkes process converges weakly to a Feller diffusion and that the rescaled point process converges weakly to the integrated diffusion. Under a heavy-tailed condition they  proved that the rescaled point process converges weakly to the integral of a rough fractional diffusion; see \cite{JaissonRosenbaum2016}. Their result provides a microscopic foundation for the rough Heston model; see  \cite{ElEuchRosenbaum2019a,ElEuchRosenbaum2019b}.

 Motivated by the recent empirical works on stochastic volatility models with self-exciting jumps, we provide a unified microscopic foundation for stochastic volatility models with price/volatility co-jumps based on Hawkes processes. Many of the existing jump diffusion stochastic volatility models including the classical Heston model \cite{Heston1993}, the Heston model with jumps \cite{Bates1996, DuffiePanSingleton2000, Pan2002}, the OU-type volatility model \cite{Barndorff-NielsenShephard2001}, the multi-factor model with self-exciting volatility spikes  \cite{Bates2019} and the alpha Heston model \cite{JiaoMaScottiZhou2018} are obtained as scaling limits under different scaling regimes.  As such, our work contributes to the rich literature on scaling limits for financial market models\footnote{Much of the earlier work including \cite{Foellmer1994, FoellmerSchweizer1993,Horst2005} focussed on the temporary occurrence and bubbles, due to imitation and contagion effects rather than volatility. More recently, the focus seems to have shifted to order books models and volatility.} as well as to the growing literature on Hawkes processes by establishing novel scaling limits for Hawkes systems.

Our analysis uses the link between Hawkes processes and continuous-state branching processes (CB-processes). The extinction behavior of CB-processes as analyzed in Grey \cite{Grey1974} is of particular interest to us as it allows us to study the impact duration of external shocks on order flow. Branching processes are a particular class of affine processes. Affine processes have been widely used in the financial mathematics literature. Duffie et al. \cite{DuffieFilipovicSchachermayer2003} define an affine process as a time-homogeneous Markov process, whose characteristic function is the exponential of an affine function of the state vector under a regularity assumption. They showed that this type of process unifies the concepts of continuous-state branching processes with immigration (CBI-processes) and Ornstein-Uhlenbeck type processes (OU-processes). They also provide a rigorous mathematical approach to affine processes, including the characterization of affine processes in terms of admissible parameters (comparable to the characteristic triplet of a L\'evy process). Dawson and Li \cite{DawsonLi2006} provide a construction of affine process as the unique strong solution to a system of stochastic differential equations with non-Lipschitz coefficients and Poisson-type integrals over some random sets. Keller-Ressel \cite{Keller-Ressel2011} considers the long-term behavior of affine stochastic volatility models, including an expression for the invariant distribution. He also provided explicit expressions for the time at which a moment of given order becomes infinite. We shall repeatedly draw on the results in \cite{DawsonLi2006,DuffieFilipovicSchachermayer2003,Keller-Ressel2011}.

We consider a microstructure model of a financial market with two types market orders that we refer to as {\sl exogenous orders} and {\sl induced orders}, respectively. We think of {exogenous orders} as exogenous shocks; they arrive at an exogenous Poisson dynamics. Exogenous orders generate a random environment for the arrival of induced orders. Induced orders arrive according to a marked Hawkes process with exponential kernel. The marks represent the magnitudes by which the orders change prices as well as their impact on the arrival rate of future induced order flow. In particular, jumps in prices and/or volatility may trigger cascades of child-jumps and hence volatility clusters. Jump cascades are particularly likely to occur after large exogenous shocks. Figure 2 shows the evolution of the CBOE VIX index from 1990 to 2019; the time series clearly displays the occasional occurrence exogenous shocks. Apart from the already mentioned clustering of jumps during the hight of the global financial crisis there seem to be further jump clusters, for instance after the 1998 Russian and the 2011 Eurozone debt crisis. 

\begin{figure}
\begin{center}
	\includegraphics[width=16cm]{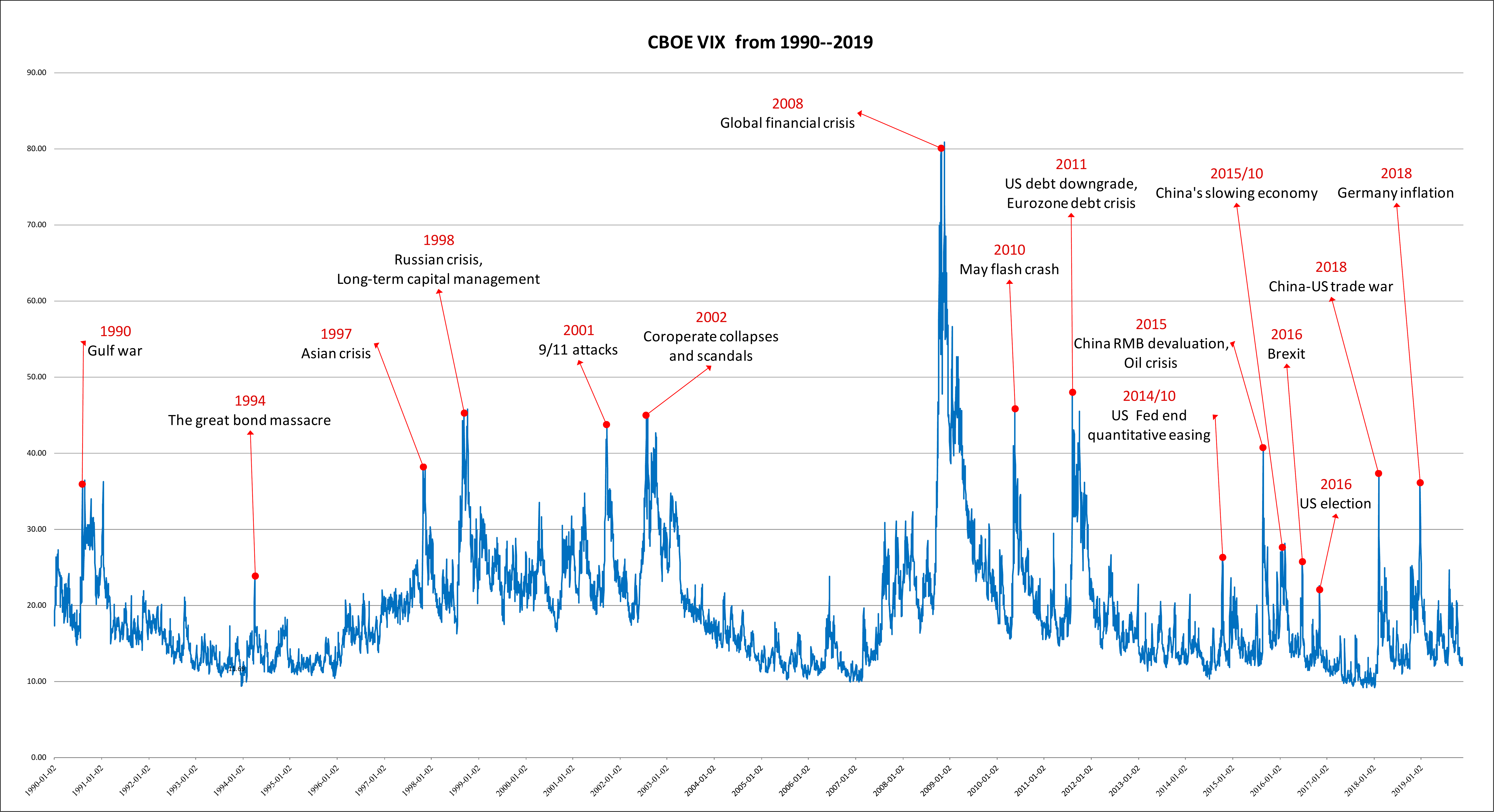}
	\caption{CBOE VIX index based on daily closing values from 1999-2019.}
\end{center}
\end{figure}

Our first main contribution is to analyze the genealogical decomposition of the benchmark model, and to anlayze the impact duration of large external shocks on induced order flow. To this end, we provide a decomposition of the benchmark model as the sum of a sequence of independent self-enclosed sub-models that describe the impact of exogenous shocks on induced order flow. We then provide four regimes for the long-run impact of an exogenous shock to the market dynamics. The impact of an exogenous shock will last forever with positive probability in the supercritical case and vanish at an exponential rate in the subcritical case. In the critical case the impact duration of external shocks is heavy-tailed despite the exponential decay of individual events on future dynamics.

 Our second main contribution is a scaling limit for the benchmark model when the frequency of order arrivals tends to infinity and the impact of an individual order on the market dynamics tends to zero. Depending on the choice of scaling parameters, various well-known jump-diffusion stochastic volatility models are obtained in the scaling limit. Different to the arguments in \cite{JaissonRosenbaum2015} on the convergence of nearly unstable Hawkes processes, we give sufficient conditions on the model parameters that guarantee the existence of a non-degenerate scaling limit based on the link between Hawkes proceses and branching particle systems. Loosely speaking, we require the convergence of the sequence of branching mechanisms. From this, we conclude that the sequence of generators converges to a limiting generator when restricted to exponential functions. Since the linear span of exponential functions is not dense in the domain of the limit generator, methods and technique based in the convergence of generators can not be applied to establish the convergence of the rescaled sequence of market models. Instead, we use general convergence results for infinite dimensional stochastic integrals established in Kurtz and Protter \cite{KurtzProtter1996}. Their methods have previously been applied to prove diffusion approximations for limit order book models in \cite{HorstKreher2019}. We prove that the rescaled sequence of market models converges in distribution to the unique solution of a stochastic differential equation driven by two independent Gaussiam white noise processes, a Poisson random measure that describes the arrivals of large exogenous shocks and an independent Poisosn random measure that describes the dynamics of endogenously induced self-excited jumps.

Our third main contribution is to analyze the genealogical decomposition of the limiting jump-diffusion volatility model. We provide an economically intuitive decomposition in terms of three sub-models. The first sub-model is self-enclosed and captures the impact of all events prior to time $0$ on future order flow.  The second sub-model  describes the cumulative impact of the exogenous shocks of positive magnitude on the market dynamics; this sub-model can be further decomposed into a sequence of self-enclosed sub-models that capture the impact of individual shocks. The third sub-model is the most interesting one. This self-enclosed sub-model describes the impact of exogenous shocks ``of insignificant magnitude'' on the market dynamics. Specifically, in the scaling limit large exogenous shocks translate into jumps while vanishingly small exogenous shocks translate in a well defined sense into a non-trivial mean-reversion level of the stochastic volatility process that keeps the volatility bounded away from zero at all times. Due to the dependence of the jump arrivals on the volatility process, this shows in a mathematically rigorous manner how many small \textsl{exogenous} events may trigger \textsl{endogenous} jump cascades. Our decomposition is very different from that in \cite{JiaoMaScottiZhou2018}. They decompose the volatility process into a truncated variance process plus a variance process that captures all jumps larger than some threshold. Unlike ours, their sub-models are not self-enclosed and do not classify jumps by their origin.

\begin{figure}
\begin{center}
	\includegraphics[width=16cm]{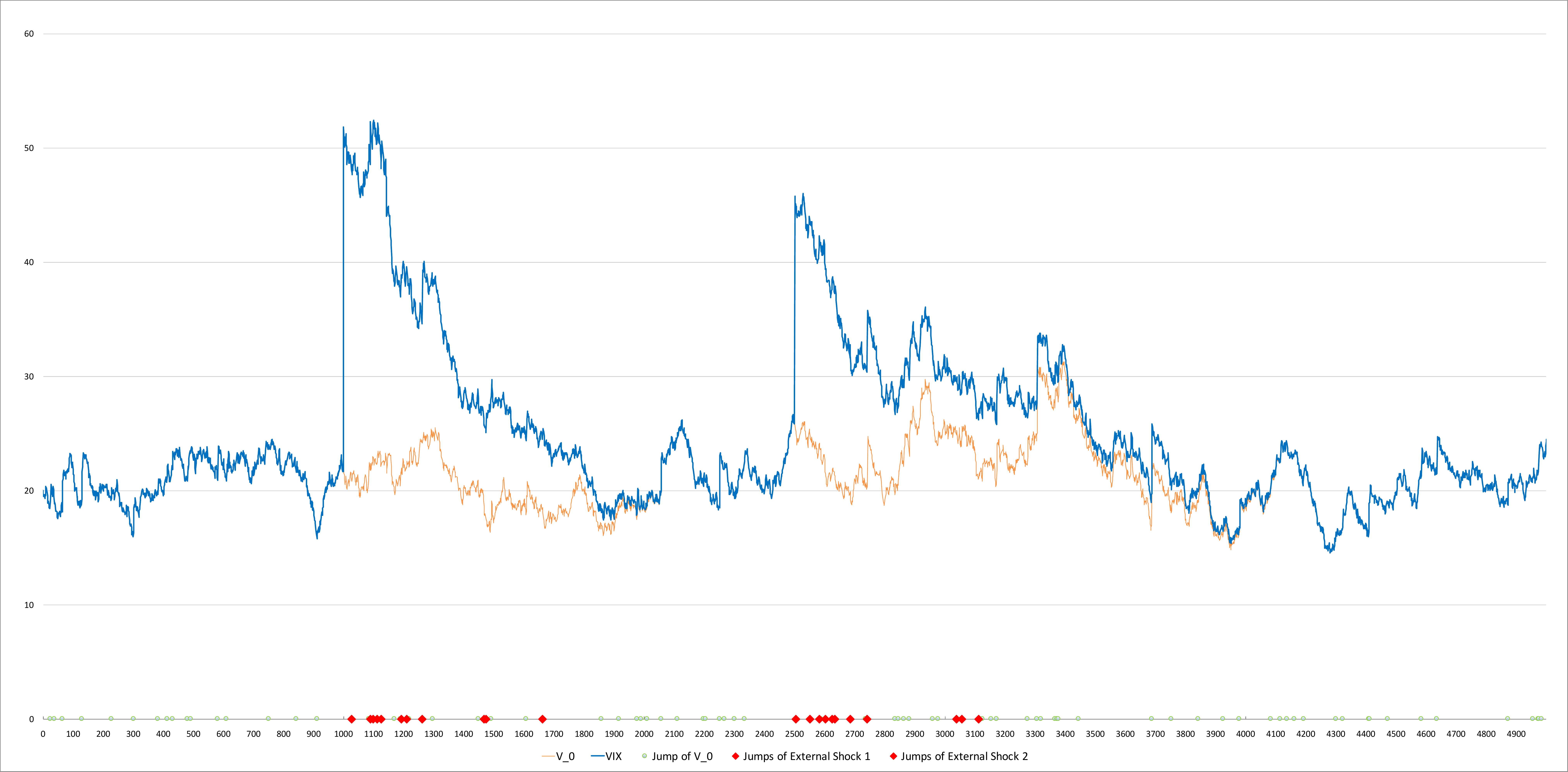}
	\caption{Path simulation from our limit model with (blue trajectory) and without (orange trajectory) external shocks. Red crosses indicate jumps triggered by the first external shocks; light blue dots indicate jumps triggered endogenously.}
\end{center}
\end{figure}

Finally, we analyze the distribution of jumps of different magnitudes and different origins in the scaling limit. We give an explicit expression for the joint distribution of the number and magnitudes of jumps induced by an exogenous shocks of given size and the time of the last induced jump in terms of the unique continuous solution to a Riccati equation with singular initial condition. We also show that exogenously (by exogenous shocks) and endogenously (by the volatility) triggered jump cascades share an important characteristic. In both cases, on average the \textsl{proportions of the total number of jumps} triggered by a given time is the same\footnote{In other words, in both cases, on avarege 80\% of the total number of shocks triggered are triggered by the same time}. The \textsl{number} of jumps triggered by an external shock, however, scales linearly in the magnitude of that shock. As a result, jump cascades triggered by external shocks will usually comprise more jumps than those triggered endogenously and thus trigger more dense jump clusters. Figure 3 illustrates this effect. It shows the evolution of a sample path of our modle with (blue) and without (orange) external shocks. There are two external shocks; the times of the  shocks originating from the external shocks are indicated by red dots.  Light blue dots times of indicate endogenously triggered jumps. For our choice of parameters they are more evenly distributed across time. 

The remainder of this paper is organized as follows. Our benchmark financial market model is introduced in Section \ref{LFG}. Its genealogical structure is analyzed in Section \ref{genealogy}. The scaling result is established in Section \ref{HFHM}. In this section we also show how various well studied stochastic volatility models can be obtained as scaling limits. The genealogical structure of the scaling is analyzed in Section \ref{gen-limit}.

\section{Hawkes market model} \label{LFG}
\setcounter{equation}{0}

 In this section, we introduce a benchmark stochastic volatility model for which we drive a scaling limit in a later section. There are two types of buy/sell orders in our model that we refer to as {\sl exogenous orders} and {\sl induced orders}, respectively. Exogenous orders as orders that arrive at an exogenous Poisson dynamics (``exogenous shocks''). They generate a random environment for the arrival of induced orders. Induced orders will arrive at much higher frequencies than exogenous orders and follow a self-exciting dynamics.

In what follows all random variables are defined on a common probability space $(\Omega,\mathscr{F}, \mathbf{P})$ endowed with filtration $\{\mathscr{F}_t:t\geq 0\}$ that satisfies the usual hypotheses.

 \subsection{The benchmark model}

 The arrivals of exogenous/induced market buy/sell orders are recorded by $(\mathscr{F}_t)$-random point processes
 $\{N^{\mathtt{e}/\mathtt{i},\mathtt{b}/\mathtt{s}}_t:t\geq 0\}$ with respective arrival times $\{\tau_k^{\mathtt{e}/\mathtt{i},\mathtt{b}/\mathtt{s}}:k=1,2,\cdots\}$. We denote by $\{J^{\mathtt{e}/\mathtt{i},\mathtt{b}/\mathtt{s}}_k:k=1,2\cdots\}$ the sequences of price changes (in ticks) resulting from exogenous/induced market buy/sell orders.   For any time $t\geq 0$, the (logarithmic) price $P_t$ is given by
 \beqlb\label{eqn2.01}
 P_t\ar=\ar P_0+
 \sum_{k=1}^{N_t^{\mathtt{e}, \mathtt{b}}}\delta\cdot J_k^{\mathtt{e},  \mathtt{b}}
 -\sum_{k=1}^{N_t^{\mathtt{e}, \mathtt{s}}}\delta \cdot J_k^{\mathtt{e}, \mathtt{s}}
 +\sum_{k=1}^{N_t^{\mathtt{i}, \mathtt{b}}}\delta\cdot  J_k^{\mathtt{i}, \mathtt{b}}
 -\sum_{k=1}^{N_t^{\mathtt{i}, \mathtt{s}}}\delta \cdot J_k^{\mathtt{i}, \mathtt{s}},
 \eeqlb
 where $P_0$ is the price at time $0$ and $\delta$ is the tick size, i.e. the minimum price movement.

 We now formulate three assumptions on the order flow dynamics that greatly simplify the subsequent analysis but that do not change our main results on the occurrence of jumps and jump-cascades. First, we assume that price increments are conditionally independent.

 \begin{assumption}
 Price changes are described by independent sequences $\{J^{\mathtt{e}/\mathtt{i},\mathtt{b}/\mathtt{s}}_k:k=1,2,\cdots\}$ of i.i.d. $\mathbb{Z}_+$-valued random variables.
 \end{assumption}

Second, we assume that price increments are uncorrelated. 

 \begin{assumption}
 For any $l,l'\in \{\mathtt{e},\mathtt{i}\}$ and $j,j'\in\{\mathtt{b},\mathtt{s}\}$,
 \beqnn
 \mathbf{E}[dN_t^{l,j}\cdot dN^{l',j'}_t|\mathscr{F}_{t-}]=0,\quad \mbox{if }l\neq l' \mbox{\ or } j\neq j'.
 \eeqnn
 \end{assumption}

Third, we assume that exogenous buy/sell orders arrive according to independent Poisson processes.

 \begin{assumption}\label{AssumptionPoisson}
 $\{N^{\mathtt{e},\mathtt{b}}_t:t\geq 0\}$ and $\{N^{\mathtt{e},\mathtt{s}}_t:t\geq 0\}$   are two independent Poisson processes with rates $p_{\mathtt{e}}^\mathtt{b}$ and $p_{\mathtt{e}}^\mathtt{s} = 1-p_{\mathtt{e}}^\mathtt{b}$ respectively.
 \end{assumption}

Induced orders arrive according to Hawkes processes with $\beta$-exponential kernel for some $\beta>0$.  We assume that both past exogenous and past induced orders increase the arrival intensity of induced orders.

 \begin{assumption}\label{AssumptionHawkes}
The processes  $\{N^{\mathtt{i},\mathtt{b}}_t:t\geq 0\}$
 and $\{N^{\mathtt{i},\mathtt{s}}_t:t\geq 0\}$ are marked Hawkes processes
 with intensities $p_{\mathtt{i}}^\mathtt{b} V_{t-}dt$ and $p_{\mathtt{i}}^\mathtt{s} V_{t-}dt$ respectively, where $p_{\mathtt{i}}^\mathtt{b} + p_{\mathtt{i}}^\mathtt{s} = 1$, and where the intensity process $\{V_t:t\geq 0\}$ is given by
 \beqlb\label{eqn2.02}
 V_t\ar=\ar\mu_t+\sum_{ j\in\{\mathtt{b},\mathtt{s}\}}\sum_{k= 1}^{N_t^{\mathtt{e},j}}X_k^{\mathtt{e},j}e^{-\beta(t-\tau^{\mathtt{e},j}_k)}+\sum_{ j\in\{\mathtt{b},\mathtt{s}\}}\sum_{k= 1}^{N_t^{\mathtt{i},j}}X_k^{\mathtt{i},j}e^{-\beta(t-\tau^{\mathtt{i},j}_k)}, \quad t \geq 0.
 \eeqlb
Here,  $\{\mu_t:t\geq 0\}$ is an $\mathscr{F}_0$-measurable functional-valued random variable that represents the impact of all the orders that arrived prior to time $0$ on the arrival rate of future orders, and $\{X_k^{\mathtt{e}/\mathtt{i},\mathtt{b}/\mathtt{s}}:k=1,2,\cdots\}$ are sequences of nonnegative random variables that represent the impact of each order arrival on the intensity process.
 \end{assumption}

We now provide a stochastic integral representation of the price process in terms of marked Hawkes point measures as introduced in Horst and Xu \cite{HorstXu2019b}. To this end, we associate with each order a mark from the mark space $\mathbb{U}=\mathbb{R}\times \mathbb{R}_+$. A mark comprises the amount by which the order changes the price (in ticks) along with its impact on the arrival intensity of future orders. Specifically, associated with the exogenous/induced order arrival process is a sequence $\{\xi^{\mathtt{e/i}}_k:k=1,2,\cdots\}$ of independent and identically distributed random variables, where $\xi^{\mathtt{e/i}}_k:=(\xi^{\mathtt{e/i}}_{k,P},\xi^{\mathtt{e/i}}_{k,V})$. The quantity $\xi^{\mathtt{e/i}}_{k,P}$ specifies the movement of the price caused by the $k$-th exogenous/induced order, and $\xi^{\mathtt{e/i}}_{k,V}$ specifies the contribution to the intensity process. The law of $\xi^{\mathtt{e/i}}_k$ is given by
 \beqlb\label{eqn2.03}
 \nu_\mathtt{e/i}(du)\ar:=\ar   p_{\mathtt{e/i}}^\mathtt{b}\cdot \mathbf{P}\left\{(J^{\mathtt{e/i},\mathtt{b}},X^{\mathtt{e/i},\mathtt{b}})\in (du_1,du_2)\right\}
 +p_{\mathtt{e/i}}^\mathtt{s}\cdot \mathbf{P}\left\{(-J^{\mathtt{e/i},\mathtt{s}},X^{\mathtt{e/i},\mathtt{s}})\in (du_1,du_2)\right\}.
 \eeqlb
Let $\{\tau^{\mathtt{e/i}}_k:k=1,2,\cdots  \}:= \{\tau^{\mathtt{e/i},\mathtt{b}}_i:i=1,2,\cdots  \}\cup\{\tau^{\mathtt{e/i},\mathtt{s}}_i:i=1,2,\cdots  \}$ be the arrival times of exogenous/induced (buy and sell) orders.
In view of Assumption~\ref{AssumptionPoisson}, we can associate with the sequence  $\{(\tau_k^{\mathtt{e}},\xi^{\mathtt{e}}_k):k=1,2,\cdots\}$ a Poisson point measure
 \beqlb\label{eqn2.04}
 N_{\mathtt{e}}(dt,du) := \sum_{k=1}^\infty \mathbf{1}_{\{ \tau_k^{\mathtt{e}}\in dt, \xi^{\mathtt{e}}_k\in du  \}}
 \eeqlb
 on $[0,\infty)\times \mathbb{U}$ with intensity $dt\nu_\mathtt{e}(du)$. Likewise, associated with the sequence $\{(\tau_k^{\mathtt{i}},\xi^{\mathtt{i}}_k):k=1,2,\cdots\}$ is an $(\mathscr{F}_t)$-random point measure
 \beqlb\label{eqn2.05}
 N_{H}(dt,du) := \sum_{k=1}^\infty \mathbf{1}_{\{ \tau_k^{\mathtt{i}}\in dt, \xi^{\mathtt{i}}_k\in du  \}}
 \eeqlb
 on $[0,\infty)\times \mathbb{U}$ with intensity $V_{t-}dt\nu_\mathtt{i}(du)$. From Horst and Xu \cite{HorstXu2019b}, we know that $N_{H}(dt,du)$ is a \textit{marked Hawkes point measure with homogeneous immigration}. In particular, on an extension of the original probability space we can define a time-homogeneous Poisson random measure $N_\mathtt{i}(ds,du,dz)$  on $(0,\infty)\times \mathbb{U}\times \mathbb{R}_+$ with intensity $ds\nu_\mathtt{i}(du)dz$ that is independent of $N_\mathtt{e}(ds,du)$ and satisfies
 \beqlb\label{eqn2.07}
 \int_0^t \int_{\mathbb{U}} f(u)  N_H(ds,du)\ar=\ar \int_0^t \int_{\mathbb{U}}\int_0^{V_{s-}} f(u)N_\mathtt{i}(ds,du,dx),\quad f\in B(\mathbb{U}),
 \eeqlb
 {where $B(\mathbb{U})$ is the collection of bounded functions on $\mathbb{U}$.}
 As a result, the price process $\{ P_t:t\geq 0  \}$ and the intensity process $\{ V_t:t\geq  0 \}$  can be represented as
 \beqlb
 P_t\ar=\ar P_0 + \int_0^t \int_\mathbb{U}  \delta\cdot u_1  N_\mathtt{e}(ds,du) + \int_0^t \int_{\mathbb{U}}\int_0^{V_{s-}}\delta\cdot u_1  N_\mathtt{i}(ds,du,dx),\label{eqn2.08} \\
 V_t\ar=\ar \mu_t + \int_0^t \int_\mathbb{U}  u_2 \cdot e^{-\beta(t-s)}N_\mathtt{e}(ds,du) + \int_0^t \int_{\mathbb{U}}\int_0^{V_{s-}}u_2 \cdot e^{-\beta(t-s)} N_\mathtt{i}(ds,du,dx). \label{eqn2.09}
 \eeqlb

The preceding integral representation can be further simplified if we assumes that the impact of all the orders that arrived prior to time $0$ on the arrival rate of future events decreases exponentially.

 \begin{assumption}
The process $\{\mu_t : t \geq 0\}$ satisfies $\mu_t:=V_0 \cdot e^{-\beta t}$ for any $t \geq 0$.
 \end{assumption}

 Under the above assumptions, and using the fact that  $ e^{-\beta t}= 1-\int_0^t  \beta e^{-\beta s}ds$,
the model (\ref{eqn2.08})-(\ref{eqn2.09}) can be rewritten into
 \beqlb
 P_t\ar=\ar P_0+\int_0^t \int_{\mathbb{U}}\delta\cdot  u_1N_{\mathtt{e}}(ds,du)+\int_0^t \int_{\mathbb{U}} \int_0^{V_{s-}}\delta\cdot  u_1N_\mathtt{i}(ds,du,dx),\label{eqn2.10}\\
 V_t\ar=\ar V_0-\int_0^t \beta V_sds+\int_0^t\int_{\mathbb{U}} u_2N_{\mathtt{e}}(ds,du)+\int_0^t\int_{\mathbb{U}} \int_0^{V_{s-}} u_2N_\mathtt{i}(ds,du,dx).\label{eqn2.11}
 \eeqlb
 By Theorem~6.2 in \cite{DawsonLi2006}, there exists a unique $\mathbb{U}$-valued strong solution $\{ (P_t,V_t):t\geq 0  \}$
 to (\ref{eqn2.10})-(\ref{eqn2.11}). We call this solution \textit{Hawkes market model} with parameter $(\delta,\beta;\nu_{\mathtt{e}/\mathtt{i}})$. The solution is a strong Markov process whose infinitesimal generator
  $\mathcal{A}_\delta$ acts on any function $f\in C^2(\mathbb{U})$ according to
 \beqlb\label{eqn2.12}
 \mathcal{A}_\delta f(p,v)
 \ar=\ar
 -v\cdot\beta\frac{\partial }{\partial v}f(p,v)
 +v\cdot\int_{\mathbb{U}}\big[f(p+\delta\cdot u_1,v+u_2)-f(p,v)\big]\nu_\mathtt{i}(du) \cr
 \ar\ar
 +\int_{\mathbb{U}}\big[f(p+\delta\cdot u_1,v+u_2)-f(p,v)\big]\nu_\mathtt{e}(du).
 \eeqlb


 \subsection{Examples}

We now consider three specific examples. Within each example we clarify when co-jumps in prices and volatilities are negatively correlated. We revisit all three examples when anlayzing scaling limits.

 We say that an $(\mathbb{Z}_+\times\mathbb{R}_+) $-valued random variable $\xi=(\xi_1,\xi_2)$
 has {\it bivariate exponential distribution} ${\rm BVE}(\boldsymbol{\lambda})$
 with parameter $\boldsymbol{\lambda}:=(\lambda_1,\lambda_2,\lambda_{12})\in\mathbb{R}_+^3$
 if for any  $(k,x)\in \mathbb{Z}_+\times\mathbb{R}_+$,
 \beqnn
 \mathbf{P}\{\xi_1\geq k, \xi_2\geq x\}= \exp\{-\lambda_1 (k-1)-\lambda_2 x-\lambda_{12}((k-1)\vee x)\}.
 \eeqnn
 The first moment $\mathrm{M}^\mathbf{e}_{1}(\boldsymbol{\lambda}):=(\mathrm{M}^\mathbf{e}_{1,k}(\boldsymbol{\lambda}))_{k=1,2}$ and the second moment $\mathrm{M}^\mathbf{e}_{2}(\boldsymbol{\lambda}):=(\mathrm{M}^\mathbf{e}_{2,jk}(\boldsymbol{\lambda}))_{j,k=1,2}$ are given by
 \beqlb
 \mathrm{M}^\mathbf{e}_{1,1}(\boldsymbol{\lambda}):= \mathbf{E}[\xi_1]= \frac{1}{1-e^{-\lambda_1-\lambda_{12}}},
 \quad
 \mathrm{M}^\mathbf{e}_{1,2}(\boldsymbol{\lambda}):= \mathbf{E}[\xi_2]= \frac{1}{\lambda_2+\lambda_{12}}
 \eeqlb
 and
 \beqlb
 \ar\ar\mathrm{M}^\mathbf{e}_{2,11}(\boldsymbol{\lambda}):= \mathbf{E}[|\xi_1|^2]= \frac{e^{-\lambda_1-\lambda_{12}}+(2e^{-\lambda_1-\lambda_{12}}-1)^3}{(1-e^{-\lambda_1-\lambda_{12}})^2},
 \quad
 \mathrm{M}^\mathbf{e}_{2,22}(\boldsymbol{\lambda}):= \mathbf{E}[|\xi_2|^2]=\frac{2}{\lambda_2+\lambda_{12}},\cr
 \ar\ar\mathrm{M}^\mathbf{e}_{2,12}(\boldsymbol{\lambda}):= \mathbf{E}[\xi_1\xi_2]
 =\frac{1/\lambda_2}{1-e^{-(\lambda_1+\lambda_{12})}}
 + \frac{1/\lambda_2-1/(\lambda_2+\lambda_{12})}{1-e^{-(\lambda_1+\lambda_{12}+\lambda_{12})}}.
 \eeqlb
This implies that
 \beqlb\label{eqn2.26}
 \mathrm{C}_\mathbf{e} (\boldsymbol{\lambda}):={\rm Cov}(\xi_1,\xi_2)
 \ar=\ar \frac{1/\lambda_2-1/(\lambda_2+\lambda_{12})}{1-e^{-(\lambda_1+\lambda_{12})}}
 + \frac{1/\lambda_2-1/(\lambda_2+\lambda_{12})}{1-e^{-(\lambda_1+\lambda_{12}+\lambda_{12})}} \geq 0
 \eeqlb
 with equality if $\lambda_{12}=0$, which holds if and only $\xi_1$ and $\xi_2$ are independent.

 \begin{example}[\rm Exponential market model]\label{Ecample-Exp}
 For $j\in\{\mathtt{e},\mathtt{i} \}$ and some constants $\boldsymbol{\lambda}^\mathtt{b}_j,\boldsymbol{\lambda}^\mathtt{s}_j\in\mathbb{R}_+^3$, let
 \beqlb\label{eqn2.28}
 (J^{j,\mathtt{b}},X^{j,\mathtt{b}}) \overset{\rm d}= {\rm BVE}(\boldsymbol{\lambda}_j^\mathtt{b})
 \quad\mbox{and}\quad
 (J^{j,\mathtt{s}},X^{j,\mathtt{s}}) \overset{\rm d}={\rm BVE}(\boldsymbol{\lambda}_j^\mathtt{s}).
 \eeqlb
 In this case, we call the market model (\ref{eqn2.10})-(\ref{eqn2.11})  {\rm exponential market model} with parameter
 $(\delta,\beta, p^{\mathtt{b}/\mathtt{s}}_{\mathtt{e}/\mathtt{i}}; \boldsymbol{\lambda}^{\mathtt{b}/\mathtt{s}}_{\mathtt{e}/\mathtt{i}})$.
 From (\ref{eqn2.26}), the jumps in prices and volatilities are negatively correlated
 if $p^\mathtt{b}_j\mathrm{C}_\mathbf{e}(\boldsymbol{\lambda}^\mathtt{b}_j)< p^\mathtt{s}_j\mathrm{C}_\mathbf{e}(\boldsymbol{\lambda}^\mathtt{s}_j)$
 for $j\in\{\mathtt{e},\mathtt{i} \}$.
 \end{example}

 We say that an $(\mathbb{Z}_+\times\mathbb{R}_+) $-valued random variable $\xi=(\xi_1,\xi_2)$ has a {\it bivariate Pareto distribution} $\mathcal{P}(\alpha,\boldsymbol{\theta})$ with parameters $\alpha>0$ and $\boldsymbol{\theta}=(\theta_1,\theta_2)\in (0,\infty)^2$ if
 for any $(k,x)\in \mathbb{Z}_+\times\mathbb{R}_+ $,
 \beqnn
 \mathbf{P}\{\xi_1\geq k,\xi_2\geq x\}=\Big( 1+\frac{k-1}{\theta_{1}} +\frac{x}{\theta_{2}}\Big)^{-\alpha}.
 \eeqnn
 The probability law of $\xi=(\xi_1,\xi_2)$ is multivariate regularly varying with index $\alpha$ and $\mathbf{E}[\|\xi\|^\kappa]<\infty$, for any $\kappa<\alpha$.
 When $\alpha>2$,
 the first moment $\mathrm{M}^\mathbf{p}_{1}(\alpha,\boldsymbol{\theta}):=(\mathrm{M}^\mathbf{p}_{1,k}(\alpha,\boldsymbol{\theta}))_{k=1,2}$
 and the second moment $\mathrm{M}^\mathbf{p}_{2}(\alpha,\boldsymbol{\theta}):=(\mathrm{M}^\mathbf{p}_{2,jk}(\alpha,\boldsymbol{\theta}))_{j,k=1,2}$ have the following representation:
 \beqlb
 \mathrm{M}^\mathbf{p}_{1,1}(\alpha,\boldsymbol{\theta}):= \mathbf{E}[\xi_1]= \sum_{k=0}^\infty  \Big( 1+\frac{k}{\theta_{1}}\Big)^{-\alpha},
 \quad
 \mathrm{M}^\mathbf{p}_{1,2}(\alpha,\boldsymbol{\theta}):= \mathbf{E}[\xi_2]= \frac{\theta_2}{\alpha-1}
 \eeqlb
 and
 \beqlb
 \mathrm{M}^\mathbf{p}_{2,11}(\alpha,\boldsymbol{\theta})\ar:=\ar \mathbf{E}[|\xi_1|^2]
 = \sum_{k=0}^\infty  (2k+1)\Big( 1+\frac{k}{\theta_{1}}\Big)^{-\alpha},
 \quad
 \mathrm{M}^\mathbf{p}_{2,22}(\alpha,\boldsymbol{\theta}):= \mathbf{E}[|\xi_2|^2]
 =\frac{2\theta_2^2}{(\alpha-1)(\alpha-2)},\cr
 \mathrm{M}^\mathbf{p}_{2,12}(\alpha,\boldsymbol{\theta})\ar:=\ar \mathbf{E}[\xi_1\xi_2]
 =\frac{\theta_2}{\alpha-1}\sum_{k=0}^\infty \Big( 1+\frac{k}{\theta_{1}}\Big)^{-\alpha+1}.
 \eeqlb
This implies that
 \beqlb\label{eqn2.27}
 \mathrm{C}_\mathbf{p}(\alpha,\boldsymbol{\theta}):={\rm Cov}(\xi_1,\xi_2)
 = \frac{\theta_2}{\alpha-1}\sum_{k=0}^\infty \frac{k}{\theta_{1}}\Big( 1+\frac{k}{\theta_{1}}\Big)^{-\alpha}> 0.
 \eeqlb

 \begin{example}[\rm Pareto market model]\label{Ecample-Pareto}
 For $j\in\{\mathtt{e},\mathtt{i} \}$ and some constants $\alpha_j>0$, $\boldsymbol{\theta}_j^\mathtt{b}, \boldsymbol{\theta}_j^\mathtt{s}\in (0,\infty)^2$,
let
 \beqnn
 (J^{j,\mathtt{b}},X^{j,\mathtt{b}}) \overset{\rm d}= \mathcal{P}(\alpha_j,\boldsymbol{\theta}^\mathtt{b}_j)
 \quad\mbox{and}\quad
 (J^{j,\mathtt{s}},X^{j,\mathtt{s}}) \overset{\rm d}= \mathcal{P}(\alpha_j,\boldsymbol{\theta}^\mathtt{s}_j).
 \eeqnn
 In this case, we call the market model (\ref{eqn2.10})-(\ref{eqn2.11})  {\rm Pareto market model} with parameter
 $(\delta,\beta, p^{\mathtt{b}/\mathtt{s}}_{\mathtt{e}/\mathtt{i}}; \alpha_{\mathtt{e}/\mathtt{i}},\boldsymbol{\theta}^{\mathtt{b}/\mathtt{s}}_{\mathtt{e}/\mathtt{i}})$.
 From (\ref{eqn2.27}), when $\alpha_{\mathtt{e}/\mathtt{i}}>2$, the jumps in prices and volatilities are negatively correlated if $p^\mathtt{b}_j\mathrm{C}_\mathbf{p}(\alpha_j,\boldsymbol{\theta}^\mathtt{b}_j)< p^\mathtt{s}_j\mathrm{C}_\mathbf{p}(\alpha_j,\boldsymbol{\theta}^\mathtt{s}_j)$ for all $j\in\{\mathtt{e},\mathtt{i} \}$.
 \end{example}

 We say that a $(\mathbb{Z}_+\times\mathbb{R}_+) $-valued random variable $\xi=(\xi_1,\xi_2)$ has an exponential-Pareto mixing distribution with parameter $(\boldsymbol{\lambda}, \alpha, \boldsymbol{\theta})$ if for any $(k,x)\in \mathbb{Z}_+\times\mathbb{R}_+ $, and some $q \in (0,1)$
 \beqnn
 \mathbf{P}\{\xi_1\geq k,\xi_2\geq x\}=
 q \Big( 1+\frac{k-1}{\theta_{1}} +\frac{x}{\theta_{2}}\Big)^{-\alpha}
+(1-q) \exp\{-\lambda_1 (k-1)-\lambda_2 x-\lambda_{12}((k-1)\vee x)\}.
 \eeqnn

 \begin{example}[\rm Exponential-Pareto mixing market model] \label{Ecample-Exp-Pareto}
 If the mark of each event has an exponential-Pareto mixing distribution, then we call the market model (\ref{eqn2.10})-(\ref{eqn2.11})
 {\rm exponential-Pareto mixing market model} with parameter
 $(\delta,\beta, p^{\mathtt{b}/\mathtt{s}}_{\mathtt{e}/\mathtt{i}}; \boldsymbol{\lambda}^{\mathtt{b}/\mathtt{s}}_{\mathtt{e}/\mathtt{i}};\alpha_{\mathtt{e}/\mathtt{i}},\boldsymbol{\theta}^{\mathtt{b}/\mathtt{s}}_{\mathtt{e}/\mathtt{i}})$
 and selecting mechanism {$q_{\mathtt{e}/\mathtt{i}}$}.
 \end{example}


\section{The genealogy of market dynamics}\label{genealogy}

In this section, we analyze the genealogical structure of the market dynamics and establish a representation of the dynamics in terms of independent and identically distributed {\sl self-enclosed} sub-models. Self-enclosed sub-models correspond to market models with no exogenous orders, except initial ones; they describe the impact of exogenous shocks on induced orders. Specifically, we call our market model self-enclosed if $\nu_\mathtt{e}(\mathbb{U})=0$; in this case, we denote the model by $\{ (P_{0,t},V_{0,t}):t\geq 0\}$. In view of (\ref{eqn2.10})-(\ref{eqn2.11}) it satisfies the following dynamics:
 \beqlb
 P_{0,t}\ar=\ar P_0+\int_0^t \int_{\mathbb{U}} \int_0^{V_{0,s-}}\delta\cdot  u_1N_\mathtt{i}(ds,du,dx),\label{eqn2.17}\\
 V_{0,t}\ar=\ar V_0-\int_0^t \beta V_{0,s}ds +\int_0^t\int_{\mathbb{U}} \int_0^{V_{0,s-}} u_2N_\mathtt{i}(ds,du,dx).\label{eqn2.18}
 \eeqlb
From the cluster representation of Hawkes process, we can decompose {the Hawkes market model} into a sum of  self-enclosed models
 $\{(P_{k,t},V_{k,t}):t\geq 0\}_{k\geq 1}$. Associated to the sequences of arrival times and marks $\{ (\tau^{\mathtt{e}}_k, \xi_k^{\mathtt{e}}): k=1,2,\cdots  \}$, these sub-models satisfy $(P_{k,t},V_{k,t})=(0,0)$ if $t<\tau_k^{\mathtt{e}}$, and for $t\geq \tau^{\mathtt{e}}_k$,
 \beqlb
 P_{k,t}\ar=\ar \xi_{k,P}^{\mathtt{e}}+\int_{\tau^{\mathtt{e}}_k}^t \int_{\mathbb{U}} \int_{ \sum_{j=0}^{k-1}V_{i,s-}}^{\sum_{j=0}^{k-1}V_{i,s-}+V_{k,s-}}\delta\cdot  u_1N_\mathtt{i}(ds,du,dx),\label{eqn2.24}\\
 V_{k,t}\ar=\ar \xi_{k,V}^{\mathtt{e}}-\int_{\tau^{\mathtt{e}}_k}^t \beta V_{k,s}ds +\int_{\tau^{\mathtt{e}}_k}^t \int_{\mathbb{U}} \int_{\sum_{j=0}^{k-1}V_{i,s-}}^{\sum_{j=0}^{k-1}V_{i,s-}+V_{k,s-}} u_2N_\mathtt{i}(ds,du,dx).\label{eqn2.25}
 \eeqlb

 \begin{theorem}\label{Thm206}
 The Hawkes market model $\{(P_t,V_t):t\geq 0\}$ defined by (\ref{eqn2.10})-(\ref{eqn2.11}) admits the following decomposition:
 \beqnn
 \{ (P_t,V_t):t\geq 0  \} \overset{\rm a.s.}= \Big\{ \sum_{k= 0}^\infty (P_{k,t},V_{k,t})  :t\geq 0 \Big\}.
 \eeqnn
 \end{theorem}
 \proof
 It suffices to prove that the infinite sum is well defined and equals $\{ (P_t,V_t):t\geq 0  \}$. For any $K\geq 1$, let $(P^K_{t},V_{t}^K):=\sum_{k= 0}^K (P_{k,t},V_{k,t})$, which solves
 \beqnn
 P^K_{t}\ar=\ar P_0+ \sum_{k=1}^K \xi_{k,P}^{\mathtt{e}}\mathbf{1}_{\{t\geq \tau^{\mathtt{e}}_{k}\}}+\int_{\tau^{\mathtt{e}}_k}^t \int_{\mathbb{U}} \int_0^{V^K_{s-}}\delta\cdot  u_1N_\mathtt{i}(ds,du,dx),\\
 V^K_{t}\ar=\ar V_0 +\sum_{k=1}^K \xi_{k,V}^{\mathtt{e}}\mathbf{1}_{\{t\geq \tau^{\mathtt{e}}_{k}\}}-\int_{\tau^{\mathtt{e}}_k}^t \beta V^K_{s}ds +\int_{\tau^{\mathtt{e}}_k}^t \int_{\mathbb{U}} \int_0^{V^K_{s-}} u_2N_\mathtt{i}(ds,du,dx).\\
 \eeqnn
 For any $T\geq 0$, the set $\{k \in \mathbb N : \tau^{\mathtt{e}}_{k}\leq T\}$ is a.s.~finite. Hence $\{(P^K_{t},V_{t}^K): t\in[0,T]\}\overset{\rm a.s.}\to \{(P_{t},V_{t}): t\in[0,T]\}$ as $K\to\infty$. \hfill  \qed

The sub-models $\{(P_{k,t},V_{k,t}):t\geq 0\}_{k\geq 0}$ are self-enclosed, mutually independent and identically distributed:
for any $j,k\geq 1$,
 \beqnn
 \{(P_{j,t+\tau^{\mathtt{e}}_j},V_{j,t+\tau^{\mathtt{e}}_j}):t\geq 0\}\overset{\rm d}= \{(P_{k,t+\tau^{\mathtt{e}}_k},V_{k,t+\tau^{\mathtt{e}}_k}):t\geq 0\}.
 \eeqnn
 Conditioned on $(\xi_{1,P}^{\mathtt{e}},\xi_{1,V}^{\mathtt{e}})=(P_0,V_0)$, the model $\{(P_{1,t+\tau^{\mathtt{e}}_1},V_{1,t+\tau^{\mathtt{e}}_1}):t\geq 0\}$ equals $\{(P_{0,t},V_{0,t}):t\geq 0\}$ in law. As a result, the impact of exogenous orders on the market dynamics can be analyzed by analyzing the model $\{(P_{0,t},V_{0,t}):t\geq 0\}$.  By \cite[Theorem 1.1]{KawazuWatanabe1971} the volatility process $\{V_{0,t}:t\geq 0 \}$ is a continuous-state branching process. Arguments given by Grey \cite{Grey1974} show that it tends to either $0$ or $\infty$ as $t \to \infty$. We show that this implies that the price process $\{P_{0,t}:t\geq 0 \}$ either settles down or fluctuates strongly as $t \to \infty$.

\begin{remark}
The case {$V_{0,t} \to 0$} is economically very intuitive. In the absence of exogenous shocks it seems reasonable to assume that prices settle down in the {long run}.
\end{remark}

In order to study the long-run behavior of the price process $\{P_{0,t}:t\geq 0 \}$, we denote by $\mathcal{T}_0$  its last jump time and for $0\leq a\leq b \leq\infty$ we denote the number of jumps in the time interval $[a,b)$ by $\mathcal{J}_a^b$. That is,
 \beqnn
 \mathcal{T}_0:= \sup\{t\in[0,\infty): |P_{0,t}-P_{0,t-}|>0\}\quad \mbox{and}\quad \mathcal{J}_a^b:=\# \{t\in[a,b): |P_{0,t}-P_{0,t-}|>0\}.
 \eeqnn
The following theorem analyzes the joint distribution of $( \mathcal{T}_0,  \mathcal{J}^\infty_0)$ in terms of the function
\beqnn
g_0(x)\ar :=\ar \beta x+  \int_\mathbb{U} [e^{-x\cdot u_2 }-1]\nu_\mathtt{i}(du),\quad x\geq 0.
\eeqnn

 \begin{theorem}
 For any $\lambda, t\geq 0$,
 \beqlb\label{eqn2.15}
 \mathbf{E}\big[e^{-\lambda \cdot\mathcal{J}_0^\infty }, \mathcal{T}_0\leq  t \big]  =  \exp\big\{- \phi_t(\lambda)\cdot V_0 \big\},
 \eeqlb
 where $t\mapsto\phi_t(\lambda)$ solves the Riccatti equation
 \beqlb\label{eqn2.16}
 \phi_t(\lambda)\ar=\ar \frac{1}{\beta} -\int_0^t g_0(\phi_s(\lambda)) ds -(e^{-\lambda}-1) \int_0^t ds \int_\mathbb{U}  \exp\{ -\phi_s(\lambda)\cdot u_2\} \nu_\mathtt{i}(du).
 \eeqlb
 \end{theorem}
 \proof
 From (\ref{eqn2.17}), we have
 \beqnn
 \mathbf{E}\big[e^{-\lambda\cdot \mathcal{J}_0^\infty }, \mathcal{T}_0\leq t \big] \ar=\ar  \mathbf{E} \Big[\exp\Big\{  -\lambda \int_0^t \int_{\mathbb{U}} \int_0^{V_{0,s-} }  N_\mathtt{i}(ds,du,dz)\Big\}, \int_t^\infty \int_{\mathbb{U}} \int_0^{V_{0,s-} }  N_\mathtt{i}(ds,du,dz)=0 \Big] \cr
 \ar=\ar  \mathbf{E} \Big[\exp\Big\{  -\lambda \int_0^t \int_{\mathbb{U}} \int_0^{V_{0,s-} }  N_\mathtt{i}(ds,du,dz)\Big\}, \mathbf{P}_{\mathscr{F}_t}\Big\{\int_t^\infty \int_{\mathbb{U}} \int_0^{V_{0,s-} }  N_\mathtt{i}(ds,du,dz)=0 \Big\} \Big].
 \eeqnn
 From the properties of generalized Poisson processes, we have
 \beqnn
 \mathbf{P}_{\mathscr{F}_t}\Big\{\int_t^\infty \int_{\mathbb{U}} \int_0^{V_{0,s-} }  N_\mathtt{i}(ds,du,dz)=0 \Big\} \ar=\ar \mathbf{E}_{\mathscr{F}_t}\Big[ \exp\Big\{ -\int_t^\infty V_{0,s} ds \Big\} ; \mathcal{T}_0\leq t \Big].
 \eeqnn
 Moreover, conditioned on $\{\mathcal{T}_0\leq t\}$, we have $V_{0,s}= V_{0,t}e^{-\beta(s-t)}$ for any $s\geq t$, and
 \beqnn
 \mathbf{E}_{\mathscr{F}_t}\Big[ \exp\Big\{ -\int_t^\infty V_{0,s} ds \Big\} ; \mathcal{T}_0\leq t \Big]= \mathbf{E}_{\mathscr{F}_t}\big[ \exp\big\{ -V_{0,t}/\beta\big\}  \big].
 \eeqnn
 Putting all the above results together, we have
 \beqnn
 \mathbf{E}\big[ e^{-\lambda\cdot\mathcal{J}_0^\infty}, \mathcal{T}_0\leq t \big] \ar=\ar  \mathbf{E} \Big[\exp\Big\{  -\int_0^t \int_{\mathbb{U}} \int_0^{V_{0,s} }  \lambda N_\mathtt{i}(ds,du,dz)-V_{0,t}/\beta\Big\} \Big]=  \mathbf{E}\big[e^{-Y_t}\big],
 \eeqnn
 where
 \beqnn
 Y_t= V_0/\beta-\int_0^t  V_{0,s}ds +\int_0^t\int_{\mathbb{U}} \int_0^{V_{0,s-}}\big( \lambda+ u_2/\beta \big)N_\mathtt{i}(ds,du,dz).
 \eeqnn
 Applying It\^o's formula to $\exp\{-Y_t-(\phi_{t-s}(\lambda)-1/\beta)V_{0,t}\}$, we have
 \beqnn
 e^{-Y_t}
 \ar=\ar
 e^{-\phi_{t}(\lambda) V_0} +\int_0^t\int_{\mathbb{U}} \int_0^{V_{0,s-}}
 e^{-Y_{s-}-(\phi_{t-s}(\lambda)-1/\beta)V_{0,s-}}\big[e^{-( \frac{u_2}{\beta}+\lambda) -(\phi_{t-s}(\lambda)-1/\beta)u_2} -1\big]\tilde{N}_\mathtt{i}(ds,du,dz) ,
 \eeqnn
 where $\tilde{N}_\mathtt{i}(ds,du,dz):= N_\mathtt{i}(ds,du,dz)-ds\nu_\mathtt{i}(du)dz$.
 Taking expectations on both sides yields (\ref{eqn2.15}).
 \qed

 \begin{corollary}\label{Thm207}
 We have $\mathcal{T}_0<\infty$ if and only if $|\mathcal{J}_0^\infty|<\infty$. Moreover,  for any $\lambda\geq 0$, the limit $\phi_\infty(\lambda):=\lim_{t\to\infty} \phi_t(\lambda)$ exists, is the largest root of the function
 \beqlb\label{eqn2.21}
 g_\lambda(x):= g_0(x)+(e^{-\lambda}-1)\int_\mathbb{U} e^{-x\cdot u_2 }\nu_\mathtt{i}(du), \qquad x \geq 0,
 \eeqlb
and
 \beqlb\label{eqn2.20}
 \mathbf{E}\big[e^{-\lambda \cdot\mathcal{J}_0^\infty }, \mathcal{T}_0<\infty\big]  =  \exp\big\{- V_0 \cdot \phi_\infty(\lambda)\big\}
 \quad \mbox{and}\quad
 \mathbf{P}\{\mathcal{T}_0<\infty \}=\exp\{-V_0 \cdot \phi_\infty(0)\}.
 \eeqlb
 \end{corollary}
  \proof
  	The first result follows directly from the second. Indeed, $|\mathcal{J}_0^\infty|<\infty$ means that only finite jumps happen and so $\mathcal{T}_0<\infty$. Conversely,  from (\ref{eqn2.15}) and the continuity of $\phi_\infty(\lambda)$, we have
  \beqnn
  \mathbf{P}\big\{\mathcal{J}_0^\infty<\infty \big| \mathcal{T}_0<\infty \big\}=\lim_{\lambda\to 0+}	\mathbf{E}\big[e^{-\lambda\cdot\mathcal{J}_0^\infty} \big|  \mathcal{T}_0<\infty\big]  = \lim_{\lambda\to 0+} e^{- (\phi_\infty(\lambda)- \phi_\infty(0)) V_0 }=1.
  \eeqnn

  We now prove the second result. For any $\lambda\geq 0$, the {function} $\{ \phi_t(\lambda):t\geq 0 \}$ is {non-negative} and non-increasing. Hence, the limit $\phi_\infty(\lambda):=\lim_{t\to\infty} \phi_t(\lambda)$ exists.
  Applying the dominated convergence theorem to (\ref{eqn2.15}) yields
  \beqnn
  \mathbf{E}\big[e^{-\lambda\cdot\mathcal{J}_0^\infty}, \mathcal{T}_0<\infty \big]
  =	\lim_{t\to\infty}\mathbf{E}\big[e^{-\lambda\cdot\mathcal{J}_0^\infty}, \mathcal{T}_0\leq  t \big]
  = \exp\big\{- V_0 \cdot \lim_{t\to\infty}\phi_t(\lambda) \big\}
  = \exp\big\{- V_0\cdot \phi_\infty(\lambda) \big\}.
  \eeqnn

  By (\ref{eqn2.16}), the process $\{ \phi_t(\lambda):t\geq 0 \}$ satisfies the semigroup property $\phi_{t+s}(\lambda) =  \phi_t( \phi_s(\lambda))$ for any $s,t\geq 0$.
Hence, for any $t\geq 0$
  \beqnn
  \phi_\infty(\lambda) =\lim_{s\to\infty} \phi_{t+s}(\lambda) =  \lim_{s\to\infty}\phi_t( \phi_s(\lambda))=\phi_t( \phi_\infty(\lambda)).
  \eeqnn
  Taking this back into (\ref{eqn2.16}), we have
  \beqnn
  \int_0^tg_0(\phi_\infty(\lambda))ds+(e^{-\lambda}-1)\int_0^tds\int_\mathbb{U} \exp\{-\phi_\infty(\lambda) \cdot u_2 \}\nu_\mathtt{i}(du)    \equiv 0, 
  \eeqnn
  which means that
  \beqnn
  g_0(\phi_\infty(\lambda)) +  (e^{-\lambda}-1)\int_\mathbb{U} \exp\{ -\phi_\infty(\lambda)\cdot u_2\} \nu_\mathtt{i}(du)=0.
  \eeqnn

 It remains to show that $\phi_\infty(\lambda)$ is the largest root. The function $g_\lambda$ is smooth, strictly convex as $\nu_{\mathtt{i}}$ is supported on $\mathbb Z_+ \times \mathbb R_+$, and  $g_\lambda(x) \to \infty$ if $x \to \infty$. If $\lambda > 0$, then $g_\lambda(0) < 0$ and hence $g_\lambda$ has a unique root. If $\lambda = 0$, then $x=0$ is a root but there is a second one if $g'_0(0) < 0$. In this case, it follows from the continuity of $g_\lambda(x)$ in $(\lambda,x)$ that $\phi_\infty(\lambda)$  decrease to $\phi_\infty(0)$ continuously as $\lambda\to 0+$. This shows that $\phi_\infty(0)$ is the largest root of $g_0$.
  \qed

For the remainder of this section, we always assume that $\int_\mathbb{U}\|u\|\nu_{\mathtt{i}}(du)<\infty$.
 Since $\mathbf{P}\{\mathcal{T}_0<\infty \}=\exp\{- \phi_\infty(0) V_0\}$ we see that $\mathbf{P}\{|\mathcal{J}_0^\infty| =\infty \}= \mathbf{P}\{ \mathcal{T}_0=\infty \}>0$ if and only if $\phi_\infty(0) > 0$.
 The proof of Corollary~\ref{Thm207} shows that this is the case if and only if the function $g_0$ has a strictly positive root and that this is the case if and only if
\begin{equation} \label{beta-tilde}
	\tilde\beta := g'_0(0) = \beta -\int_{\mathbb{U}} u_2\nu_{\mathtt{i}}(du) < 0.
\end{equation}
 Economically, $\tilde \beta$ describes the net decay in the long run of the impact of induced orders on the volatility: the impact of past orders is discounted at a rate $\beta$ while new impact is added at a rate $\int_{\mathbb{U}} u_2\nu_{\mathtt{i}}(du)$.

 \begin{corollary}\label{Thm209}
 We have $\mathbf{P}\{\mathcal{J}_0^\infty =\infty \}= \mathbf{P}\{ \mathcal{T}_0=\infty \}>0$
 if and only if $\tilde\beta<0$.
 In this case, conditioned on $\{ \mathcal{T}_0=\infty \}$,
 the following holds:
 \begin{enumerate}
 \item[(1)] The number of price changes tends to infinity as time increases, i.e.
  $\limsup_{k\to\infty} \mathcal{J}_k^{k+1}\to \infty $ a.s.
  	
 \item[(2)] The price drifts to $\infty$, $-\infty$ or  oscillates
  if and only if  $\int_\mathbb{U}u_1\nu_{\mathtt{i}}(du)>0$, $<0$ or $=0$.
  	
 \end{enumerate}
 \end{corollary}
 \proof
 The first assertion has already been established.
 The second result follows from the fact that the intensity process $V_{0,t}$ tends to either $0$ or $\infty$ a.s.
 Indeed, conditioned on $\{\mathcal{T}_0=\infty\}$, we have $\limsup_{t\to\infty} V_{0,t}>0$
 and hence $V_{0,t}\to \infty$ almost surely.
 Denote by $\tau_1<\tau_2<\cdots$ the jump times of the price process. Then $\tau_k\to\infty$ a.s. as $k\to\infty$ and
it suffices to prove that the random walk $ P_{\tau_k}=P_0+\sum_{j=1}^{k} \eta_{j,P}^{\mathtt{i}}$
 drifts to $\infty$, $-\infty$ or oscillates
 if and only if $\mathbf{E}[\eta_{1,P}^{\mathtt{i}}]>0$, $<0$ or $=0$.
This, however, follows from, e.g.~Theorem~4 in \cite[p.203]{Feller1971}.
 \qed

 \begin{proposition}
 We have $\mathbf{E}[ \mathcal{J}_0^\infty ]<\infty $ if and only if $\tilde\beta>0$.
 In this case, $\mathbf{E}[\mathcal{J}_0^\infty ]=V_0/\tilde\beta$.
 \end{proposition}
  \proof
  Taking expectation on both sides of (\ref{eqn2.18}), we have
  \beqnn
  \mathbf{E}[V_{0,t}]\ar=\ar V_0 -\int_0^t \tilde\beta \cdot \mathbf{E}[V_{0,s}]ds.
  \eeqnn
  Solving this equation, we get $\mathbf{E}[V_{0,t}]= V_0 \cdot e^{-\tilde{\beta}t}$.
  From the definition of $\mathcal{J}_0^\infty$,
  \beqnn
  \mathbf{E}[\mathcal{J}_0^\infty]
  \ar=\ar \mathbf{E}\Big[ \int_0^\infty \int_{\mathbb{U}}\int_0^{V_{0,s-}}N_\mathtt{i}(ds,du,dx) \Big]
  = \int_0^\infty \mathbf{E}[V_{0,s}]ds = V_0 \int_0^\infty e^{-\tilde{\beta}s}ds,
  \eeqnn
  which is finite if and only if $\tilde\beta>0$.
  In this case, we have  $\mathbf{E}[\mathcal{J}_0^\infty]=V_0/\tilde\beta$.
  \qed

 The following corollary provides four regimes for the long-run impact of an exogenous shock of magnitude $(P_0,V_0)$ to the market dynamics.
 As pointed out above, the economically interesting case is $\tilde \beta \geq 0$.
 The critical case $\tilde \beta = 0$ is particularly relevant.
 In this case, the impact of exogenous orders on induced ones is slowly decaying.
 In the scaling limit it corresponds to a loss of mean-reversion of the volatility process.

 \begin{corollary}\label{Thm210}
 There are the four regimes for the long-run impact of exogenous shocks of magnitude $(P_0,V_0)$ on induced order flow.
 \begin{enumerate}
 \item[(1)] If $\tilde\beta <0$, then as $t\to\infty$,
   \beqnn
   \mathbf{P}\{ \mathcal{T}_0\geq  t \} \to 1-e^{- \phi_\infty(0) V_0}\in(0,1).
   \eeqnn

 \item[(2)] If $\tilde\beta >0$,
   then for any $t>0$,
    \beqnn
    \mathbf{P}\{ \mathcal{T}_0\geq  t \} \leq \frac{V_0}{\beta}\cdot e^{- \tilde\beta  \cdot t }.
    \eeqnn

 \item[(3)] If $\tilde\beta = 0$
   and $\nu_\mathtt{i}(|u_2|^2):=\frac{1}{2}\int_{\mathbb{U}}|u_2|^2\nu_{\mathtt{i}}(du)<\infty$,
   then as $t\to\infty$,
   \beqnn
    \mathbf{P}\{  \mathcal{T}_0\geq  t \} \sim \frac{V_0}{\nu_\mathtt{i}(|u_2|^2)}\cdot t^{-1}.
   \eeqnn

 \item[(4)] If $\tilde\beta = 0$
   and $\nu_{\mathtt{i}}(\mathbb{Z}\times [x,\infty))\sim C(1+x)^{-1-\alpha}$ as $x\to \infty$ for some $\alpha\in(0,1)$,
   then as $t\to \infty$,
   \beqnn
    \mathbf{P}\{  \mathcal{T}_0\geq  t \} \sim  C\cdot V_0\cdot t^{-1/\alpha}.
   \eeqnn
   	
 \end{enumerate}
 \end{corollary}
 \proof
 The first regime follows from Corollary~\ref{Thm207} and \ref{Thm209}.
 For other three regimes, {from (\ref{eqn2.15}) we have} 
 
 \beqnn
 \mathbf{P}\{\mathcal{T}_0>t \} =1-e^{-V_0 \phi_t(0)} \sim V_0 \cdot  \phi_t(0).
 \eeqnn
 Thus, it suffices to consider the asymptotic behavior of $\phi_t(0)$ as $t\to\infty$.
 By (\ref{eqn2.16}) with $\lambda=0$, we have
 \beqnn
 \phi_t(0)= \frac{1}{\beta}e^{-\tilde\beta t} - \int_0^t e^{-\tilde\beta (t-s)} ds \int_\mathbb{U} [e^{ -u_2\phi_s(0)}-1+u_2\phi_s(0)]\nu_\mathtt{i}(du).
 \eeqnn
 Thus, (2) follows from the fact that the integral term above is nonnegative.
 We now prove (3).
 If $\int_{\mathbb{U}}|u_2|^2\nu_{\mathtt{i}}(du)<\infty$, then for  $\lambda\to 0+$,
 \beqnn
 \int_\mathbb{U} [e^{ -u_2\lambda}-1+u_2\lambda]\nu_\mathtt{i}(du)\sim  \nu_\mathtt{i}(|u_2|^2)\cdot \lambda^2.
 \eeqnn
 From the monotonicity of $\{\phi_t(0):t\geq 0\}$, for any $\epsilon>0$ there exists $t_0>0$ large enough such that for any $t\geq 0$,
 \beqnn
 \phi_{t_0+t}(0)\ar= \ar \phi_{t_0}(0)  - \int_0^t  ds \int_\mathbb{U} [e^{ -u_2\phi_{t_0+s}(0)}-1+u_2\phi_{t_0+s}(0)]\nu_\mathtt{i}(du)\cr
 \ar\leq\ar \phi_{t_0}(0)  -  [\nu_\mathtt{i}(|u_2|^2)+\epsilon]\cdot\int_0^t |\phi_{t_0+s}(0)|^2 ds .
 \eeqnn
 Applying the nonlinear Gr\"onwall's inequality, we have
 \beqnn
 \frac{1}{\phi_{t_0+t}(0)}-\frac{1}{\phi_{t_0}(0)}\geq  [\nu_\mathtt{i}(|u_2|^2)+\epsilon]\cdot t,
 \eeqnn
 which implies that
 \beqnn
  \phi_{t+t_0}(0)\leq \frac{1}{\frac{1}{\phi_{t_0}(0)}+ [\nu_\mathtt{i}(|u_2|^2)+\epsilon]\cdot t}
  \quad\mbox{and}\quad
  \limsup_{t\to\infty}\ (t+t_0)\phi_{t+t_0}(0)\leq \frac{1}{\nu_\mathtt{i}(|u_2|^2)+\epsilon}.
 \eeqnn
 Similarly, we also have for any $\epsilon\in(0,\nu_\mathtt{i}(|u_2|^2))$,
 \beqnn
  \phi_{t+t_0}(0)\geq \frac{1}{\frac{1}{\phi_{t_0}(0)}+ [\nu_\mathtt{i}(|u_2|^2)-\epsilon]\cdot t}
  \quad\mbox{and}\quad
  \liminf_{t\to\infty}\ (t+t_0)\phi_{t+t_0}(0)\geq \frac{1}{\nu_\mathtt{i}(|u_2|^2)-\epsilon}.
 \eeqnn
 This shows (3) as $\epsilon$ is arbitrary.
 For (4), from \cite[Theorem 8.1.6]{BinghamGoldieTeugels1987}, we have as $\lambda\to 0+$,
 \beqnn
 \int_\mathbb{U} [e^{ -u_2\lambda}-1+u_2\lambda]\nu_\mathtt{i}(du)
 \sim C\lambda^{\alpha+1}.
 \eeqnn
  As before, we also have as $t\to \infty$,
  \beqnn
  \phi_t(0)\sim Ct^{-1/\alpha}.
  \eeqnn
  \qed

 \section{The scaling limit}\label{HFHM}
 \setcounter{equation}{0}

 In this section, we consider the weak convergence of a sequence of rescaled market models. Our scaling limit provides a microscopic foundation for an array of stochastic volatility models including the classic Heston model, the Heston model with jumps, the jump-diffusion stochastic volatility model in \cite{DuffiePanSingleton2000} and the alpha Heston model in \cite{JiaoMaScottiZhou2018}.

 \subsection{Assumptions and asymptotic results}

 For each $n \in \mathbb{N}$, we consider a market model $\{(P_t^{(n)},V_t^{(n)}):t\geq 0\}$ of the form (\ref{eqn2.10})-(\ref{eqn2.11}) with initial state $(P_0^{(n)},V_0^{(n)})$ and parameter $(1/n,\beta^{(n)}; \nu_{\mathtt{e}/\mathtt{i}}^{(n)})$.
 Without loss of generality, we assume that all models are defined on the common filtered probability space $(\Omega,\mathscr{F}, \mathscr{F}_t,\mathbf{P})$.
 We are interested in the weak convergence of the rescaled market models $\{(P^{(n)}(t), V^{(n)}(t)):t\geq 0 \}$ defined by
 \beqlb\label{eqn3.02}
 \Big(\begin{array}{c}P^{(n)}(t)\cr V^{(n)}(t) \end{array}  \Big) \ar=\ar
 \Big(\begin{array}{c}P^{(n)}(0)\cr V^{(n)}(0) \end{array}  \Big) -\int_0^t \Big(\begin{array}{c} 0\cr \gamma_n\beta^{(n)}\end{array}  \Big)  V^{(n)}(s)ds
 +\int_0^t\int_{\mathbb{U}} \frac{u}{n}N_{\mathtt{e}}^{(n)}(d\gamma_ns,du) \cr
 \ar\ar +\int_0^t\int_{\mathbb{U}} \int_0^{V^{(n)}(s-)} \frac{u}{n}N^{(n)}_\mathtt{i}(d\gamma_ns,du,dnx)
 \eeqlb
where $\{ \gamma_n \}_{n\geq 1}$ is a sequence of positive numbers that converges to infinity as $n\to\infty$. The market models are driven by the Poisson random measures $N_\mathtt{e}^{(n)}$ and $N_\mathtt{i}^{(n)}$ whose respective compensators are given by
 \[
 	\hat{N}_\mathtt{e}^{(n)}(d\gamma_ns,du):=\gamma_n ds\nu^{(n)}_\mathtt{e}(du) \quad \mbox{and} \quad
	\hat{N}_\mathtt{i}^{(n)}(d\gamma_ns,du,dnx):=n\gamma_n ds\nu^{(n)}_\mathtt{i}(du)dx.
\]
In particular, the arrival rate of exogenous orders in the $n$-th market model is $\gamma_n$, and the arrival rate of induced orders is $n\gamma_n$\footnote{Since order sizes are scaled by a factor $\frac{1}{n}$, this suggests that exogenous orders will not generate a diffusive behavior in the limit; they only generate a drift and/or jumps. Induced orders, on the other hand, may well generate diffusive behavior.}. This justifies our interpretation of induced orders as originating from high-frequency trading.

The third term on the right side of equation (\ref{eqn3.02}) is a two-dimensional, compound Poisson process. Its weak convergence has been extensively studied in the literature; see \cite[Chapter VII]{JacodShiryaev2003} or \cite[Corollary 15.20]{Kallenberg2002}. 
The following condition is necessary for its weak convergence.

 \begin{assumption}\label{MainCondition00}
 There exists a constant $\gamma\in [0,\infty)$ such that $ \gamma_n\sim \gamma \cdot n$ as $n\to\infty$\footnote{We emphasize that $\gamma =0$ is allowed. In Section 3 we consider the case $\gamma_n=n$ as well as the case $\gamma_n = n^\alpha$ for some $\alpha \in (0,1)$.}.
 \end{assumption}

 We now give sufficient conditions on the parameters $\beta^{(n)}$ and $\nu_{\mathtt{e}/\mathtt{i}}^{(n)}$ that guarantee the existence of a non-degenerate scaling limit. Our arguments are based on the link between the benchmark model and general branching particle systems. By (\ref{eqn2.12}) the infinitesimal generator $\mathcal{A}^{(n)}$ of $\{(P^{(n)}(t),V^{(n)}(t)):t\geq 0\}$ acts on functions $f\in C^2(\mathbb{U})$ according to,
 \beqlb\label{eqn3.01}
 \mathcal{A}^{(n)} f(p,v)
 \ar=\ar
 -v\cdot\gamma_n\beta^{(n)}\frac{\partial }{\partial v}f(p,v)
 +v\cdot n\gamma_n\int_{\mathbb{U}}\big[f((p,v)+u/n)-f(p,v)\big]\nu^{(n)}_\mathtt{i}(du) \cr
 \ar\ar
 +\gamma_n\int_{\mathbb{U}}\big[f((p,v)+u/n)-f(p,v)\big]\nu^{(n)}_\mathtt{e}(du).
 \eeqlb
By Corollary~8.9 in \cite[p.232]{EthierKurtz1986}, the sequence of $\{(P^{(n)}(t),V^{(n)}(t)):t\geq 0\}$ is weakly convergent if there exits an infinitesimal generator $\mathcal{A}$ with domain $\mathscr{D}(\mathcal{A})$ such that for any $f\in \mathscr{D}(\mathcal{A})$,
 \beqlb\label{eqn3.28}
 \mathcal{A}^{(n)} f(p,v)\to \mathcal{A} f(p,v),\quad \mbox{as }n\to\infty.
 \eeqlb
 In what follows, we identify a candidate limit generator by analyzing the limit of the sequence $\mathcal{A}^{(n)}f$ for the special case where $f$ is the exponential function. To this end, let $\mathbb{U}_*:= \mathrm{i}\mathbb{R}\times \mathbb{C}_-$ with $ \mathrm{i}\mathbb{R}:=\{\mathrm{i}\cdot x:\ x\in\mathbb{R}\} $ and $\mathbb{C}_-:=\{x+\mathrm{i}\cdot y:\ (x,y)\in\mathbb{R}_+\times\mathbb{R}\}$, where  $\mathrm{i} := \sqrt{-1}$.
 For any $z=(z_1,z_2)\in \mathbb{U}_*$,
 \beqlb\label{eqn3.27}
 \mathcal{A}^{(n)}\exp\{z_1p+z_2v\}
 \ar=\ar \exp\{z_1p+z_2v\}\cdot \Big[\gamma_n\int_{\mathbb{U}} u_1\nu_\mathtt{i}^{(n)}(du)\cdot vz_1
   +\gamma_n\int_{\mathbb{U}} (u_2-\beta^{(n)})\nu_\mathtt{i}^{(n)}(du) \cdot vz_2  \cr
 \ar\ar +\gamma_n\int_{\mathbb{U}}\big( e^{\frac{1}{n}\langle z,u \rangle } -1\big)\nu_\mathtt{e}^{(n)}(du)
 +n\gamma_n \int_{\mathbb{U}}\big( e^{\frac{1}{n}\langle z,u \rangle }  -1-\frac{\langle z,u \rangle}{n}\big)\nu_\mathtt{i}^{(n)}(du)\cdot v\Big].
 \eeqlb
Convergence as $n\to\infty$ holds if all the four terms in the above sum are convergent.
 The first two terms converge if and only if the following condition holds.

 \begin{assumption}\label{MainCondition01}
 Assume that $\int_\mathbb{U}\|u\|\nu_{\mathtt{i}}^{(n)}(du)<\infty$ for any $n\geq 1$ and  that there exists a constant $b:=(b_1,b_2)\in\mathbb{R}^2$ such that
 \beqnn
 -\gamma_n\int_{\mathbb{U}} u_1\nu_\mathtt{i}^{(n)}(du)\to b_1
 \quad \mbox{and}\quad
 \gamma_n\int_{\mathbb{U}} (\beta^{(n)}-u_2)\nu_\mathtt{i}^{(n)}(du)\to b_2.
 \eeqnn
 \end{assumption}

 We split the last two terms in (\ref{eqn3.27}) into four parts according to the direction of price movements. Specifically, let $\mathbb{U}_\pm=\mathbb{R}_\pm \times\mathbb{R}_+$, and for $j\in\{+,-\}$ and $z \in \mathbb{U}_j$ let
 \beqnn
 G_{j}^{(n)}(z)
 \ar:=\ar n\gamma_n \int_{\mathbb{U}_{j}}\Big( e^{-\frac{1}{n}\langle z,u\rangle}-1 +\frac{\langle z,u\rangle }{n}\Big)\nu_\mathtt{i}^{(n)}(du)
 \quad \mbox{and} \quad
 H_{j}^{(n)}(z)
 := \gamma_n \int_{\mathbb{U}_j}\big(e^{-\frac{1}{n}\langle z,u\rangle}-1\big)\nu_\mathtt{e}^{(n)}(du) .
 \eeqnn
The next condition guarantees that the remaining terms on the right hand side of equation (\ref{eqn3.27}) converge.

 \begin{condition}\label{MainCondition02}
 There exists a constant $C>0$ such that $ \sup_{n \geq 1} n\gamma_n \int_{\mathbb{U}_+}  \|u\|^2 \wedge \|u\| \nu_\mathtt{i}^{(n)}(dnu)\leq C$.
 Moreover, $ G_{\pm}^{(n)}(\cdot)$ and $ H_{\pm}^{(n)}(\cdot)$ converge to continuous functions $G_{\pm}(\cdot)$ and $H_{\pm}(\cdot)$ respectively, as $n\to\infty$.
 \end{condition}

The following proposition provides an exact representation of the limit functions $G_\pm(\cdot)$ and $H_{\pm}(\cdot)$.

 \begin{proposition}\label{Representation}
 Under Condition~\ref{MainCondition02}, the limit functions $G_\pm(\cdot)$ and $H_{\pm}(\cdot)$ have the following representations: for any  $z=(z_1,z_2)\in \mathbb{U}_{\pm}$,
 \beqnn
 G_{\pm}(z)
 \ar=\ar \langle z,\sigma^{\pm} z\rangle
 +\int_{\mathbb{U}_{\pm}} (e^{-\langle z,u\rangle}-1+\langle z,u\rangle)\hat\nu_\mathtt{i}(du)
 \quad \mbox{and}\quad
 H_\pm(z)
 =\langle z,a^{\pm}\rangle
 +\int_{\mathbb{U}_\pm} (e^{-\langle z,u\rangle}-1)\hat\nu_\mathtt{e}(du),
 \eeqnn
 where $a^{\pm}\in\mathbb{U}_\pm$, $\sigma^{\pm}:=(\sigma^{\pm}_{j,l})_{j,l=1,2}$ is a symmetric, nonnegative definite matrix,
 and $(\|u\|\wedge \|u\|^2)\hat{\nu}_\mathtt{i}(du)$ and $(\|u\|\wedge 1)\hat{\nu}_\mathtt{e}(du)$ are finite measures on $\mathbb{U}\setminus \{ 0 \}$.
 \end{proposition}
 \proof
 We prove the result for the function $G_+(\cdot)$. All other results can be established in the same way.
 From Theorem~8.1 in \cite{Sato1999}, the representation of $G_+(\cdot)$ in terms of $\sigma^+$ and $\hat\nu_\mathtt{i}(du)$ is unique { if it exists.}
 For any $n\geq 1$, let
 \beqlb\label{eqn3.04}
 \theta_\mathtt{i}^{(n)}
 := n\gamma_n\int_{\mathbb{U}_+}(\|u\|^2\wedge 1) \nu_\mathtt{i}^{(n)}(dnu)
 \quad \mbox{and}\quad
 \mathbf{P}^{(n)}_\mathtt{i}(du):=  n\gamma_n\frac{\|u\|^2\wedge 1}{\theta^{(n)}_\mathtt{i}}\cdot \mathbf{1}_{\{u\in\mathbb{U}_+\}}\cdot\nu_\mathtt{i}^{(n)}(dnu),
 \eeqlb
 which is a probability law on the compact space $\overline{\mathbb{U}}_+:= \mathbb{U}_+\cup\{\infty\}$.
 Thus, we can always find a subsequence that converges weakly to some limit probability law $\mathbf{P}_\mathtt{i}$ on $\overline{\mathbb{U}}_+$. We notice that $\mathbf{P}_\mathtt{i}$ may have a point mass in $0$. Among other things, we need to prove that $\mathbf{P}_\mathtt{i}\{+\infty\}=0$.

 Let $\mathcal{C}:= \big\{r\geq 0: \mathbf{P}_\mathtt{i}\{\|u\| =r \}=0 \big\};$ its complement is at most countable. For any $r\in\mathcal{C}$, we put $\mathbb{B}_r:=\{ u\in  \mathbb{U} : \|u\|\leq r \}$ and decompose the function $G^{(n)}_+$ as
 \beqlb\label{eqn3.03}
 {G^{(n)}_+(z)}
 \ar=\ar \theta^{(n)}_\mathtt{i} \left( \langle\hat{\beta}^{(n)},z\rangle
 +\sum_{j,l=1}^2\sigma^{+(n)}_{j,l}(r) z_jz_l+ \mathcal{J}^{(n)}(z,r) + \mathcal{E}^{(n)}(z,r) \right),
 \eeqlb
where
 \beqlb
 \hat{\beta}^{(n)} \ar:= \ar \int_{\overline{\mathbb{U}}_+} (u-u\wedge 1)\frac{\mathbf{P}^{(n)}_\mathtt{i}(du)}{\|u\|^2\wedge 1},\\
\mathcal{J}^{(n)}(z,r) \ar:= \ar \int_{\overline{\mathbb{U}}_+\setminus\mathbb{B}_r}\big[e^{-\langle z,u\rangle}-1+\langle z,u\wedge 1\rangle \big]\frac{\mathbf{P}^{(n)}_\mathtt{i}(du)}{\|u\|^2\wedge 1}, \label{eqn3.05} \\
 \sigma^{+(n)}_{j,l}(r)
 \ar:=\ar \frac{1}{2}\int_{\mathbb{B}_r}  (u_j\wedge 1) (u_l\wedge 1)\frac{\mathbf{P}^{(n)}_\mathtt{i}(du)}{\|u\|^2\wedge 1}\in[-1,1],\quad j,l=1,2, \label{eqn3.06}\\
 \mathcal{E}^{(n)}(z,r)
 \ar:=\ar \int_{\mathbb{B}_r}\Big[e^{-\langle z,u\rangle}-1+\langle z,u\wedge 1\rangle -\frac{1}{2}\sum_{j,l=1}^2 (u_j\wedge 1) (u_l\wedge 1)z_jz_l\Big]\frac{\mathbf{P}^{(n)}_\mathtt{i}(du)}{\|u\|^2\wedge 1}.\label{eqn3.07}
 \eeqlb
The weak convergence of $\{\mathbf{P}^{(n)}_\mathtt{i}\}_{n\geq 1}$ along with the fact that $\mathbf{P}_\mathtt{i}(\partial \mathbb{B}_r)=0$  $(r \in \mathcal{C})$ implies that

  \beqlb
\lim_{n \to \infty} \sigma^{+(n)}_{j,l}(r)
 \ar=\ar \frac{1}{2}\int_{\mathbb{B}_r}  (u_j\wedge 1) (u_l\wedge 1)\frac{\mathbf{P}_\mathtt{i}(du)}{\|u\|^2\wedge 1} =: \sigma^+_{j,l}(r) \nonumber, \\
 \lim_{n\to\infty}\mathcal{E}_{n}(z,r)
 \ar =\ar \int_{\mathbb{B}_r}\Big[e^{-\langle z,u\rangle}-1+\langle z,u\wedge 1\rangle -\frac{1}{2}\sum_{j,l=1}^2 (u_j\wedge 1) (u_l\wedge 1)z_jz_l\Big]\frac{\mathbf{P}_\mathtt{i}(du)}{\|u\|^2\wedge 1} =: \mathcal{E}(z,r). \nonumber
 \eeqlb
By Condition~\ref{MainCondition02}, the sequences $\{\theta^{(n)}_\mathtt{i}, n \in \mathbb{N} \}$ and $\{ \hat\beta^{(n)}, n \in \mathbb{N}\}$ are bounded. Hence, we may w.l.o.g.~assume that
 \[
 	(\mathbf{P}^{(n)}_\mathtt{i}, \theta^{(n)}_\mathtt{i},\hat{\beta}^{(n)}) \to
	(\mathbf{P}_\mathtt{i},\theta_\mathtt{i},\hat{\beta})
 \]
 as $n \to \infty$. If $\theta_\mathtt{i} = 0$, the convergence of $G^{(n)}_z$ implies $G_+=0$, due to the moment condition on the measures $\nu^{(n)}_\mathtt{i}$. Hence, we may assume that $\theta_\mathtt{i}>0$. Thus, as $n \to \infty$
  \beqnn
 \int_{\overline{\mathbb{U}}_+\setminus\mathbb{B}_r}\frac{\mathbf{P}^{(n)}_\mathtt{i}(du)}{\|u\|^2\wedge 1}\to c_0(r)
  \quad \mbox{and} \quad
 \int_{\overline{\mathbb{U}}_+\setminus\mathbb{B}_r}
 (u\wedge 1)\cdot\frac{\mathbf{P}^{(n)}_\mathtt{i}(du)}{\|u\|^2\wedge 1}\to {c}(r)
 \eeqnn
for some non-negative numbers $c_0(r)$ and $c(r)$. As a result, for any $z \in \mathbb{R}^2_+$ (noting that $u \geq 0$) it follows from (\ref{eqn3.03}) that
 \beqnn
 \lim_{n \to \infty} \int_{\overline{\mathbb{U}}_+\setminus\mathbb{B}_r}e^{-\langle z,u\rangle}\frac{\mathbf{P}^{(n)}_\mathtt{i}(du)}{\|u\|^2\wedge 1} \ar = \ar
 \int_{\overline{\mathbb{U}}_+\setminus\mathbb{B}_r}e^{-\langle z,u\rangle}\frac{\mathbf{P}_\mathtt{i}(du)}{\|u\|^2\wedge 1} \\
\ar = \ar \frac{G_+(z)}{\theta_i}- \langle\hat{\beta},z\rangle -\sum_{j,l=1}^2 \hat\sigma^+_{j,l}(r)z_jz_l-\mathcal{E}(z,r)+c_0(r)-\langle \mathbf{c}(r),z\rangle.
 \eeqnn
Since $G_+(\cdot)$ is continuous by assumption and  $\mathcal{E}(\cdot,r)$ is continuous by the dominated convergence theorem,
the function $ \int_{\overline{\mathbb{U}}_+\setminus\mathbb{B}_r}e^{-\langle \cdot,u\rangle}\frac{\mathbf{P}_\mathtt{i}(du)}{\|u\|^2\wedge 1}$ is continuous and hence $\mathbf{P}_\mathtt{i}$ is a probability law on $\mathbb{U}_+$, i.e. $\mathbf{P}_\mathtt{i}\{ \infty\}=0$. In particular,
 \beqnn
  \lim_{n \to \infty} \hat\beta^{(n)} = \hat{\beta}= \int_{\mathbb{U}_+} (u-u\wedge 1)\frac{\mathbf{P}_\mathtt{i}(du)}{\|u\|^2\wedge 1}.
 \eeqnn
Moreover,
 $\lim_{r\to 0+}\lim_{n\to\infty} \sigma^{+(n)}_{j,l}(r)=\hat\sigma^+_{j,l}$ and
  $\lim_{r\to 0+}\lim_{n\to\infty} \mathcal{J}^{(n)}(z,r)
 = \int_{\mathbb{U}_+\setminus\{ 0\}} \big[e^{-\langle z,u\rangle}-1+\langle z,u\wedge 1\rangle\big]\frac{\mathbf{P}_\mathtt{i}(du)}{\|u\|^2\wedge 1}$.
 Taking these back into (\ref{eqn3.03}), we have
 \beqnn
 G_+(z)
 \ar=\ar \sum_{j,l=1}^2 \sigma^+_{j,l} z_jz_l
 + \int_{\mathbb{U}_+\setminus\{ 0\}} \big[e^{-\langle z,u\rangle}-1+\langle z,u \rangle \big]\hat\nu_\mathtt{i}(du),
 \eeqnn
 where $\sigma^+_{j,l}:=\theta_\mathtt{i}\hat\sigma^+_{j,l}$ and
 $\hat\nu_\mathtt{i}(du)
 := \theta_\mathtt{i}\cdot \frac{\mathbf{P}_\mathtt{i}(du)}{\|u\|^2\wedge 1}$ on $\mathbb{U}_+\setminus\{ 0\}$.
 \qed

Let $a:=(a_1,a_2)= a^++a^-$, and $\sigma:=(\sigma_{jl})= \sigma^++\sigma^-$, and for any $z=(z_1,z_2)\in \mathbb{U}_*$, let
 \beqlb
 G(z)\ar:=\ar -\langle b,z \rangle+ G_+(-z)+G_-(-z)= -\langle b,z \rangle + \langle z,\sigma z\rangle
 +\int_{\mathbb{U}} (e^{\langle z,u\rangle}-1-\langle z,u\rangle)\hat\nu_\mathtt{i}(du), \label{eqn3.25}\\
 H(z)\ar:=\ar H_+(-z)+H_-(-z)= \langle a, z\rangle
 +\int_{\mathbb{U}} (e^{\langle z,u\rangle}-1)\hat\nu_\mathtt{e}(du).\label{eqn3.26}
 \eeqlb
Under Conditions~\ref{MainCondition01} and \ref{MainCondition02} we thus have that
 \beqlb\label{eqn3.22}
 \mathcal{A}^{(n)}\exp\{z_1p+z_2v\} \to  \exp\{z_1p+z_2v\}\cdot \big[H(z) + v\cdot G(z)\big],\quad (p,v)\in\mathbb{U},\ z\in  \mathbb{U}_*
 \eeqlb
as $n\to\infty$. It will turn out that the infinitesimal generator $\mathcal{A}$ of the limit process does indeed act on $f\in C^2(\mathbb{U})$ according to
 \beqlb\label{eqn3.31}
 \mathcal{A} f(p,v)
 \ar=\ar \langle(a-bv), \nabla f(p,v)\rangle +v\cdot\langle \nabla , \sigma \nabla f(p,v)  \rangle
 +  \int_{\mathbb{U}}\big[f((p,v)+u)-f(p,v)\big]\hat\nu_\mathtt{e}(du) \cr
 \ar\ar
 +v\cdot \int_{\mathbb{U}}\big[f((p,v)+u)-f(p,v)-\langle u, \nabla f(p,v)\rangle\big]\hat\nu_\mathtt{i}(du).
 \eeqlb

\begin{remark}\label{remark01}
As pointed out in the proof of the previous proposition the measure $\mathbf{P}_{\mathtt{i}}$ may has a point-mass in $0$. Likewise, the corresponding measure $\mathbf{P}_{\mathtt{e}}$ arising in the analysis of the functions $H_\pm$ may have a point mass in $0$ as well. If there are no point measures in zero, then $\sigma^\pm$, respectively $a^\pm$ are zero. Loosely speaking, the quantities $\sigma$ and $a$ account for the arrival of infinitely many induced, respectively exogenous orders of ``insignificant magnitude''. This turns out to be very important for the analysis of the jump dynamics of the scaling limit.
\end{remark}

 Before we prove the weak convergence of the rescaled market models, we provide an alternative
 link between the parameters $(1/n,\beta^{(n)}; \nu_{\mathtt{e}/\mathtt{i}}^{(n)})$ and $(a,b,\sigma,\hat{\nu}_{\mathtt{e}/\mathtt{i}})$ that clarifies their interpretation as the drift, diffusion and jump measure of the limiting model. 

 \begin{proposition}\label{MainAssumptionNew}
 Condition~\ref{MainCondition02} holds if and only if
 $ \sup_{n\geq 1}n\gamma_n \int_{\mathbb{U}_+}  \|u\|^2 \wedge \|u\| \nu_\mathtt{i}^{(n)}(dnu)<\infty$ and there exist parameters $(a^\pm,\sigma^\pm, \hat{\nu}_{\mathtt{e}/\mathtt{i}})$ such that the following holds.
 \begin{enumerate}
 \item[(a)] As $n\to\infty$,
 \beqnn
 \gamma_n\int_{\mathbb{U}_\pm}\Big(\frac{u}{n}\wedge 1 \Big)\nu^{(n)}_\mathtt{e}(du)
 \to
 a^{\pm}+\int_{\mathbb{U}_\pm}(u\wedge 1)\hat{\nu}_\mathtt{e}(du)
 \eeqnn
 and
 \beqnn
 n\gamma_n\int_{\mathbb{U}_\pm} \Big(\frac{u}{n}\wedge 1\Big)^{\mathrm{T}}\Big(\frac{u}{n}\wedge 1\Big)\nu_{\mathtt{i}}^{(n)}(du)
 \to
 2\sigma^\pm +\int_{\mathbb{U}_\pm}(u\wedge 1)^{\mathrm{T}}(u\wedge 1)  \hat\nu_{\mathtt{i}}(du);
 \eeqnn

 \item[(b)] For any $f_1,f_2\in C_b(\mathbb{R}^2)$ that satisfy $f_k(u)=O(\|u\|^k)$ as $\|u\|\to 0$ for $k=1,2$, if $n\to\infty$,
 \beqnn
 \gamma_n\int_{\mathbb{U}} f_1\Big(\frac{u}{n}\Big)\nu_\mathtt{e}^{(n)}(du)
 \to
 \int_{\mathbb{U}} f_1(u)\hat\nu_{\mathtt{e}}(du)
 \quad\mbox{and}\quad
 n\gamma_n\int_{\mathbb{U}} f_2\Big(\frac{u}{n}\Big)\nu_\mathtt{i}^{(n)}(du)
 \to
 \int_{\mathbb{U}} f_2(u)\hat\nu_\mathtt{i}(du).
 \eeqnn
 \end{enumerate}
 \end{proposition}
 \proof

A direct computation shows that conditions (a) and (b) imply the convergence of the terms (\ref{eqn3.05})-(\ref{eqn3.07}) to continuous functions, which immediately implies the convergence of $G^{(n)}_+$ to $G_+$. Using the same arguments as in the proof of the previous lemma, the converse follows from direct computation using the convergence along a subsequence of $\{(\mathbf{P}^{(n)}_\mathtt{i}, \theta^{(n)}_\mathtt{i})\}_{n\geq 1}$.
 \qed


 \subsection{Weak convergence}

 Since the linear span of the set $\{ e^{\langle z,\cdot\rangle}:z\in\mathbb{U}_* \}$ is not dense in the space of continuous functions on $\mathbb{U}$ vanishing at infinity, which is a subspace of $\mathscr{D}(\mathcal{A})$, we cannot prove (\ref{eqn3.28}) directly. Instead, we prove the weak convergence of the rescaled market models using the general convergence results for infinite dimensional stochastic integrals established by Kurtz and Protter \cite{KurtzProtter1996}. To this end, we consider the separable Banach space $L^{1,2}(\mathbb{R}_+):=L^{1}(\mathbb{R}_+)\cap L^{2}(\mathbb{R}_+)$, endowed with the norm $\|\cdot\|_{L^{1,2}}:= \|\cdot\|_{L^{1}} \vee \|\cdot\|_{L^{2}}$ and the Haar basis $\{\varphi_j^k:j\geq -1,k\geq 0 \}$. The Haar basis is defined by $\varphi_{-1}^k(x)=\mathbf{1}_{[k,k+1)}(x)$ and
 \beqnn
 \varphi_j^k(x)
 :=
 \left\{\begin{array}{ll}
 2^{j/2}, & x\in [k\cdot 2^{-j},(k+1/2)\cdot 2^{-j}); \vspace{5pt} \cr
 -2^{j/2}, & x\in [(k+1/2)\cdot 2^{-j},(k+1)\cdot 2^{-j});  \vspace{3pt}\cr
 0, &\mbox{else},
 \end{array}
 \right. \quad j,k\geq 0.
 \eeqnn
Let the separable Banach space $\mathbb{H}:= \mathbb{R}^2 \times L^{1,2}(\mathbb{R}_+)$ be endowed with norm $\|\cdot\|_{\mathbb{H}}$ defined by
 \beqnn
 \|Z\|_{\mathbb{H}}\ar:=\ar |Z_1|+ |Z_2| +\|Z_3\|_{L^{1,2}}, \quad  \mbox{for any} \quad Z:=(Z_1,Z_2,Z_3)\in \mathbb{H}.
 \eeqnn
Following Kurtz and Protter \cite{KurtzProtter1996}, we say that an $(\mathscr{F}_t)$-adapted process $\{\mathbf{Y}(t):= (\mathbf{Y}_1(t),\mathbf{Y}_2(t),\mathbf{Y}_3(t)):t\geq 0\}$ is a {\it $\mathbb{H}^\#$-semimartingale} if $\{\mathbf{Y}_1(t):t\geq 0\}$ and $\{\mathbf{Y}_2(t):t\geq 0\}$ are two $\mathbb{R}$-semimartingales, and  $\{\mathbf{Y}_3(t):t\geq 0\}$ is an $L^{1,2}(\mathbb{R}_+)^\#$-semimartingale random measure on $\mathbb{R}_+$, i.e.
 \begin{enumerate}
  \item[(1)] for any $f \in L^{1,2}(\mathbb{R}_+)$, $\{\mathbf{Y}_3(f,t):t\geq 0\}$ is a c\'adl\'ag, $(\mathscr{F}_t)$ semimartingale with $\mathbf{Y}_3(f,0)=0$;

  \item[(2)] for any $f_1,\cdots f_m\in L^{1,2}(\mathbb{R}_+)$ and $w_1,\cdots, w_m\in\mathbb{R}$, $\mathbf{Y}_3(\sum_{i=1}^m w_i f_i,t)=\sum_{i=1}^mw_i\mathbf{Y}_3(f_i,t)$ a.s. for any $t\geq 0$.
 \end{enumerate}

Let us recall the definition of stochastic integrals w.r.t.~the $\mathbb{H}^\#$-semimartingale $\mathbf{Y}$. To this end, let
 $\mathcal{S}^0_{\mathbb{H}}$ be the collection of $\mathbb{H}$-valued, simple processes of the form
 \beqlb\label{eqn3.11}
 \mathbf{X}(t):=(\mathbf{X}_1(t),\mathbf{X}_2(t),\mathbf{X}_3(t))=\Big(\sum \xi_{1,k}\cdot\mathbf{1}_{[t_k,t_{k+1})}(t),\sum \xi_{2,k}\cdot\mathbf{1}_{[t_k,t_{k+1})}(t),\sum \xi_{3,k,i,j}\cdot\mathbf{1}_{[t_k,t_{k+1})}(t)\cdot\varphi_i^j \Big),
 \eeqlb
 where $0= t_1<t_2<\cdots$, and $\{\xi_{1,k}\}$, $\{\xi_{2,k}\},\{\xi_{2,k,i,j}\}$ are sequences of bounded, $\mathbb{R}$-valued, $(\mathscr{F}_{t_k})$-adapted random variables, all but finitely many of which being zero.
 For any $\mathbf{X} \in \mathcal{S}^0_{\mathbb{H}}$, the stochastic integral w.r.t.~$\mathbf{Y}$ is defined as
 \beqnn
 \int_0^t \mathbf{X}(s-)\cdot \mathbf{Y}(ds) \ar=\ar \sum_{k=1}^\infty \xi_{1,k}[\mathbf{Y}_1(t_{k+1}\wedge t)- \mathbf{Y}_1(t_{k}\wedge t)] + \sum_{k=1}^\infty \xi_{2,k}[\mathbf{Y}_2(t_{k+1}\wedge t)- \mathbf{Y}_2(t_{k}\wedge t)]\cr
 \ar\ar +\sum_{k=1}^\infty\sum_{i=-1}^\infty\sum_{j=0}^\infty \xi_{3,k,i,j}\cdot[\mathbf{Y}_3(t_{k+1}\wedge t, \varphi_i^j)- \mathbf{Y}_3(t_{k}\wedge t,\varphi_i^j)].
 \eeqnn
 Let $\hat{\mathbb{H}}$ be the completion of the linear space $\mathcal{S}^0_{\mathbb{H}}$ with respect to the norm $\|\cdot\|_\mathbb{H}$, i.e. for any $\{\mathbf{X}(t):t\geq 0\}\in\hat{\mathbb{H}}$, there exists a sequence of simple processes $\{\mathbf{X}_n(t):t\geq 0\}\in \mathcal{S}^0_{\mathbb{H}}$ such that for any $T\geq 0$,
 \beqnn
 \lim_{n\to\infty} \int_0^T \| \mathbf{X}(t)-\mathbf{X}_n(t) \|_\mathbb{H}dt =0.
 \eeqnn
 The stochastic integral for $\mathbf{X}(\cdot)\in\hat{\mathbb{H}}$ is then defined as
 \beqnn
 \int_0^t \mathbf{X}(s-)\cdot \mathbf{Y}(ds) := \lim_{n\to\infty}\int_0^t \mathbf{X}_n(s-)\cdot \mathbf{Y}(ds).
 \eeqnn
 We say that a $(\mathscr{F}_{t})$-adapted process $\{\mathbf{Y}(t):t\geq 0  \}$ is a {\it standard $\mathbb{H}^\#$-semimartingale} if $\int_0^\cdot \mathbf{X}(s-)\cdot \mathbf{Y}(ds)\in\mathbf{D}([0,\infty);\mathbb{R}^2)$ for any $\mathbf{X}\in \mathcal{S}^0_{\mathbb{H}}$ and
 \beqlb\label{eqn3.10}
 \mathcal{H}_t^0:= \left\{ \Big\|\int_0^t \mathbf{X}(s-)\cdot \mathbf{Y}(ds)  \Big\|: \mathbf{X}\in \mathcal{S}^0_{\mathbb{H}},\  \sup_{s\leq t} \|\mathbf{X}(s)\|_{\mathbb{H}}\leq 1\right\}
 \eeqlb
 is stochastically bounded for each $t\geq 0$.

We are now ready to provide an alternative representation of the market model (\ref{eqn3.02}).
 For any $t\geq 0$ and $n\geq 1$, let us define $\mathbf{Y}^{(n)}(t):= (\mathbf{Y}^{(n)}_1(t),\mathbf{Y}^{(n)}_2(t),\mathbf{Y}^{(n)}_3(t))$ by
 \beqnn
 \mathbf{Y}^{(n)}_1(t):=\gamma_n\left(  \int_{\mathbb{U}}u\nu^{(n)}_{\mathtt{i}}(du)-
   \left( \begin{array}{c} 0\\ \beta_n
   \end{array}  \right)\right) \cdot t, \quad
 \mathbf{Y}^{(n)}_2(t):= \int_0^t\int_{\mathbb{U}} \frac{u}{n}N^{(n)}_{\mathtt{e}}(d\gamma_n s,du)
 \eeqnn
 and
 \beqnn
 \mathbf{Y}^{(n)}_3(t):=\int_0^t \int_{\mathbb{U}} \frac{u}{n}\tilde{N}^{(n)}_{\mathtt{i}}(d\gamma_n s,du,dnx).
 \eeqnn
 {The process $\{\mathbf{Y}^{(n)}_1(t):t\geq 0\}$ is a real-valued process, $\{\mathbf{Y}^{(n)}_2(t):t\geq 0\}$ is a compound Poisson process, and
 $\{ \mathbf{Y}^{(n)}_3(t):t\geq 0\}$ is an $L^{1,2}(\mathbb{R}_+)^\#$-martingale random measure on $\mathbb R_+$. }
 We can rewrite the market model (\ref{eqn3.02}) as
 \begin{equation} \label{SDE}
 \Big( \begin{array}{c}
 \mathbf P^{(n)}(t)\\ \mathbf V^{(n)}(t)
 \end{array} \Big)
 = \Big( \begin{array}{c}
 \mathbf P^{(n)}(0)\\ \mathbf V^{(n)}(0)
 \end{array} \Big)+ \int_0^t \mathbf{F}(\mathbf V^{(n)}(s-)) \cdot \mathbf{Y}^{(n)}(ds),
 \end{equation}
 where the function $\mathbf{F}: \mathbb{R}_+ \to (\mathbb{R}_+^{2}\times L^{1,2}(\mathbb{R}_+))^2$ is defined by
 \[
 	\mathbf{F}(v) := \left( \begin{array}{lll} v & 1 &  \mathbf{1}_{\{\cdot< v\}} \\ v & 1 &  \mathbf{1}_{\{\cdot< v\}}
	\end{array} \right).
 \]
In the next subsection, we prove the weak convergence of the sequence of integrators $\{\mathbf{Y}^{(n)}(t):t\geq 0\}_{n\geq 1} $ in the space $\mathbf{D}([0,\infty);\mathbb{H}^\#)$. Subsequently, we show that this implies convergence of the market models.

\subsubsection{Convergence of $\mathbf{Y}^{(n)}$}

Since the elements of $\{\mathbf{Y}^{(n)}(t):t\geq 0\}$ are mutually independent, it suffices to prove the weak convergence of the processes $\{\mathbf{Y}^{(n)}_i(t):t\geq 0\}_{n\geq 1} $ $(i=1,2,3)$ separately.
 The convergence of $\{\mathbf{Y}^{(n)}_1(t):t\geq 0\}_{n\geq 1} $ and $\{\mathbf{Y}^{(n)}_2(t):t\geq 0\}_{n\geq 1}$ follows from \cite[Theorem 3.4]{JacodShiryaev2003}.

  \begin{lemma}\label{Convergence01}
 Under Condition~\ref{MainCondition01} and \ref{MainCondition02}, we have $\{(\mathbf{Y}^{(n)}_1(t),\mathbf{Y}^{(n)}_2(t)):t\geq 0\}_{n\geq 1} $ converges weakly to $\{(\mathbf{Y}_1(t),\mathbf{Y}_2(t)):t\geq 0\}$ in $\mathbf{D}([0,\infty);\mathbb{R}^2)$ as $n\to\infty$, where

 \beqnn
 \mathbf{Y}_1(t)=-bt
 \quad \mbox{and}\quad
 \mathbf{Y}_2(t)= at+ \int_0^t \int_{\mathbb{U}} u N_1(ds,du),
 \eeqnn
 and $N_1(ds,du)$ is a Poisson random measure on $(0,\infty)\times \mathbb{U}$ with intensity $ds\hat\nu_{\mathtt{e}}(du)$.
 \end{lemma}

We now turn to the convergence of the sequence $\{\mathbf{Y}^{(n)}_3(t):t\geq 0\}_{n\geq 1}$. Following Kurtz and Protter \cite{KurtzProtter1996} we say that this sequence converges weakly to $\{\mathbf{Y}_3(t):t\geq 0\}$ as $n \to \infty$ and write {$\mathbf{Y}_3^{(n)} \Rightarrow \mathbf{Y}_3$} if
 \[
 	 \left( \mathbf{Y}^{(n)}_3(f_1,\cdot),\cdots, \mathbf{Y}^{(n)}_3(f_m,\cdot) \right) \to
	 \left( \mathbf{Y}_3(f_1,\cdot),\cdots, \mathbf{Y}_3( f_m,\cdot) \right)
 \]
weakly in {$\mathbf{D}([0,\infty),\mathbb R^{2\times m})$} for any $f_1, \cdots, f_m \in L^{1,2}(\mathbb R_+)$ and any $m \in \mathbb N$. The following lemma establishes the weak convergence. It uses the fact that $2\sigma$ is symmetric and nonnegative-definite so that the square root $\sqrt{2\sigma}$ is well defined.

 \begin{lemma}\label{Convergence}
 Under Condition~\ref{MainCondition01} and \ref{MainCondition02}, we have that $\mathbf{Y}_3^{(n)} \Rightarrow \mathbf Y_3$ where
 $\{\mathbf{Y}_3(t):t\geq 0\}$ is an $L^{1,2}(\mathbb{R}_+)^\#$-martingale with the following representation:
 \beqnn
 \mathbf{Y}_3(t)=
 \int_0^t \sqrt{2\sigma} \Big(\begin{array}{c}
 W_1(ds,dx)\\ W_2(ds,dx)
 \end{array} \Big) + \int_0^t \int_{\mathbb{U}} u \tilde{N}_0(ds,du,dx).
 \eeqnn
Here, $W_1(dt,dx)$ and $W_2(dt,dx)$ are orthogonal Gaussian white noises on $(0,\infty)^2$ with intensities $dsdx$, $N_0(ds,du,dx)$ is a Poisson random measure on $(0,\infty)\times \mathbb{U} \times \mathbb{R}_+$ with intensity $ds\hat\nu_\mathtt{i}(du)dx$, and $$\tilde{N}_0(ds,du,dx):=N_0(ds,du,dx)-ds\hat\nu_{\mathtt{i}}(du)dx.$$
 \end{lemma}
 \proof
 We prove the weak convergence of $\{\mathbf{Y}_3^{(n)}(f,t):t\geq 0\}_{n\geq 1}$ for any $f \in L^{1,2}(\mathbb R_+)$; the general case be proved in the same way. From Kurtz's criterion \cite[p. 137]{EthierKurtz1986} tightness of this sequence follows from the following estimate: for any $t\in[0,1] $ and $\epsilon>0$,
 \beqlb\label{eqn3.35}
 \mathbf{E}_{\mathscr{F}_t}\big[\|\mathbf{Y}_3^{(n)}(f,t+\epsilon)-\mathbf{Y}_3^{(n)}(f,t)\|\big] \leq 2(\epsilon+\sqrt{\epsilon}).
 \eeqlb
 In order to verify this inequality, we first apply Jensen's inequality to get
 \beqlb\label{eqn3.36}
  \mathbf{E}_{\mathscr{F}_t}\big[\|\mathbf{Y}_3^{(n)}(f,t+\epsilon)-\mathbf{Y}_3^{(n)}(f,t)\| \big]
  \ar=\ar \mathbf{E}\Big[\Big\|\int_t^{t+\epsilon} \int_{\mathbb{U}} \int_0^\infty f(x) \frac{u}{n}\tilde{N}^{(n)}_{\mathtt{i}}(d\gamma_n s,du,dnx)\Big\|\Big]\cr
  \ar\leq \ar\mathbf{E}\Big[\Big\|\int_t^{t+\epsilon} \int_{\|u\|\leq 1} \int_0^\infty f(x) \frac{u}{n}\tilde{N}^{(n)}_{\mathtt{i}}(d\gamma_n s,du,dnx)\Big\|^2\Big]^{1/2}\cr
  \ar\ar +\mathbf{E}\Big[\Big\|\int_t^{t+\epsilon} \int_{\|u\|> 1} \int_0^\infty f(x) \frac{u}{n}\tilde{N}^{(n)}_{\mathtt{i}}(d\gamma_n s,du,dnx)\Big\|\Big].
 \eeqlb
 Applying the Burkholder-Davis-Gundy inequality to the first expectation on the right side of the last inequality, this term can be bounded by
 \beqnn
 C\mathbf{E}\Big[\int_t^{t+\epsilon} \int_{\|u\|\leq 1} \int_0^\infty |f(x)|^2 \frac{\|u\|^2}{n^2}\tilde{N}^{(n)}_{\mathtt{i}}(d\gamma_n s,du,dnx)\Big]
 \ar\leq\ar C\cdot \epsilon\cdot \|f\|_{L^2}  dx \cdot n\gamma_n\int_{\|u\|\leq 1}\|u\|^2\nu_\mathtt{i}^{(n)}(dnu).
 \eeqnn
 Moreover, the second term on the right side of the last inequality in (\ref{eqn3.36}) can be bounded by
 \beqnn
 2\epsilon \cdot \|f\|_{L^1}  n\gamma_n\int_{\|u\|> 1} \|u\|\nu^{(n)}_{\mathtt{i}}(dnu).
 \eeqnn
 Taking these two upper estimates back into (\ref{eqn3.36}) and using Condition~\ref{MainCondition02} yields (\ref{eqn3.35}). We may hence assume that   $\{\mathbf{Y}^{(n)}_3(f,t):t\geq 0\}_{n\geq 1}$ converges weakly in $\mathbf{D}([0,\infty);\mathbb R^2)$
 along a subsequence. By Skorokhood's representation theorem, we may actually assume almost sure convergence and need to prove that

 \beqnn
  \mathbf{Y}_3^{(n)}(f,t)   \ar\to\ar \int_0^t\int_0^\infty f(x)\sqrt{2\sigma}\Big(
  \begin{array}{c}
  W_1(ds,dx)\\ W_2(ds,dx)
  \end{array}\Big)
  +\int_0^t\int_{\mathbb{U}} \int_0^\infty f(x) u\tilde{N}_0(ds,du,dx)=:\mathbf{Y}_3(f,t).
 \eeqnn

The next step, then, consists in proving that the limit is a local martingale; subsequently we apply a general representation result for local martingales to conclude.  In order to prove the local martingale property of the limit process, let $\mathcal{A}^{(n)}_f$ be the infinitesimal generator of the $\mathbb{R}^2$-valued martingale $\{\mathbf{Y}_3^{(n)}(f,t):t\geq 0\}$. It acts on $\phi\in C_b^2(\mathbb{U})$ according to
 \beqlb\label{eqn3.17}
 \mathcal{A}^{(n)}_f\phi(z)\ar=\ar n\gamma_n\int_0^\infty dx\int_{\mathbb{U}} \big[\phi(z+f(x)u)-\phi(z)-\langle \nabla\phi(z),f(x)u \rangle \big] \nu^{(n)}_\mathtt{i}(dnu)\cr
 \ar=\ar n\gamma_n\int_0^\infty dx\int_{\mathbb{U}_+} \big[\phi(z+f(x)u)-\phi(z)-\langle \nabla\phi(z),f(x)u \rangle \big] \nu^{(n)}_\mathtt{i}(dnu)\cr
 \ar\ar + n\gamma_n\int_0^\infty dx\int_{\mathbb{U}_-} \big[\phi(z+f(x)u)-\phi(z)-\langle \nabla\phi(z),f(x)u \rangle \big] \nu^{(n)}_\mathtt{i}(dnu).
 \eeqlb
 We can rewrite the two terms on the right side of the last equality as
 \beqnn
 \lefteqn{n\gamma_n\int_0^\infty dx\int_{\mathbb{U}_\pm} \big[\phi(z+f(x)u)-\phi(z)-\langle \nabla\phi(z),f(x)u \rangle \big] \nu^{(n)}_\mathtt{i}(dnu)}\ar\ar\cr
 \ar=\ar
 \frac{1}{2}\sum_{j,l=1}^2\frac{\partial^2 \phi(z)}{\partial z_j\partial z_l} n\gamma_n\int_0^\infty|f(x)|^2 dx\int_{\mathbb{U}_\pm}  (u_j\wedge 1)(u_l\wedge 1) \nu^{(n)}_\mathtt{i}(dnu)\cr
 \ar\ar +
 n\gamma_n\int_0^\infty dx\int_{\mathbb{U}_\pm} \Big[\phi(z+f(x)u)-\phi(z)-\langle \nabla\phi(z),f(x)u \rangle\cr
  \ar\ar\quad - \frac{1}{2}\sum_{j,l=1}^2\frac{\partial^2 \phi(z)}{\partial z_j\partial z_l} |f(x)|^2(u_j\wedge 1)(u_l\wedge 1)\Big] \nu^{(n)}_\mathtt{i}(dnu).
 \eeqnn
  From Proposition~\ref{MainAssumptionNew}, as $n\to\infty$ the first term on the right side of the equality above converges to
 \beqnn
  \sum_{j,l=1}^2 \frac{\partial^2 \phi(z)}{\partial z_j\partial z_l}\int_0^\infty |f(x)|^2 dx
  \Big[ \sigma^+_{j,l} +\frac{1}{2}
\int_{\mathbb{U}_\pm}  (u_j\wedge 1)(u_l\wedge 1) \hat\nu_\mathtt{i}(du)\Big],
 \eeqnn
 and the second term converges to
 \beqnn
  \int_0^\infty dx\int_{\mathbb{U}_\pm} \Big[\phi(z+f(x)u)-\phi(z)-\langle \nabla\phi(z),f(x)u \rangle - \frac{1}{2}\sum_{j,l=1}^2\frac{\partial^2 \phi(z)}{\partial z_j\partial z_l} |f(x)|^2(u_j\wedge 1)(u_l\wedge 1)\Big] \hat\nu_\mathtt{i}(du).
 \eeqnn
 Hence, the generator $\mathcal{A}^{(n)}_f$ converges to
 the linear operator $\mathcal{A}_f$ defined by
 \beqlb\label{eqn3.55}
 \mathcal{A}_f\phi(x)\ar:=\ar  \sum_{j,l=1}^2 \sigma_{j,l}\frac{\partial^2 \phi(z)}{\partial z_j \partial z_l}\int_0^\infty |f(x)|^2 dx + \int_0^\infty dx\int_{\mathbb{U}} \Big[\phi(z+f(x)u)-\phi(z)-\langle \nabla\phi(z),f(x)u \rangle \Big] \hat\nu_\mathtt{i}(du).
 \eeqlb
 Applying It\^o's formula to the function $e^{\mathrm{i}\langle z,\mathbf{Y}_3^{(n)}(f,\cdot)\rangle}$, we see that
  \beqnn
 \mathcal{M}^{(n)}_t(f):=e^{\mathrm{i}\langle z,\mathbf{Y}_3^{(n)}(f,t)\rangle}-\int_0^t \mathcal{A}^{(n)}_f e^{\mathrm{i}\langle z,\mathbf{Y}_3^{(n)}(f,s)\rangle}ds
 \eeqnn
 is an $(\mathscr{F}_t)$-local martingale. Since
 \beqlb \label{eqn4.57}
 \sup_{s\in[0,t]}|\mathcal{A}^{(n)}_f e^{\mathrm{i}\langle z,\mathbf{Y}_3^{(n)}(f,s)\rangle}|
 \ar=\ar\sup_{s\in[0,t]}|e^{\mathrm{i}\langle z,\mathbf{Y}_3^{(n)}(f,s)\rangle}|\cdot
 n\gamma_n\int_0^\infty dx\int_{\mathbb{U}} \big| e^{\mathrm{i}\langle z,f(x)u\rangle} -1-\mathrm{i}\langle z,f(x)u \rangle \big| \nu^{(n)}_\mathtt{i}(dnu)\cr
 \ar\leq\ar n\gamma_n\int_0^\infty dx\int_{\mathbb{U}} |\langle z,f(x)u \rangle|^2\wedge  |\langle z,f(x)u \rangle| \nu^{(n)}_\mathtt{i}(dnu)\cr
 \ar\leq\ar \int_0^\infty\|f(x)z\|^2\vee\|f(x)z\| dx \cdot n\gamma_n \int_{\mathbb{U}}
 \|u \|^2\wedge  \|u\| \nu^{(n)}_\mathtt{i}(dnu),
 \eeqlb
 and because the last term is uniformly bounded in $n$, due to Condition~\ref{MainCondition02}, the process $\{\mathcal{M}^{(n)}_t(f):t\geq 0 \}$ is in fact a true martingale.

 By the almost sure convergence in the Skorohood topology we know that $\mathbf{Y}_3^{(n)}(f,t)\to \mathbf{Y}_3(f,t)$ a.s.~for all but countably many $t > 0$. Thus, it follows from (\ref{eqn3.55}) and the dominated convergence theorem that
\[
	\int_0^t \mathcal{A}^{(n)}_f e^{\mathrm{i}\langle z,\mathbf{Y}_3^{(n)}(f,s)\rangle}ds \to
	\int_0^t \mathcal{A}_f e^{\mathrm{i}\langle z,\mathbf{Y}_3(f,s)\rangle}ds
\]
 almost surely in the space $\mathbf{C}([0,\infty), \mathbb C)$. This implies that for any $z\in\mathbb{R}^2$,
 \beqnn
 \mathcal{M}^{(n)}_t(f) \to \mathcal{M}_t(f) := e^{\mathrm{i}\langle z,\mathbf{Y}_3(f,t)\rangle}-\int_0^t \mathcal{A}_f e^{\mathrm{i}\langle z,\mathbf{Y}_3(f,s)\rangle}ds
 \eeqnn
 almost surely in the Skorohood space ${\mathbf D}([0,\infty), \mathbb C)$ and hence that $\mathcal{M}^{(n)}_t(f) \to \mathcal{M}_t(f)$ almost surely for almost all $t > 0$.
 In view of (\ref{eqn4.57}) we also have convergence in $L^1$ for almost all $t >0$.
 Since $\{\mathcal{M}_t(f): t \geq 0\}$ is right-continuous, the martingale property of $\{\mathcal{M}^{(n)}_t(f): t \geq 0\}$ carries over to $\{\mathcal{M}_t(f): t \geq 0\}$.
 By \cite[Theorem 2.42]{JacodShiryaev2003} this is equivalent to the local martingale property of $\{\mathbf{Y}_3(f,t):t\geq 0\}$.

 We now prove the desired representation. The local martingale  $\{\mathbf{Y}_3(f,t):t\geq 0\}$ admits the the canonical representation
 \beqlb\label{eqn3.18}
 \mathbf{Y}_3(f,t)\ar=\ar M^c_t(f)+ \int_0^t \int_{\mathbb{U}} u' \tilde{N}_f^d(ds,du'),
 \eeqlb
 where ${N}_f^d(ds,du')$ is an integer-valued random measure on $[0,\infty)\times\mathbb{U}$
 with compensator $ds \nu^d_\mathtt{i}(du') $ and
 \beqnn
 \nu^d_\mathtt{i}(du')= \int_0^\infty dx\int_\mathbb{U} \mathbf{1}_{\{f(x)u\in du'\}} \hat\nu_\mathtt{i}(du),
 \eeqnn
 and $\{ M^c_t(f):t\geq 0\}$ is a continuous, $\mathbb{R}^2$-valued local martingale with quadratic covariation process
 \beqnn
 \langle {M}_j^c(f),{M}_l^c(f)\rangle_t
 =t\cdot 2\sigma_{j,l}\int_0^\infty |f(x)|^2 dx,\quad j,l=1,2.
 \eeqnn
 Similarly, for any $f,g\in L^{1,2}(\mathbb{R}_+)$, we have
 \beqnn
 \langle {M}_j^c(f),{M}_l^c(g)\rangle_t
 =t\cdot 2\sigma_{j,l}\int_0^\infty f(x)g(x) dx, \quad j,l=1,2.
 \eeqnn
 By \cite[Theorem III-7]{ElKaroui1990}, there exist two orthogonal Gaussian white noise $W_1(ds,dx)$ and $W_2(ds,dx)$ on $(0,\infty)^2$ with density $dsdx$ such that for any $f\in L^{1,2}(\mathbb{R}_+)$,
 \beqnn
 {M}_j^c(f)= \int_0^t \int_{\mathbb{R}_+} f(x)\sqrt{2\sigma}_{j1}W_1(ds,dx) +\int_0^t \int_{\mathbb{R}_+} f(x) \sqrt{2\sigma}_{j2}W_2(ds,dx),\quad j=1,2.
 \eeqnn
 By \cite[Theorem 7.4]{IkedaWatanabe1989}, there exists a Poisson random measure $N_0(ds,du,dx)$ on $(0,\infty)\times \mathbb{U} \times \mathbb{R}_+$ with intensity $ds\hat\nu_\mathtt{i}(du)dx$ such that
 \beqnn
  \int_0^t \int_{\mathbb{U}} u \tilde{N}_f^d(ds,du)= \int_0^t \int_{\mathbb{U}}\int_0^\infty f(x) \cdot u \tilde{N}_0(ds,du,dx).
 \eeqnn
 Taking these two representations back into (\ref{eqn3.18}) yields the desired result.
 \qed

 \subsubsection{Uniform tightness and weak convergence of market models}

 The next steps towards the proof of the weak convergence of the market models is to prove that $\{ \mathbf{Y}^{(n)}(t):t\geq 0 \}_{n\geq 1}$ is a sequence of \textit{uniformly tight} standard $\mathbb{H}^\#$-semimartingales. That is, the sequence $\{\mathcal{H}_t^{(n)}\}_{n\geq1}$ is uniformly stochastically bounded for any $t\geq 0$, where $\mathcal{H}_t^{(n)}$ is defined as in (\ref{eqn3.10}) with $\mathbf{Y}$ replaced by $\mathbf{Y}^{(n)}$.
 \begin{lemma}\label{UniformlyTight}
 Under Conditions~\ref{MainCondition01} and \ref{MainCondition02}, $\{\mathbf{Y}^{(n)}(t):t\geq 0\}_{n\geq 1}$ is a sequence of uniformly tight $\mathbb{H}^\#$-semimartingales.
 \end{lemma}
 \proof
 We just prove that $\{\mathcal{H}_1^{(n)}\}_{n\geq1}$ is uniformly stochastically bounded; the general case can be proved similarly.
 For any simple $\mathbb{H}$-valued processes $\mathbf{X}$,
 \beqlb\label{eqn3.12}
 \int_0^1 \mathbf{X}(s-)\cdot\mathbf{Y}^{(n)}(ds)
 \ar=\ar \sum_{k=1}^3 \int_0^1 \mathbf{X}_k(s-)\cdot\mathbf{Y}_k^{(n)}(ds),
 \eeqlb
 where
 \beqnn
 \int_0^1\mathbf{X}_1(s-)\cdot \mathbf{Y}_1^{(n)}(ds)
   \ar:=\ar  \gamma_n\left(  \int_{\mathbb{U}}u\nu^{(n)}_{\mathtt{i}}(du)-
   \left( \begin{array}{c} 0\\ \beta_n
   \end{array}  \right)\right) \int_0^1 \sum \xi_{1,k}\mathbf{1}_{[t_k,t_{k+1})}(s)ds,\cr
 \int_0^1\mathbf{X}_2(s-)\cdot \mathbf{Y}_2^{(n)}(ds)
   \ar:=\ar \int_0^1\sum \xi_{2,k}\mathbf{1}_{[t_k,t_{k+1})}(s)\int_{\mathbb{U}} \frac{u}{n}N^{(n)}_{\mathtt{e}}(d\gamma_n s,du),\cr
 \int_0^1\mathbf{X}_3(s-)\cdot \mathbf{Y}_3^{(n)}(ds)
   \ar:=\ar \int_0^1\int_0^\infty \sum \xi_{3,k,i,j}\mathbf{1}_{[t_k,t_{k+1})}(s)\varphi_i^j(x)\int_{\mathbb{U}} \frac{u}{n}\tilde{N}^{(n)}_{\mathtt{i}}(d\gamma_n s,du,dnx).
 \eeqnn
 Since $\sup_{s\in[0,1]}\|X(s)\|_{\mathbb{H}}\leq 1$ a.s.,
 we have $|\xi_{1,k}|, |\xi_{2,k}|, |\xi_{3,k,i,j}|\leq 2$  a.s. for any $k\geq 1$, $i\geq -1$ and $j\geq 0$.
 By Condition~\ref{MainCondition01} there exists a constant $C>0$ such that for any $n\geq 1$,
 \beqnn
 \gamma_n\left\| \int_{\mathbb{U}}u\nu^{(n)}_{\mathtt{i}}(du)-
   \left( \begin{array}{c} 0\\ \beta_n
   \end{array}  \right)\right\|\leq C
 \eeqnn
 and from the Markov inequality,
 \beqlb
 \mathbf{P}\Big\{\Big\|\int_0^1\mathbf{X}_1(s-)\cdot \mathbf{Y}_1^{(n)}(ds)\Big\|\geq K  \Big\}
 \leq \frac{C}{K}\int_0^1\sum \mathbf{E}[ |\xi_{1,k}|]\mathbf{1}_{[t_k,t_{k+1})}(s)ds\leq \frac{2C}{K}. \label{eqn3.14}
 \eeqlb
 For the second term in (\ref{eqn3.12}), we have
 \beqnn
 \Big\|\int_0^1\mathbf{X}_2(s-)\cdot\mathbf{Y}_2^{(n)}(ds)\Big\|
 \leq 2\int_0^1 \int_{\mathbb{U}} \frac{\|u\|}{n}N^{(n)}_{\mathtt{e}}(d\gamma_n s,du).
 \eeqnn
 For any $M>0$, we have
 \beqlb\label{eqn3.13}
 \mathbf{P} \Big\{ \int_0^1 \int_{\mathbb{U}} \frac{\|u\|}{n}N^{(n)}_{\mathtt{e}}(d\gamma_n s,du)\geq K \Big\}\ar\leq\ar \mathbf{P} \Big\{ \int_0^1 \int_{\mathbb{U}} \Big(\frac{\|u\|}{n}\wedge M\Big)N^{(n)}_{\mathtt{e}}(d\gamma_n s,du)\geq K \Big\}\cr
 \ar\ar + \mathbf{P} \Big\{\int_0^1\int_{\mathbb{U}}\Big(\frac{\|u\|}{n}\vee M\Big)\mathbf{1}_{\{\|u\|\geq nM\}} N^{(n)}_{\mathtt{e}}(d\gamma_n s,du)\geq K\Big\}.
 \eeqlb
 Applying Jensen's inequality to the first term on the right side of this inequality, it can be bounded by
 \beqnn
  \frac{1}{K}\mathbf{E} \Big[\int_0^1\int_{\mathbb{U}}\Big(\frac{\|u\|}{n}\wedge M\Big) N^{(n)}_{\mathtt{e}}(d\gamma_n s,du)\Big]
 \ar\leq\ar \frac{1}{K}\cdot\gamma_n \int_{\mathbb{U}}\Big(\frac{\|u\|}{n}\wedge M\Big) \nu_{\mathtt{e}}^{(n)}(du).
 \eeqnn
 Moreover, the second term on the right side of (\ref{eqn3.13}) can be bounded by
 \beqnn
 \mathbf{P} \Big\{\int_0^1\int_{\mathbb{U}} \mathbf{1}_{\{\|u\|\geq nM\}} N^{(n)}_{\mathtt{e}}(d\gamma_n s,du)\geq 1\Big\}
 \leq 1-\exp\{ -\gamma_n \nu^{(n)}_{\mathtt{e}}(\|u\|\geq nM)\}\leq C \cdot \gamma_n \nu^{(n)}_{\mathtt{e}}(\|u\|\geq nM).
 \eeqnn
 By Proposition~\ref{MainAssumptionNew}(b), there exist constants  $C>0$ independent of $M$ and  $C_M>0$ such that for any $n\geq 1$,
 \beqnn
 \gamma_n \int_{\mathbb{U}}\Big(\frac{\|u\|}{n}\wedge M\Big) \nu_{\mathtt{e}}^{(n)}(du) \leq C_M
 \quad \mbox{and}\quad
 \gamma_n \nu^{(n)}_{\mathtt{e}}(nM,\infty) \leq C \cdot\hat{\nu}_{\mathtt{e}}(\|u\|\geq M).
 \eeqnn
 Altogether, this yields
 \beqlb\label{eqn3.15}
 \mathbf{P}\Big\{  \Big\|\int_0^1\mathbf{X}_2(s-)\cdot\mathbf{Y}_2^{(n)}(ds)\Big\|\geq K \Big\}\leq \frac{C_M}{K}+ C \hat{\nu}_{\mathtt{e}}(\|u\|\geq M),
 \eeqlb
 which goes to $0$ when first sending $K$ and then $M$ to $\infty$.
 We now consider the third term on the right side of (\ref{eqn3.12}).
 Applying Jensen's inequality again, we have
 \beqlb\label{eqn3.29}
 \mathbf{P}\Big\{\Big\| \int_0^1\mathbf{X}_3(s-)\cdot \mathbf{Y}_3^{(n)}(ds) \Big\|\geq K \Big\}
 \ar\leq\ar
 \mathbf{P}\big\{\big\|\mathbf{M}_{\leq 1}\big\|\geq K/2 \big\}  + \mathbf{P}\big\{\big\|\mathbf{M}_{> 1}\big\|\geq K/2 \big\} \cr
 \ar\leq\ar \frac{4}{K^2}\mathbf{E}\Big[\big\|\mathbf{M}_{\leq 1}\big\|^2\Big] +\frac{2}{K}\mathbf{E}\Big[\big\|\mathbf{M}_{> 1}\big\| \Big],
 \eeqlb
 where
 \beqnn
 \mathbf{M}_{\leq 1}\ar:=\ar \int_0^1\int_0^\infty\int_{\|u\|\leq n} \sum \xi_{3,k,i,j}\mathbf{1}_{[t_k,t_{k+1})}(s)\varphi_i^j(x) \frac{u}{n} \tilde{N}^{(n)}_{\mathtt{i}}(d\gamma_n s,du,dnx),\cr
 \mathbf{M}_{>1}\ar:=\ar \int_0^1\int_0^\infty\int_{\|u\|> n} \sum \xi_{3,k,i,j}\mathbf{1}_{[t_k,t_{k+1})}(s)\varphi_i^j(x) \frac{u}{n} \tilde{N}^{(n)}_{\mathtt{i}}(d\gamma_n s,du,dnx).
 \eeqnn
 Applying the Burkholder-Davis-Gundy inequality to the first term on the right side of the last inequality in (\ref{eqn3.29}), we have
 \beqnn
 \mathbf{E}\big[ \big\|\mathbf{M}_{\leq 1}\big\|^2  \big]
 \ar\leq\ar \int_0^1\sum \mathbf{E}[|\xi_{3,k,i,j}|^2]\mathbf{1}_{[t_k,t_{k+1})}(s)ds\int_0^\infty|\varphi_i^j(x)|^2dx \cdot n\gamma_n\int_{\frac{\|u\|}{n}\leq 1 } \frac{\|u\|^2}{n^2} \hat{\nu}^{(n)}_{\mathtt{i}}(du)\cr
 \ar=\ar  \|\mathbf{X}_3\|_{L^2}^2\cdot n\gamma_n\int_{\frac{\|u\|}{n}\leq 1 } \frac{\|u\|^2}{n^2} \nu^{(n)}_{\mathtt{i}}(du)   \leq C n\gamma_n\int_{\frac{\|u\|}{n}\leq 1 } \frac{\|u\|^2}{n^2} \nu^{(n)}_{\mathtt{i}}(du).
 \eeqnn
 Moreover, we also have
 \beqnn
 \mathbf{E}\big[ \big\|\mathbf{M}_{> 1}\big\| \big]
 \ar\leq\ar  2 \int_0^1\sum |\xi_{3,k,i,j}|\mathbf{1}_{[t_k,t_{k+1})}(s)ds\int_0^\infty|\varphi_i^j(x)|dx\cdot n\gamma_n \int_{\frac{\|u\|}{n}> 1 } \frac{\|u\|}{n} {\nu}^{(n)}_{\mathtt{i}}(du)\cr \cr
 \ar=\ar 2\|\mathbf{X}_3\|_{L^1}\cdot n\gamma_n \int_{\frac{\|u\|}{n}> 1 } \frac{\|u\|}{n} {\nu}^{(n)}_{\mathtt{i}}(du)
 \leq C\cdot n\gamma_n \int_{\frac{\|u\|}{n}> 1 } \frac{\|u\|}{n} {\nu}^{(n)}_{\mathtt{i}}(du).
 \eeqnn
  Along with the assumption that $ \sup_{n\geq 1}n\gamma_n \int_{\mathbb{U}_+}  \|u\|^2 \wedge \|u\| \nu_\mathtt{i}^{(n)}(dnu)<\infty$, we have
 \beqlb\label{eqn3.16}
 \mathbf{P} \Big\{\Big\| \int_0^1\mathbf{X}_3(s-)\cdot \mathbf{Y}_3^{(n)}(ds) \Big\|\geq K  \Big\}\leq \frac{C}{K}+\frac{C}{K^2}.
 \eeqlb
 \qed

We are now ready to prove the weak convergence of the rescaled Hawkes market models.

 \begin{theorem}\label{Thm305}
 Under Condition~\ref{MainCondition01} and \ref{MainCondition02},
 if $(P^{(n)}(0),V^{(n)}(0))\to (P(0),V(0))$ in distribution,
 the rescaled processes $\{(P^{(n)}(t),V^{(n)}(t)):t\geq 0\}$ converge weakly to $\{(P(t),V(t)):t\geq 0\}$ in $\mathbf{D}([0,\infty);\mathbb{U})$ as $n\to\infty$, where $\{(P(t),V(t)):t\geq 0\}$ is the unique strong solution to the following stochastic system:
 \beqlb
 \Big( \begin{array}{c}P(t)\\ V(t) \end{array} \Big)\ar=\ar \Big( \begin{array}{c}P(0)\\ V(0) \end{array} \Big) + \int_0^t (a-bV(s))ds  + \int_0^t \int_0^{V(s)}\sqrt{2\sigma}\Big( \begin{array}{c}W_1(ds,dx)\\  W_2(ds,dx) \end{array} \Big) \cr
 \ar\ar\cr
 \ar\ar + \int_0^t\int_{\mathbb{U}} u N_1(ds,du) +\int_0^t\int_{\mathbb{U}}\int_0^{V(s-)}u\tilde{N}_0(ds,du,dx).\label{eqn3.21}
 \eeqlb
 \end{theorem}
\proof
By \cite[Theorem 2.5]{DawsonLi2012} the stochastic system (\ref{eqn3.21}) admits a unique strong solution $(P,V)$. In order to establish the convergence we recall that
 \begin{equation*}
 \Big( \begin{array}{c}
 P^{(n)}(t)\\ V^{(n)}(t)
 \end{array} \Big)
 = \Big( \begin{array}{c}
 P^{(n)}(0)\\ V^{(n)}(0)
 \end{array} \Big)+ \int_0^t \mathbf{F}(V^{(n)}(s-)) \cdot \mathbf{Y}^{(n)}(ds).
 \end{equation*}
The function $\mathbf{F}$ is unbounded and does not satisfy the requirements of \cite[Theorem 7.5]{KurtzProtter1996}. To overcome this problem, we use a standard localization argument; see \cite{KurtzProtter1991} for details. Let
\[
	\eta^c(\omega) = \inf_{t \geq 0} \{\omega(t) \vee \omega(t-) \geq c\} \quad \mbox{and} \quad \mathbf{F}^c(\omega,s-) :=  {\bf 1}_{[0,\eta^c)}(s) \mathbf{F}(\omega(s-))
\]
for $\omega \in \mathbf{D}([0,\infty), \mathbb{R})$ and $c>0$, and consider the equation
 \begin{equation*}
 \Big( \begin{array}{c}
 P^{(n),c}(t)\\ V^{(n),c}(t)
 \end{array} \Big)
 = \Big( \begin{array}{c}
 P^{(n)}(0)\\ V^{(n)}(0)
 \end{array} \Big)+ \int_0^t \mathbf{F}^c(V^{(n),c},s-) \cdot \mathbf{Y}^{(n)}(ds).
 \end{equation*}
 This system satisfies the assumptions of \cite[Theorem 7.5]{KurtzProtter1996}. Thus, as $n \to \infty$,
 \beqnn
 \big( P^{(n),c}, V^{(n),c} \big) \to  \big( P^{c}, V^{c} \big)
 \eeqnn
 weakly in $\mathbf{D}([0,\infty),\mathbb{U})$ and hence  $\eta^c( V^{(n),c})\to \eta^c(V^{c})$ weakly in $\mathbf{D}([0,\infty),\mathbb{R})$, where $\left( P^{c}, V^{c} \right)$ is the unique strong solution to (\ref{eqn3.21}), restricted to $[0,\eta^c(V^{c}))$. Uniqueness of strong solutions to (\ref{eqn3.21}) also yields $\left( P^{c}, V^{c} \right) \to \left( P, V \right)$ a.s.~as $c \to \infty$.
This shows that $(P^{(n)},V^{(n)}) \to (P,V)$ weakly in $\mathbf{D}([0,\infty),\mathbb{U})$.
\qed

The process $\{ (P(t),V(t)):t\geq 0\}$ is a strong Markov process with infinitesimal generator $\mathcal{A}$ defined by (\ref{eqn3.31}).
 Standard arguments (see \cite[Theorem 7.1' and 7.4]{IkedaWatanabe1989}) show that the limit process $\{ (P(t),V(t)):t\geq 0\}$ also can be represented as a stochastic integral driven by two Brownian motions.

 \begin{corollary}
  Let $\{B_1(t):t\geq 0 \}$ and $\{B_2(t):t\geq 0 \}$ are  two Brownian motions with correlation coefficient $\rho=\frac{\sigma_{12}+\sigma_{21}}{\sqrt{\sigma_{11}\sigma_{22}}}$. The unique strong solution to the stochastic system\footnote{The existence and  uniqueness of strong solution follows from, e.g.~\cite[Theorem 6.2]{DawsonLi2006}.}
 \beqlb
 \Big( \begin{array}{c}P(t)\\ V(t) \end{array} \Big)\ar=\ar \Big( \begin{array}{c}P(0)\\ V(0) \end{array} \Big) + \int_0^t (a-bV(s))ds  +\int_0^t \sqrt{V(s)}d\Big( \begin{array}{c}\sqrt{2\sigma_{11}}B_1(s)\\ \sqrt{2\sigma_{22}}B_2(s) \end{array} \Big) \cr
 \ar\ar\cr
 \ar\ar + \int_0^t\int_{\mathbb{U}} u N_1(ds,du) +\int_0^t\int_{\mathbb{U}}\int_0^{V(s-)}u\tilde{N}_0(ds,du,dx)\label{eqn3.30}
 \eeqlb
 is a realization for the limiting market system.
 The solution is an affine process with characteristic function
 \beqlb\label{eqn3.23}
 \mathbf{E}\big[\exp\{z_1P(t)+ z_2V(t) \}\big]=\exp\Big\{ z_1P(0)+  \psi^G_t(z)V(0)+\int_0^t H(z_1,\psi^G_s(z)) ds \Big\},\quad z\in\mathbb{U}_*,
 \eeqlb
 where
 $\{\psi^G_t(z):t\geq 0\}$ is the unique solution to the following Riccati equation
 \beqlb\label{eqn3.24}
 \psi^G_t(z)\ar=\ar z_2+ \int_0^t  G(z_1,\psi^G_s(z))ds.
 \eeqlb
 \end{corollary}


\subsection{Examples}

We now provide scaling limits for each of the examples considered in Section 2.2. We show that the rescaled market models converge to Heston-type stochastic volatility models with or without jumps, depending on the arrival frequency of large orders.

 \begin{example}[\rm Heston volatility model]\label{Ecample-Exp02}
 Let $\{(P^{(n)}_t,V^{(n)}_t):t\geq 0\}$ be an exponential market  as defined in Example~\ref{Ecample-Exp} with parameters $(1/n, \beta^{(n)}, p_{\mathtt{e}/\mathtt{i}}^{\mathtt{b}/\mathtt{s}},
 \boldsymbol{\lambda}^{\mathtt{b}/\mathtt{s}}_{\mathtt{e}},
 \boldsymbol{\lambda}^{\mathtt{b}/\mathtt{s} (n)}_{\mathtt{i}})$ satisfying
  \beqlb\label{eqn3.33}
   \boldsymbol{\lambda}^{\mathtt{b}/\mathtt{s} (n)}_{\mathtt{i}}\to \boldsymbol{\lambda}^{\mathtt{b}/\mathtt{s} }_{\mathtt{i}} ,
  \quad n\big[p_\mathtt{i}^\mathtt{s}\mathrm{M}^\mathbf{e}_{1,1}(\boldsymbol{\lambda}^{\mathtt{s}(n)}_{\mathtt{i}})- p_\mathtt{i}^\mathtt{b}\mathrm{M}^\mathbf{e}_{1,1}(\boldsymbol{\lambda}^{\mathtt{b}(n)}_{\mathtt{i}})\big]\to 0,\quad n\big[\beta^{(n)}-p_\mathtt{i}^\mathtt{s}\mathrm{M}^\mathbf{e}_{1,2}(\boldsymbol{\lambda}^{\mathtt{s}}_{\mathtt{i}})- p_\mathtt{i}^\mathtt{b}\mathrm{M}^\mathbf{e}_{1,2}(\boldsymbol{\lambda}^{\mathtt{b}}_{\mathtt{i}})\big]\to b_2>0
  \eeqlb
as $n\to\infty$. Let $\gamma_n=n$.
Then Condition~\ref{MainCondition01} holds. The conditions in Proposition~\ref{MainAssumptionNew} hold as well with limiting parameters
 $$
 \hat\nu_\mathtt{e}(\mathbb{U})=\hat\nu_\mathtt{i}(\mathbb{U})=0,
 \quad
 a^+ =p_{\mathtt{e}}^\mathtt{b}\cdot\mathrm{M}^\mathbf{e}_{1}(\boldsymbol{\lambda}^\mathtt{b}_{\mathtt{e}}),
 \quad
 a^- =-p_{\mathtt{e}}^\mathtt{s}\cdot\mathrm{M}^\mathbf{e}_{1}(\boldsymbol{\lambda}^\mathtt{s}_{\mathtt{e}}),
 \quad
 \sigma^+= \frac{p_{\mathtt{i}}^\mathtt{b}}{2} \cdot \mathrm{M}^\mathbf{e}_{2}(\boldsymbol{\lambda}^\mathtt{b}_{\mathtt{i}}),
 $$
 $$
 \sigma^-_{11}= \frac{p_{\mathtt{i}}^\mathtt{s}}{2} \cdot\mathrm{M}^\mathbf{e}_{2,11}(\boldsymbol{\lambda}^\mathtt{s}_{\mathtt{i}}),
 \quad
 \sigma^-_{22}= \frac{p_{\mathtt{i}}^\mathtt{s}}{2}\cdot \mathrm{M}^\mathbf{e}_{2,22}(\boldsymbol{\lambda}^\mathtt{s}_{\mathtt{i}}),
 \quad
 \sigma^-_{12}= \sigma^-_{21}= -\frac{p_{\mathtt{i}}^\mathtt{s}}{2}\cdot \mathrm{M}^\mathbf{e}_{2,12}(\boldsymbol{\lambda}^\mathtt{s}_{\mathtt{i}}).
 $$
Since all order size distributions are light-tailed, there are no jumps in the scaling limit, and the limit model (\ref{eqn3.30}) reduces to the standard Heston volatility model
 \beqnn
 P(t)\ar=\ar P(0)+ a_1 t  +\int_0^t \sqrt{2\sigma_{11}V(s)}d B_1(s) , \cr
 V(t)\ar=\ar V(0)+ \int_0^t b_2\Big[\frac{a_2}{b_2}-V(s)\Big]ds +\int_0^t \sqrt{2\sigma_{22}V(s)}d B_2(s).
 \eeqnn
 \end{example}

Next, we consider an exponential-Pareto mixing market model as defined in Example~\ref{Ecample-Exp-Pareto}. In this case, we obtain the jump-diffusion Heston volatility model as analyzed in \cite{DuffiePanSingleton2000,Pan2002} among many others in the scaling limit.

 \begin{example}[\rm Pareto-jump-diffusion volatility model] \label{Ecample-Pareto-Diffusion}
 Let $\{(P^{(n)}_t,V^{(n)}_t):t\geq 0\}$ be an exponential-Pareto mixing market model as defined in Example~\ref{Ecample-Exp-Pareto} with parameters
 $(1/n,\beta^{(n)}, p^{\mathtt{b/s}}_{\mathtt{e}/\mathtt{i}}; \boldsymbol{\lambda}^{\mathtt{b/s}}_{\mathtt{e}},\boldsymbol{\lambda}^{\mathtt{b/s}(n)}_{\mathtt{i}}
 \alpha_{\mathtt{e}/\mathtt{i}},\boldsymbol{\theta}^{\mathtt{b/s}(n)}_{\mathtt{e}/\mathtt{i}})$
 and selecting mechanism $q^{(n)}_{\mathtt{e}/\mathtt{i}}$ satisfying
  \beqnn
   \alpha_{\mathtt{e}}>0,\quad  \alpha_{\mathtt{i}}=0, \quad q^{(n)}_{\mathtt{i}}=\frac{1}{n^2},\quad q^{(n)}_{\mathtt{e}}=\frac{1}{n},
   \quad
   \boldsymbol{\theta}^{\mathtt{b}/\mathtt{s}(n)}_{\mathtt{e}/\mathtt{i}}
   =n\cdot\boldsymbol{\theta}^{\mathtt{b}/\mathtt{s}}_{\mathtt{e}/\mathtt{i}}
  \eeqnn
 and (\ref{eqn3.33}) as $n\to\infty$.
 Let $\gamma_n=n$. Then, Condition~\ref{MainCondition01} as well as the conditions in Proposition~\ref{MainAssumptionNew} hold
 with limit parameters $a^\pm,\sigma^\pm$ as defined in Example~\ref{Ecample-Exp02}, with $\hat{\nu}_\mathtt{i}(\mathbb{U})=0$, and
 \beqnn
 \hat{\nu}_\mathtt{e}(du)= \sum_{j\in\{\mathtt{b},\mathtt{s}\}} \mathbf{1}_{\mathbb{U}_j}(u)  \cdot p^{j}_\mathtt{e} \cdot \frac{\alpha_\mathtt{e}(\alpha_\mathtt{e}+1)}{\theta^j_{\mathtt{e},1}\theta^j_{\mathtt{e},1}} \Big(1+\frac{|u_1|}{\theta^j_{\mathtt{e},1}} +\frac{u_2}{\theta^j_{\mathtt{e},1}} \Big)^{-\alpha_\mathtt{e}-2}du_1du_2,
 \eeqnn
 where $\mathbb{U}_b=\mathbb{U}_+$ and $\mathbb{U}_s=\mathbb{U}_-$.
 In this case, the limit stochastic system (\ref{eqn3.30}) reduces to the following jump-diffusion Heston volatility model:
 \beqnn
 P(t)\ar=\ar P(0)+ a_1 t  +\int_0^t \sqrt{2\sigma_{11}V(s)}d B_1(s)+ \sum_{k=1}^{N_t} \xi_{k,P}^\mathtt{e}  , \cr
 V(t)\ar=\ar V(0)+ \int_0^t b_2\Big[\frac{a_2}{b_2}-V(s)\Big]ds +\int_0^t \sqrt{2\sigma_{22}V(s)}d B_2(s)+\sum_{k=1}^{N_t} \xi_{k,V}^\mathtt{e} .
 \eeqnn
 where $\{N_t:t\geq 0\}$ is a Poisson process with rate $1$ and $\{(\xi_{k,P}^\mathtt{e},\xi_{k,V}^\mathtt{e} ):k=1,2,\cdots\}$ is a sequence of i.i.d. $\mathbb{U}$-valued random variables with probability law $ \hat{\nu}_\mathtt{e}(du)$.
 \end{example}

 In the previous example co-jumps in prices and volatility emerged in the scaling limit as a result of occasional large exogenous  shocks. The key assumption was that $\gamma_n=n$. The following example considers the scaling limit for $\gamma_n=n^\alpha$ for some $\alpha \in (0,1)$.
  Two types of jumps emerge in our limit: jumps originating from large exogenous orders as well as self-exciting child-jumps. For the special case where the induced orders do not contribute to the intensity of order arrivals, the child-jumps drop out of the model. In this case, the limiting volatility process reduces to a non-Gaussian process of Ornstein-Uhlenbeck type as analyzed in Barndorff-Nielsen and Shephard \cite{Barndorff-NielsenShephard2001}\footnote{Constant time-changes as allowed in Barndorff-Nielsen and Shephard \cite{Barndorff-NielsenShephard2001} can easily be incorporated into our model.}.
 
 \begin{example}[\rm Stable-volatility model without diffusion]\label{Ecample-Pareto02}
 Let $\{(P^{(n)}_t,V^{(n)}_t):t\geq 0\}$ be a Pareto market model as defined in Example~\ref{Ecample-Pareto} with parameters
 $(1/n,\beta^{(n)}, p^{\mathtt{b}/\mathtt{s}}_{\mathtt{e}},p^{{\mathtt{b}/\mathtt{s}}(n)}_{\mathtt{i}}; \alpha_{\mathtt{e}/\mathtt{i}},\boldsymbol{\theta}^{\mathtt{b}/\mathtt{s}}_{\mathtt{e}/\mathtt{i}})$ satisfying that
 $\alpha_{\mathtt{e}}=\alpha_{\mathtt{i}}-1\in(0,1)$.
 Moreover, as $n\to\infty$  we have $p^{{\mathtt{b}/\mathtt{s}}(n)}_{\mathtt{i}}\to p^{\mathtt{b}/\mathtt{s}}_{\mathtt{i}}$ with $p^{\mathtt{b}}_{\mathtt{i}}+p^{\mathtt{s}}_{\mathtt{i}}=1$ and
 \beqlb\label{eqn3.34}
 n^{\alpha_{\mathtt{e}}}\big[ p^{\mathtt{s}(n)}_{\mathtt{i}}\mathrm{M}^\mathbf{p}_{1,1}(\alpha_{\mathtt{i}},\boldsymbol{\theta}^\mathtt{s}_{\mathtt{i}})- p^{\mathtt{b}(n)}_{\mathtt{i}}\mathrm{M}^\mathbf{p}_{1,1}(\alpha_{\mathtt{i}},\boldsymbol{\theta}^\mathtt{b}_{\mathtt{i}})\big]\to b_1,\quad
 n^{\alpha_{\mathtt{e}}}\big[\beta^{(n)}- p^{\mathtt{s}(n)}_{\mathtt{i}}\mathrm{M}^\mathbf{p}_{1,2}(\alpha_{\mathtt{i}},\boldsymbol{\theta}^\mathtt{s}_{\mathtt{i}})- p^{\mathtt{b}(n)}_{\mathtt{i}}\mathrm{M}^\mathbf{p}_{1,2}(\alpha_{\mathtt{i}},\boldsymbol{\theta}^\mathtt{b}_{\mathtt{i}})\big]\to b_2.
 \eeqlb
 Let $\gamma_n=n^{\alpha_{\mathtt{e}}}$.Then Condition~\ref{MainCondition01} as well as the conditions in Proposition~\ref{MainAssumptionNew} hold with the following limit parameters: $a^+ =a^- = 0$, $\sigma^+= \sigma^- = 0$, and
 \beqnn
 \hat{\nu}_i(du)= \sum_{j\in\{\mathtt{b},\mathtt{s}\}} \mathbf{1}_{\mathbb{U}_j}(u) \cdot p^{j}_i \cdot \frac{\alpha_i(\alpha_i+1)}{\theta^j_{i,1}\theta^j_{i,1}} \Big(\frac{|u_1|}{\theta^j_{i,1}} +\frac{u_2}{\theta^j_{i,1}} \Big)^{-\alpha_i-2}du_1du_2,\quad i\in\{\mathtt{e},\mathtt{i}\}.
 \eeqnn
 In this case, the limit stochastic system (\ref{eqn3.30}) reduces to the following pure-jump model:
 \beqlb\label{eqn3.32}
 \Big( \begin{array}{c}P(t)\\ V(t) \end{array} \Big)\ar=\ar \Big( \begin{array}{c}P(0)\\ V(0) \end{array} \Big) - \int_0^t  bV(s)ds + \int_0^t\int_{\mathbb{U}} u N_1(ds,du) +\int_0^t\int_{\mathbb{U}}\int_0^{V(s-)}u\tilde{N}_0(ds,du,dx).
 \eeqlb

 The dynamics (\ref{eqn3.32}) can be represented in more convenient way.  Let  $Z_{\alpha_{\mathtt{e}}}(t)$ denote the third term on the right side of (\ref{eqn3.32}). By Theorem~14.3(ii) in \cite{Sato1999}, $\{Z_{\alpha_{\mathtt{e}}}(t):t\geq 0 \}$ is an $\mathbb{U}$-valued $\alpha_{\mathtt{e}}$-stable processes.
 As for the last term, let us introduce a random measure $N_{\alpha_{\mathtt{i}}}(ds,du)$ on $[0,\infty)\times\mathbb{U}$ as follows: for any $t\geq 0$ and $U\subset\mathbb{U}$,
 \beqnn
 N_{\alpha_{\mathtt{i}}}((0,t],U)=  \int_0^t\int_{\mathbb{U}}\int_0^{V(s-)}  \mathbf{1}_{\{ u \in \sqrt[\alpha_{\mathtt{i}}]{V(s-)}\cdot U \}}N_0(ds,du,dx).
 \eeqnn
Standard computations show that its predictable compensator has the following representation:
  \beqnn
  \hat{N}_{\alpha_{\mathtt{i}}}((0,t],U)
  \ar=\ar\sum_{j\in\{\mathtt{b},\mathtt{s}\}}p^{j}_\mathtt{i} \cdot \int_0^tds\int_{\mathbb{U}_j}V(s-)  \mathbf{1}_{\{ u \in \sqrt[\alpha_{\mathtt{i}}]{V(s-)}\cdot U \}} \frac{\alpha_\mathtt{i}(\alpha_\mathtt{i}+1)}{\theta^j_{\mathtt{i},1}\theta^j_{\mathtt{i},2}} \Big( \frac{|u_1|}{\theta^j_{\mathtt{i},1}} +\frac{u_2}{\theta^j_{\mathtt{i},2}}\Big)^{-\alpha_\mathtt{i}-2}du_1du_2\cr
  \ar=\ar \sum_{j\in\{\mathtt{b},\mathtt{s}\}}p^{j}_\mathtt{i} \cdot \int_0^tds\int_{\mathbb{U}_j}   \mathbf{1}_{\{ u \in   U \}} \frac{\alpha_\mathtt{i}(\alpha_\mathtt{i}+1)}{\theta^j_{\mathtt{i},1}\theta^j_{\mathtt{i},2}} \Big( \frac{|u_1|}{\theta^j_{\mathtt{i},1}} +\frac{u_2}{\theta^j_{\mathtt{i},2}}\Big)^{-\alpha_\mathtt{i}-2}du_1du_2.
  \eeqnn
Thus, $N_{\alpha_{\mathtt{i}}}(ds, du)$ is a Poisson random measure on $\mathbb{U}$ with intensity $ds\hat\nu_{\mathtt{i}}(du)$ and the L\'evy processes $\{Z_{\alpha_{\mathtt{i}}}(t):t\geq 0\}$ defined by
  \beqnn
  Z_{\alpha_{\mathtt{i}}}(t):= \int_0^t\int_{\mathbb{U}} u\tilde{N}_{\alpha_{\mathtt{i}}}(ds,du)
  \eeqnn
  is a compensated, $\mathbb{U}$-valued $\alpha_{\mathtt{i}}$-stable process.
  We can now rewrite   (\ref{eqn3.32})  into
  \beqnn
  \Big( \begin{array}{c}P(t)\\ V(t) \end{array} \Big)\ar=\ar \Big( \begin{array}{c}P(0)\\ V(0) \end{array} \Big) + Z_{\alpha_{\mathtt{e}}}(t) - \int_0^t  bV(s)ds  +\int_0^t \sqrt[\alpha_{\mathtt{i}}]{V(s-)} dZ_{\alpha_{\mathtt{i}}}(s).
  \eeqnn
 \end{example}

 So far, we obtained jump-diffusion models with {\sl exogenous} jump dynamics as well as pure jump models with {\sl endogenous} jump dynamics as scaling limits.
 The next example combines both dynamics into a single model.
 We call this model {\sl generalized $\alpha$-stable Heston volatility model}.
 We assume that induced orders arrive at a rate $n^2$ as in Example \ref{Ecample-Pareto-Diffusion}.
 However, we now assume that large orders arrive with much higher probabilities.
 In Example \ref{Ecample-Pareto-Diffusion} large exogenous and induced orders arrive at rate one per unit time.
 Now we assume they arrive at rate $n^{\alpha}$ per unit time; despite the increased rate, their proportion among all orders will still be negligible in the limit.
 Several existing models are obtained as special cases.
 If the arrival intensity of induced orders depends on exogenous orders only, the model reduces to the stochastic volatility model studied in Barndorff-Nielsen and Shephard \cite{Barndorff-NielsenShephard2001}.
 If, in addition, there are no jumps in the volatility, the model reduces to that analyzed in Bates \cite{Bates1996}.
 The special case with no child-jumps and no exogenous jumps in prices corresponds to the model studied in Nicolato et al. \cite{NicolatoPisaniSloth2017}.
 The special case without exogenous jumps corresponds to the alpha Heston model that has recently been studied in Jiao et al. \cite{JiaoMaScottiZhou2018}.
 The multi-factor model of Bates \cite{Bates2019} with both exogenous and self-excited shocks is also contained as a special case.

 \begin{example}[\rm Generalized $\alpha$ Heston volatility model] \label{Ecample-Stable-Diffusion}
Let $\{(P^{(n)}_t,V^{(n)}_t):t\geq 0\}$ be an exponential-Pareto mixing market model as defined in Example~\ref{Ecample-Exp-Pareto} with parameters
 $(1/n,\beta^{(n)}, p^{\mathtt{b}/\mathtt{s}}_{\mathtt{e}/\mathtt{i}}; \boldsymbol{\lambda}^{\mathtt{b}/\mathtt{s}}_{\mathtt{e}}, \boldsymbol{\lambda}^{\mathtt{b}/\mathtt{s}(n)}_{\mathtt{i}};\alpha_{\mathtt{e}/\mathtt{i}},\boldsymbol{\theta}^{\mathtt{b}/\mathtt{s}}_{\mathtt{e}/\mathtt{i}})$
 and selecting mechanism $q^{(n)}_{\mathtt{e}/\mathtt{i}}$ satisfying that $\alpha_{\mathtt{e}}\in(0,1)$, $\alpha_{\mathtt{i}}\in(1,2)$, $q^{(n)}_{\mathtt{i}}=n^{\alpha_{\mathtt{i}}-2}$, $q^{(n)}_{\mathtt{e}}=n^{\alpha_{\mathtt{e}}-1}$ and  (\ref{eqn3.33}) holds as $n\to\infty$.
 Let $\gamma_n=n$, we see that Condition~\ref{MainCondition01} and conditions in Proposition~\ref{MainAssumptionNew} hold with limit parameters $a^\pm,\sigma^\pm$ defined in Example~\ref{Ecample-Exp02} and $b,\hat{\nu}_{\mathtt{e}/\mathtt{i}}(du)$ defined in Example~\ref{Ecample-Pareto02}.
 In this case, the limit stochastic system (\ref{eqn3.30}) reduces to
 \beqnn
 \Big( \begin{array}{c}P(t)\\ V(t) \end{array} \Big)\ar=\ar \Big( \begin{array}{c}P(0)\\ V(0) \end{array} \Big) +\int_0^t (a- bV(s))ds +\int_0^t \sqrt{V(s)}d\Big( \begin{array}{c}\sqrt{2\sigma_{11}}B_1(s)\\ \sqrt{2\sigma_{22}}B_2(s) \end{array} \Big)
 +Z_{\alpha_{\mathtt{e}}}(t)   +\int_0^t \sqrt[\alpha_{\mathtt{i}}]{V(s-)} dZ_{\alpha_{\mathtt{i}}}(s). \qquad
 \eeqnn
\end{example}

\section{The genealogy of the limiting market dynamics}\label{gen-limit}
	\setcounter{equation}{0}
	
In this section, we analyze the impact of exogenous shocks on the jump dynamics of the limiting model. Exogenous shocks can be split in two groups: shocks of positive magnitude and shocks of ``insignificant magnitude''. Shocks of positive magnitude are captured by the Poisson random measure $N_1$. As argued in Section \ref{HFHM} (see Remark \ref{remark01}), shocks of ``insignificant magnitude'' are captured by the drift vector $a \in \mathbb{R}\times \mathbb{R}_+$. Both types of shocks can trigger jump cascades. Cascades triggered by shocks of positive magnitude are exogenous while cascades triggered by shocks of insignificant magnitude are endogenous in nature. 

%

Rewriting the characteristic function (\ref{eqn3.23}) as

 \beqlb\label{eqn4.01}
 \mathbf{E}\left[e^{z_1P(t)+ z_2V(t) }\right]\ar=\ar\exp\big\{ z_1P(0)+  \psi^G_t(z)V(0)\big\}\cdot\exp\Big\{\int_0^tds \int_{\mathbb{U}} [e^{z_1u_1+ \psi^G_s(z)u_2}-1]\hat{\nu}_\mathtt{e}(du) \Big\} \cr
  \ar\ar
  \times
  \exp\Big\{z_1\cdot a_1t +  a_2\int_0^t\psi^G_s(z)ds\Big\}
 \eeqlb
 the limit model can be decomposed into independent, self-enclosed sub-models in terms of the strong Markov processes induced by each of the exponential functions on the right side of the above equation.

\begin{itemize}
\item[(i)]
 From the semigroup property of $\{\psi^G_t(z):t\geq 0\}$ we conclude that the first term on the right side of (\ref{eqn4.01}) induces a Markov semigroup $(Q_{0,t})_{t\geq0}$ on $\mathbb{U}$ via 
 \beqlb\label{eqn4.02}
 \int_0^\infty e^{\langle z,u\rangle}Q_{0,t}(u',du)=  \exp\big\{z_1u'_1+\psi^G_t(z)u'_2\big\}.
 \eeqlb
 Applying the Kolmogorov consistency theorem and the relationship between the solution to (\ref{eqn3.21}) and its characteristic function (\ref{eqn3.23}),  we see that the Markov process $\{(P_0(t),V_0(t)):t\geq 0 \}$ that solves
 \beqlb\label{eqn4.19}
 \Big( \begin{array}{c}P_0(t)\\ V_0(t) \end{array} \Big)\ar=\ar \Big( \begin{array}{c}P(0)\\ V(0) \end{array} \Big) - \int_0^tbV_0(s)ds  + \int_0^t \int_0^{V_0(s)}\sqrt{2\sigma}\Big( \begin{array}{c}W_1(ds,dx)\\  W_2(ds,dx) \end{array} \Big) \cr
 \ar \ar +\int_0^t\int_{\mathbb{U}}\int_0^{V_0(s-)}u\tilde{N}_0(ds,du,dx)
 \eeqlb
 is a realization of the transition semigroup $(Q_{0,t})_{t\geq0}$. This self-enclosed market model captures the impact of all events prior to time $0$ on future order flow. We emphasize that the model is independent of the drift and that the volatility mean-reverts to the level zero if $b_2>0$.

\item[(ii)] Let $\{\mathbf{P}_{e,t}(\cdot):t\geq 0\}$ be the family of probability laws induced by second term on the right side of (\ref{eqn4.01}).
The family of kernels $(Q_{e,t})_{t\geq 0}$ on $\mathbb{U}$ defined by $Q_{e,t}(u',du) :=  \mathbf{P}_{e,t}* Q_{0,t}(u',du)$ is a Markov semigroup. The Markov process $\{(P_{e}(t),V_{e}(t)):t\geq 0 \}$ that solves
 \beqlb\label{eqn4.22}
 \Big( \begin{array}{c}P_{e}(t)\\ V_{e}(t) \end{array} \Big)\ar=\ar
 \int_0^t u N_1(ds,du) + \int_0^t \int_{V_0(s)}^{V_0(s)+V_{e}(s)}\sqrt{2\sigma}\Big( \begin{array}{c}W_1(ds,dx)\\  W_2(ds,dx) \end{array} \Big) \cr
 \ar\ar - \int_0^t bV_{e}(s)ds   +\int_0^t\int_{\mathbb{U}}\int_{V_0(s-) }^{V_0(s-)+V_{e}(s-)}u\tilde{N}_0(ds,du,dx)
 \eeqlb
 is a realization of the transition semigroup $(Q_{e,t})_{t\geq0}$. We can interpret this model as describing the cumulative impact of the exogenous orders on the market dynamics. Again, this model is independent of the drift and the volatility mean-reverts to the level zero if $b_2>0$.

\item[(iii)]
Let $\{\mathbf{P}_{a,t}(\cdot):t\geq 0\}$ be the family of probability laws induced by the last term on the right side of (\ref{eqn4.01}), and $(Q_{a,t})_{t\geq 0}$ be the Markov semigroup on $\mathbb{U}$ given by $Q_{a,t}(u',du):=  \mathbf{P}_{a,t}* Q_{0,t}(u',du)$.
 The Markov process $\{(P_a(t),V_a(t)):t\geq 0 \}$ that solves
 \beqlb\label{eqn4.20}
 \Big( \begin{array}{c}P_a(t)\\ V_a(t) \end{array} \Big)\ar=\ar  \int_0^t(a-bV_a(s))ds  + \int_0^t \int_{V_0(s)+V_{e}(s)}^{V_0(s)+V_{e}(s)+V_a(s)}\sqrt{2\sigma}\Big( \begin{array}{c}W_1(ds,dx)\\  W_2(ds,dx) \end{array} \Big) \cr
 \ar\ar  +\int_0^t\int_{\mathbb{U}}\int_{V_0(s-)+V_{e}(s-)}^{V_0(s-)+V_{e}(s-)+V_a(s-)}u\tilde{N}_0(ds,du,dx)
 \eeqlb
is a realization of the transition semigroup $(Q_{a,t})_{t\geq0}$. Unlike the two other sub-models this sub-model depends on the drift $a$. We can interpret this model as describing the impact of the drift of the volatility process on the market dynamics.
\end{itemize}

By the spatial-orthogonality of the Gaussian white noises $W_1(ds,du)$, $W_2(ds,du)$  and the Poisson random measure $N_0(ds,du,dx)$, the processes defined by (\ref{eqn4.19})-(\ref{eqn4.20}) are mutually independent. Their sum equals the process (\ref{eqn3.21}).

 \begin{theorem}
 The following decomposition of the solution $\{(P(t),V(t)):t\geq 0\}$ to (\ref{eqn3.21}) holds:
 \beqnn
 \{(P(t),V(t)):t\geq 0\}\overset{\rm a.s.}=\Big\{\Big(
 \begin{array}{c}
 P_0(t)+P_{e}(t)+ P_a(t) \cr
 V_0(t)+V_{e}(t) +V_a(t)
 \end{array}\Big):t\geq 0\Big\}.
 \eeqnn
 \end{theorem}

The sub-model $\{(P_e(t),V_e(t)):t\geq 0\}$ can be decomposed into a sum of independent and identically distributed sub-models of the form (i). To this end, we denote by ${\mathbf Q}_{p,v}$ the distribution of the process (\ref{eqn4.19}) with initial state $(P(0),V(0))=(p,v)$. Then,
\[
	 ({P}_e(t),{V}_e(t)):= \int_0^t\int_{\mathbb{U}} \int_{\mathbf{D}([0,\infty),\mathbb{U})} \omega(t-s)N_e(ds,d(p,v),d\omega)
\]
 where $N_e(dt,d\omega)$ is a Poisson random measure on $(0,\infty)\times\mathbb{U}\times \mathbf{D}([0,\infty),\mathbb{U})$ with intensity $ ds \hat\nu_{\mathtt{e}}(d(p,v))\mathbf{Q}_{p,v}(d\omega) $.
 The sub-model $\{(P_a(t),V_a(t)):t\geq 0\}$ can also be decomposed into self-enclosed sub-models where the volatility process evolves as an excursion process selected by a Poisson random measure.
 In order to make this more precise, we put $\tau_0(\omega):=\inf\{ t>0: \omega_2(t)=0 \}$ for any $\omega(\cdot):= (\omega_1(\cdot),\omega_2(\cdot))\in \mathbf{D}(\mathbb{R}_+,\mathbb{U})$, denote by $\mathbf{D}_0(\mathbb{R}_+,\mathbb{U})$ the subspace of $\mathbf{D}(\mathbb{R}_+,\mathbb{U})$ defined by
 \beqnn
 \mathbf{D}_0(\mathbb{R}_+,\mathbb{U})\ar: =\ar\{ \omega\in \mathbf{D}(\mathbb{R}_+,\mathbb{U}):   \omega(0)=(0,0)\mbox{ and } \omega_2(t)=0 \mbox{ for any } t\geq \tau_0(\omega)\},
 \eeqnn
 endowed with the filtration $\mathscr{G}_t:=\sigma(\omega(s):s\in[0,t])$, for $t \geq 0$, and by $(Q_{0,t}^\circ)_{t\geq 0}$ the restriction of the semigroup $(Q_{0,t})_{t\geq 0}$ on $\mathbb{R}\times (0,\infty)$. The following result is proved in the appendix.

\begin{theorem} \label{thm-cluster}
Let $\sigma_{22} > 0$. There exists an $\sigma$-finite measure $\mathbf{Q}(d\omega)$\footnote{We construct such a measure $\mathbf{Q}(d\omega)$ in (\ref{eqn4.31}).} on $\mathbf{D}([0,\infty),\mathbb{U})$ with support $\mathbf{D}_0([0,\infty),\mathbb{U})$, and a  Poisson random measure $N_a(ds,d\omega)$ on $(0,\infty)\times \mathbf{D}_0([0,\infty),\mathbb{U})$ with intensity $a_2 ds \mathbf{Q}(d\omega)$ such that the stochastic system $\{(\hat{P}_a(t),\hat{V}_a(t)):t\geq 0 \}$ defined by
 \beqlb\label{eqn4.18}
 (\hat{P}_a(t),\hat{V}_a(t)):= \int_0^t (a_1,0)ds+ \int_0^t \int_{\mathbf{D}_0([0,\infty),\mathbb{U})} \omega(t-s)N_a(ds,d\omega)
 \eeqlb
 is Markov with respect to the filtration $(\mathscr{E}_t)_{t\geq 0}$ with $\mathscr{E}_t:=\sigma(N_a((0,s],U): s\in[0,t],\ U\in\mathscr{G}_{t-s})$ and has the same finite dimensional distributions as $\{(P_a(t),V_a(t)):t\geq 0 \}$.
 \end{theorem}

 \begin{remark}
 Let us benchmark our decomposition result against that of Jiao et al.~\cite{JiaoMaScottiZhou2018}. 
 The variance process studied in their paper is a special case of  (\ref{eqn4.20}). 
 By considering the jumps larger than some threshold $\bar{y}>0$ as immigrations of  a CB-process, they provided a decomposition for the variance processes as a sum of a truncated variance process $V^{(y)}_\cdot$ with jump threshold $\bar{y}$ and a sub-model of the form (\ref{eqn4.22}) with $N_1(ds,du)$ replaced by a Poisson random measure with intensity $V^{(y)}_{s-}ds \mathbf{1}_{u_2\geq \bar{y}} \hat\nu_{\mathtt{i}}(du)$\footnote{We emphasize that the assumption of a Poisson arrival of exogenous shocks was make for mathematical convenience and to better distinguish the effects of Poisson and Hawkes arrivals. An extension to Hawkes arrivals is not difficult}. 
 Their decomposition offers a way to refine the sub-model (\ref{eqn4.20}), which is different to our cluster representation in Theorem~\ref{thm-cluster}.
 However, their truncated variance process $V^{(y)}_\cdot$ is not self-enclosed. 
 Moreover, we believe that our decomposition results is economically more intuitive as it is based on the different origins of the jumps.
 \end{remark}


 \subsection{The sub-model $\{(P_0(t),V_0(t)):t\geq 0\}$ }

In this section we study the sub-model $\{(P_0(t),V_0(t)):t\geq 0\}$, that is, the impact of an exogenous shock of magnitude $(P(0), V(0))$ on the market dynamics.
 By \cite[Theorem 1.1]{KawazuWatanabe1971} and arguments given in \cite{Grey1974}, the volatility process $\{V_0(t):t\geq 0\}$ is a continuous-state branching process, and $\mathbf{P}\{\lim_{t\to\infty }V_0(t)\in \{0,\infty\}\}=1$.
 Let $$\mathscr{T}_{0} := \inf\{t\geq 0: V_0(t)=0 \}$$ be its first hitting time of $0$.
 Since $0$ is a trap for this process, $V_0(t) = 0$ for all $t \geq \mathscr{T}_{0}$ almost surely.
 The following lemma can be viewed as an analogue of Corollary \ref{Thm207}. It analyzes the impact duration of exogenous shocks on the market dynamics in terms of the function
 \beqlb\label{eqn4.54}
 G_0(x) \ar := \ar G(0,-x) = b_2 x + \sigma_{22} x^2 +\int_{\mathbb{U}} (e^{-u_2 x}-1+ u_2 x )\hat\nu_\mathtt{i}(du), \qquad x \geq 0.
 \eeqlb

 \begin{lemma}[\rm Grey (1974)] \label{Grey}
For any $t>0$, $\mathbf{P}\{\mathscr{T}_{0}\leq t\}>0$ if and only if there exists a constant $\vartheta>0$ such that
 \beqlb\label{eqn4.52}
 G_0(\vartheta)>0\quad \mbox{and}\quad \int_\vartheta^\infty \frac{1}{G_0(x)}dx<\infty.
 \eeqlb
 In this case, $\mathbf{P}\{\mathscr{T}_{0}\leq t\}=\exp\{-V(0) \cdot \overline{v}_t \}$, where $\{\overline{v}_t :t\geq 0 \}$ is the minimal solution to the following Riccati differential equation
 \beqnn
 \frac{d}{dt}\overline{v}_t= -G_0(\overline{v}_t)
 \eeqnn
 with singular initial condition $\overline{v}_0=\infty$.
 Moreover, $\mathbf{P}\{\mathscr{T}_{0}<\infty\}=\exp\big\{-V(0) \cdot \overline{v}_\infty \big\}$ with $\overline{v}_\infty $ being the largest root of $G_0(x)=0$, and  $\overline{v}_\infty =0$ if and only if $b_2= G'_0(0)\geq 0$.
 \end{lemma}

\begin{remark}
 The integral condition (\ref{eqn4.52}) holds if, for instance, $\sigma_{22}>0$. In the absence of jumps, this condition is also necessary.
 \end{remark}

 We now consider the distribution of the induced jumps with magnitude $(p,v)\in(0,\infty)^2$ or larger.
 To this end, let $A:= \{u\in\mathbb{U}: |u_1|\geq p,|u_2|\geq v\}$, let $$\mathcal{T}_A:= \sup \{t\geq 0: (\Delta P_0(t),\Delta V_0(t))\in A\}$$ denote the arrival time of the last jump whose magnitude belongs to $A$, and for any $T\in[0,\infty]$, let
 \beqnn
 \mathcal{J}_A(T):= \#\{t\in[0,T]: (\Delta P_0(t),\Delta V_0(t))\in A\}=\int_0^T \int_{A} \int_0^{V_0(s-)}  N_0(ds,du,dx)
 \eeqnn
 denote the number of jumps with magnitudes in $A$ during the time interval $[0,T]$. Since the process $\{V_0(t):t\geq 0\}$ is conservative, i.e. $\mathbf{P}\{\sup_{s\in[0,T]}V_0(s)<\infty\}=1$, $\mathcal{T}_A<\infty$ if and only if $\mathcal{J}_A(\infty)<\infty$.
By definition,
 \beqnn
 \{\mathcal{T}_A\leq r, \mathscr{T}_{0}\leq t\} =  \Big\{ \int_r^t \int_{A} \int_0^{V_0(s-)}  N_0(ds,du,dx)=0, V_0(t)=0  \Big\}
 \eeqnn
almost surely, and so
\[
	\mathbf{P} \{\mathcal{T}_A\leq r, \mathscr{T}_{0}\leq t\} = \mathbf{E}\left[ \exp \left\{-\hat\nu_\mathtt{i}(A) \cdot\int_r^t V_0(s) ds \right\} {\bf 1}_{\{\mathcal{T}_A\leq r, V_0(t)=0\}} \right].
\]
The indicator function is inconvenient when computing the expected value. To bypass this problem we use a result from He and Li \cite{HeLi2016}. Under the assumptions of Lemma \ref{Grey}, the event $\{\mathcal{T}_A\leq r\}$ has strictly positive probability for any $r > 0$. Conditioned on this event, the process $\{V_0(t)  :t\geq r\}$ almost surely equals the process $\{V^A_0(t)  :t\geq r\}$ defined by
 \beqnn
 V^A_0(t)  \ar=\ar   V_0(r)  - \int_r^t\Big(b_2+\int_A u_2\hat{\nu}_{\mathtt{i}}(du)\Big)V^A_0(s)ds  + \int_r^t   \sqrt{2\sigma_{22}V^A_0(s)} dB_2(s) \cr
 	\ar \ar  +\int_r^t\int_{A^{\rm c}}\int_0^{V^A_0(s-)}u_2\tilde{N}_0(ds,du,dx).
 \eeqnn
 By Theorem 2.2 in \cite{DawsonLi2012}, $V_0^A(t) \leq V_0(t)$ for all $t \geq r$.
 In particular, under the conditions of Lemma \ref{Grey}, $\mathbf{P}\{V^A_0(t) = 0\}> 0$ for all $t > r$. Using the fact that $0$ is a trap for the volatility process, straightforward modifications of arguments given in the proof of \cite[Theorem 3.1]{HeLi2016} show that for any $t > r$,
 \beqnn
 \{\mathcal{T}_A\leq r, \mathscr{T}_{0}\leq t\} =  \Big\{ \int_r^t \int_{A} \int_0^{V^A_0(s-)}  N_0(ds,du,dx)=0, V^A_0(t)=0  \Big\}.
 \eeqnn
Using the dominated convergence theorem, and the fact that $\{(V^A_0(t),\int_0^t V^A_0(w)dw):t\geq r\}$ is an affine process,
 \beqlb\label{eqn4.55}
 \mathbf{P}_{\mathscr{F}_r}\{\mathcal{T}_A\leq r,\mathscr{T}_{0}\leq t\}\ar=\ar \lim_{\lambda_0\to\infty}\mathbf{E}_{\mathscr{F}_r}\Big[ \int_r^t \int_{A} \int_0^{V^A_0(s-)}  N_0(ds,du,dx)=0,\ \exp\big\{-\lambda_0\cdot V^A_0(t)\big\}  \Big] \cr
 \ar=\ar \lim_{\lambda_0\to\infty}\mathbf{E}_{\mathscr{F}_r}\Big[ \exp\Big\{-\hat\nu_\mathtt{i}(A)\int_r^t V^A_0(s)ds  -\lambda_0\cdot V^A_0(t)\Big\}  \Big] \cr
& = & \lim_{\lambda_0\to\infty} \exp\big\{-\psi^A_{t-r}(\lambda_0) V_0(r)\big\}, 
 \eeqlb
    where $\psi^A_\cdot(\lambda_0 ): [0,\infty) \to \mathbb R_+$  is the unique positive solution to the Riccati differential equation
     \beqlb\label{eqn4.28}
      \psi^A_t(\lambda_0)=\lambda_0+\int_0^t \big[\hat\nu_\mathtt{i}(A)- G^A_0\big(\psi^A_w(\lambda_0)\big)\big]dw,
 \eeqlb
and the function $G^A_0: \mathbb R_+ \to \mathbb R$ is defined by
\[
	G^A_0(x) := G_0(x)-\int_A (e^{-xu_2}-1)\hat{\nu}_{\mathtt{i}}(du);
 \]
see Theorem~2.7 in Duffie et al.~\cite{DuffieFilipovicSchachermayer2003} for details.
The following theorem shows that the distribution of the random variable $\left(\mathscr{T}_0, \mathcal{T}_A, \mathcal{J}_A(T)\right)$ can be expressed in terms of the unique continuous solution to the Riccati equation (\ref{eqn4.28}) with singular initial condition, and the unique non-negative continuous solution to the Riccati differential equation
 \beqlb
  \frac{d}{ds}\phi^A_s(x,\lambda)
 =  -   G_0(\phi^A_s(x,\lambda)  )
  - (e^{-\lambda}-1)   \int_A \exp\{ -  \phi^A_s (x,\lambda) \cdot u_2 \}  \hat{\nu}_{\mathtt{i}}(du)   \label{eqn4.44}
 \eeqlb
 with initial condition $\phi^A_0(x,\lambda) = x$ for $x,\lambda\geq 0$.

 \begin{theorem}
 Suppose that (\ref{eqn4.52}) holds for some $\vartheta>0$. In the class of continuous functions there exists a minimal positive solution to the Riccati equation
 \begin{equation}
  \frac{d}{ds}
 \bar\psi^A_s =  \hat\nu_\mathtt{i}(A) - G^A_0(\bar\psi^A_s), \label{eqn4.25}
 \end{equation}
 with singular initial condition $\bar\psi^A_0 = \infty$. The function is finite on $(0,\infty)$, and for any $t> r\geq 0$ and $\lambda\geq 0$,
 \beqlb\label{eqn4.27}
 \mathbf{E}\Big[\exp\big\{-\lambda \mathcal{J}_A(T)  \big\} ;\mathcal{T}_A\leq r,\mathscr{T}_{0}\leq t\Big]
 =\exp\Big\{  - \phi^A_{r\wedge T}\Big(-\psi^G_{(r-T)^+}\big(0,-\bar\psi^A_{t-r} \big),\lambda\Big)\cdot V(0) \Big\}.
 \eeqlb
 \end{theorem}
 \proof
    By \cite[Proposition 6.1]{DuffieFilipovicSchachermayer2003}, the solution $\psi^A_\cdot(\cdot)$ to the Riccati equation (\ref{eqn4.28}) is continuous in both variables. From this and the uniqueness of the solution, we conclude that the ODE satisfies a comparison principle. In particular, $\psi^A_t(\lambda_0)$ is increasing in $\lambda_0$ for any $t \geq 0$, and hence the limit $\bar{\psi}^A_t:=\lim_{\lambda_0\to \infty}\psi^A_t(\lambda_0)$ exists in $[0,\infty]$. In order to prove that $\bar{\psi}^A_t < \infty$ for any $t > 0$, we first conclude from (\ref{eqn4.55}) that
    \beqnn
	\exp\big\{-\bar\psi^A_{t-r} V_0(r)\big\} 
	\ar=\ar \lim_{\lambda_0 \to \infty} \mathbf{E}_{\mathscr{F}_r} \Big[ \exp\Big\{-\hat\nu_\mathtt{i}(A)\int_r^t V^A_0(s)ds  -\lambda_0 \cdot V^A_0(t)\Big\}  \Big] \\
	\ar=\ar \mathbf{E} \Big[ \exp\Big\{-\hat\nu_\mathtt{i}(A)\int_r^t V^A_0(s)ds   \Big\};V^A_0(t)=0  \Big].
    \eeqnn
Since $\{V_0(t): t \geq 0\}$ is conservative, so is $\{ V^A_0(t): t \geq r\}$ and hence $\int_r^t V^A_0(s)ds<\infty.$ Thus, the expectation on the right side of the last equality is positive since $\mathbf{P}\{V^A_0(t) = 0\} > 0$ for all $t > r$. This shows that $\bar\psi^A_{t-r}<\infty$  for any $t>r$. Stochastic continuity of $\{V^A_0(t) : t \geq r\}$ yields continuity of $\bar\psi_\cdot^A$ on $(0,\infty)$.

We now show that $\{\bar\psi^A_{t}:t>0\}$ solves (\ref{eqn4.25}) with singular initial condition.
Using the semigroup property $\psi^A_{t+s}(\lambda_0)= \psi^A_t(\psi^A_s(\lambda_0))$ for any $t,s> 0$ and letting $\lambda_0 \to\infty$ shows that $\bar{\psi}^A_{t+s}= \psi^A_t(\bar{\psi}^A_s)$, from which we deduce that $t\mapsto\bar{\psi}^A_t$ is a continuous solution to (\ref{eqn4.25}). Moreover, for any sequence  $\{s_n:n\geq 1\}$ that satisfies that $s_n\to 0+$, it follows from $\sup_{n\geq 1}\bar\psi^A_{s_n}\geq \sup_{n\geq 1}\psi^A_{s_n}(\lambda_0)= \lambda_0$ that $\lim_{t\to0+} \bar\psi^A_t =\infty$.
Finally, the constructed solution is also the minimal solution in the class of continuous functions. Indeed, if $\psi(\cdot)$ is another continuous solution to (\ref{eqn4.25}) with $\psi(0+)=\infty$, then the semigroup property yields $\psi(t)=\lim_{s\to 0+}\psi(t+s)=\lim_{s\to 0+}\psi^A_t(\psi(s))\geq \psi^A_t(\lambda_0)$ for any $t>0$ and all $\lambda_0\geq 0$, from which the assertion follows.

 We finally prove (\ref{eqn4.27}). For any $t>r\geq 0$, it follows from (\ref{eqn4.55}) that
 \beqnn
 \lefteqn{\mathbf{E}\big[\exp\{-\lambda \mathcal{J}_A(T)  \} ;\mathcal{T}_A\leq r,\mathscr{T}_{0}\leq t\big]
  =\mathbf{E}\big[\exp\{-\lambda \mathcal{J}_A(T\wedge r)  \} \times\mathbf{E}_{\mathscr{F}_{r\wedge T}}\big[\exp\big\{-\bar\psi^A_{t-r} V_0(r)\big\}\big]\big]}\qquad \qquad\qquad\ar\ar\cr
  \ar\ar\cr
 \ar=\ar \mathbf{E}\Big[\exp\Big\{-\int_0^{r\wedge T}\int_A\int_0^{V_0(s-)} \lambda N_0(ds,du,dx)+\psi^G_{(r-T)^+}\big(0,-\bar\psi^A_{t-r}\big)\cdot V_0(r\wedge T) \Big\} \Big].
 \eeqnn
 Following the standard arguments given in, e.g.~proof of \cite[Theorem 2.12]{DuffieFilipovicSchachermayer2003}, we have
 \beqnn
 \lefteqn{\exp\Big\{  -\int_0^{r\wedge T}\int_A\int_0^{V_0(s-)} \lambda N_0(ds,du,dx)+ \psi^G_{(r-T)^+}\big(0,-\bar\psi^A_{t-r}\big) \cdot V_0(r\wedge T) \Big\}}\qquad\ar\ar\cr
 \ar\ar\cr
 \ar=\ar
 \exp\Big\{  - \phi^A_{r\wedge T}\Big(-\psi^G_{(r-T)^+}\big(0,-\bar\psi^A_{t-r}\big),\lambda\Big)\cdot V(0) \Big\} + \mbox{ Martingale .}
 \eeqnn
 Taking expectations on both sides of above equation, yields the desired result.
 \qed

From (\ref{eqn4.52}), we see that $G^A_0(\infty)=\infty$. Hence, the right inverse $G_0^{A,-1}(y):= \inf\big\{x\geq 0: G^{A}_0(x)> y \big\}$ of $G_0^A$
is well defined. This allows us to obtain the distribution of the random variable $\left( \mathcal{J}_A(T), \mathcal{T}_A \right)$.

 \begin{corollary}\label{Thm407}
 Suppose that (\ref{eqn4.52}) holds for some $\vartheta>0$.
 As $t\to\infty$, the function $t \mapsto \bar\psi^A_t$ decreases to $ \bar\psi^A_\infty := G^{A,-1}_0\big(\hat\nu_\mathtt{i}(A)\big)$.
 Moreover, for any $r \geq 0$ and $\lambda\geq 0$,
 \beqlb\label{eqn4.50}
 \mathbf{E}\big[\exp\{-\lambda \mathcal{J}_A(T)  \} ;\ \mathcal{T}_A\leq  r\big]= \exp\Big\{  - \phi^A_{r\wedge T}\Big(-\psi^G_{(r-T)^+}\big(0,-\bar\psi^A_\infty \big),\lambda\Big)\cdot V(0) \Big\}.
\eeqlb
 \end{corollary}
 \proof From (\ref{eqn4.27}), for any $\lambda \geq 0$, we see that $\phi^A_{r\wedge T}\big(-\psi^G_{(r-T)^+}\big(0,-\bar\psi^A_{t-r} \big),\lambda \big)$ is decreasing in $t$. Moreover,
 by continuity and the uniqueness of the solution to (\ref{eqn4.44}), the equation satisfies a comparison principle, and so the mapping $x \mapsto \phi^A_t(x,0)$ is increasing. As a result, the mapping $t \mapsto -\psi^G_{(r-T)^+}\big(0,-\bar\psi^A_{t-r} \big)$ is decreasing.
 Moreover, from (\ref{eqn3.24}),
 $-\psi^G_t(0,-x)=x- \int_0^t  G_0(\psi^G_s(0,-x))ds$, which is increasing in $x$. The last two results imply that the mapping $t \mapsto \bar\psi^A_{t-r}$ is decreasing. In particular, the limit  $\bar\psi^A_\infty:=\lim_{t\to\infty} \bar\psi^A_{t-r}$ exists.  Since $\bar{\psi}^A_{s+t}= \psi^A_t(\bar{\psi}^A_s)$, as $s\to\infty$ we have $\bar{\psi}^A_\infty= \psi^A_t(\bar{\psi}^A_\infty)$. Taking this into (\ref{eqn4.28}) we conclude that $\int_0^t [\hat\nu_\mathtt{i}(A) -G^A_0\big(\bar\psi^A_\infty\big)]ds\equiv0$ and hence
 \beqnn
 \bar\psi^A_\infty = G^{A,-1}_0\left(\hat\nu_\mathtt{i}(A)\right).
 \eeqnn
 \qed

The following corollary is the analogue of Corollary \ref{Thm209}. It shows that the quantity $b_2$ is the analogue of the quantity $\tilde \beta$ introduced in equation (\ref{beta-tilde}); it determines whether or not the impact duration of an external shock is almost surely finite.

 \begin{corollary}\label{Thm408}
 We have $\mathbf{P}\{\mathcal{T}_A<\infty \}=\mathbf{P}\{\mathcal{J}_A(\infty)<\infty \}=\exp\{-V(0)\cdot \bar{v}_\infty\}$, where $\bar v_\infty$ is the largest root of $G_0(x)=0$. Moreover,
$\mathbf{P}\{\mathcal{T}_A<\infty \}=\mathbf{P}\{\mathcal{J}_A(\infty)<\infty \}=1$ if and only if $b_2=G_0'(0) \geq 0$.
 \end{corollary}
 \proof
 By equation (\ref{eqn3.23}) and Corollary~\ref{Thm407}, $\mathbf{P}\{\mathcal{T}_A\leq r \} =\exp\big\{  - \phi^A_{r}( \bar\psi^A_\infty,0)\cdot V(0) \big\}$. This implies that $\phi^A_{r}( \bar\psi^A_\infty,0)$ decreases to some limit $\phi^A_\infty(\bar\psi^A_\infty,0)$ as $r\to\infty$.
 From (\ref{eqn4.44}), we see that semigroup property holds for $\phi^A_{r}(x,0)$ for any $x\geq 0$ and hence $G_0(\phi^A_\infty(x,0))=0$.  We now show that $\phi^A_\infty(x,0) \equiv \bar{v}_\infty$.
 From (\ref{eqn4.54}), we see that $G_0(0) = 0$ and that $G_0$ is strictly convex. Hence, $G_0(x)>0$ for any $x > 0$ if and only if $b_2=G_0'(0) \geq 0$. In this case, $\bar{v}_\infty=0$ is the only root of $G_0(x)=0$ and $\mathbf{P}\{\mathcal{T}_A<\infty \}=\mathbf{P}\{\mathcal{J}_A(\infty)<\infty \}=1$.
 If $b_2=G_0'(0) < 0$, since $G_0(x)\to\infty$ as $x\to\infty$, there is only one positive root $\bar{v}_\infty$ of  $G_0(x)=0$ and  $G_0(y)<0$ for $y\in(0,\bar{v}_\infty)$. It suffices to prove $\phi^A_\infty(x,0)>0$, which follows directly from the fact that $\phi^A_t(x,0)>0$ continuously decreases to $\phi^A_\infty(x,0)>0$ as $t\to\infty$.
 \qed

 Taking expectation on both sides of (\ref{eqn4.19}), we have $\mathbf{E}[V_0(t)]=V(0) \exp\{-b_2 t\}$ for any $t\geq 0$.
 From the definition of $\mathcal{J}_A(T)$, we have
 \beqnn
 \mathbf{E}\big[\mathcal{J}_A(T)\big]\ar=\ar \mathbf{E}\Big[ \int_0^T \int_{A} \int_0^{V_0(s-)} N_0(ds,du,dx)  \Big] =\hat\nu_\mathtt{i}(A) \int_0^T \mathbf{E}[V_0(s)] ds = \hat\nu_\mathtt{i}(A)\cdot \frac{V(0)}{b_2}\cdot (1-e^{-b_2 T}).
 \eeqnn
In particular, $\mathbf{E}[\mathcal{J}_A(T)] = \hat\nu_\mathtt{i}(A) V(0) T+ O(T^2)$ for small $T >0$ and $\mathbf{E}[\mathcal{J}_A(T)]$ converges as $T\to\infty$ if and only if $b_2 > 0$. In this case, the number of induced jumps of a given magnitude following a large shock is finite as shown above and the expected number of shocks is proportional to the shock size. Specifically, we have the following corollary.

 \begin{corollary}\label{cor-cluster}
 We have $\mathbf{E}[\mathcal{J}_A(\infty)]<\infty$ if and only if $b_2>0$.
 In this case, $\mathbf{E}[\mathcal{J}_A(\infty)]= \hat\nu_\mathtt{i}(A)\cdot V(0)/b_2$.
 \end{corollary}

The following corollary is the analogue of Corollary \ref{Thm210}. It provides four regimes for the impact duration of exogenous shocks on the market dynamics. If the volatility strictly mean-reverts to $0$, then the impact decreases exponentially. In the critical case $b_2=0$, the impact decays only slowly.

 \begin{corollary}
 We have the following four regimes:
 \begin{enumerate}
  \item[(1)] If $b_2<0$, we have as $t\to\infty$,
  \beqnn
  \mathbf{P}\{\mathcal{T}_A> t\} \sim \mathbf{P}\{\mathscr{T}_{0}> t\}\to  1- \exp\big\{-V(0) \cdot \bar{v}_\infty\big\};
  \eeqnn

  \item[(2)] If $b_2>0$, there exists a constant $C>0$ such that for any $t\geq 1$,
    \beqnn
   \mathbf{P}\{\mathcal{T}_A> t\}\leq  \mathbf{P}\{\mathscr{T}_{0}> t\} \leq  C\cdot V(0)\cdot e^{-b_2t};
    \eeqnn

    \item[(3)] If $b_2=0$ and $\hat\nu_\mathtt{i}(|u_2|^2):=\frac{1}{2}\int_\mathbb{U}|u_2|^2 \hat\nu_\mathtt{i}(du)<\infty$, we have $t\to\infty$,
        \beqnn
     \mathbf{P}\{\mathcal{T}_A> t\} \sim \mathbf{P}\{\mathscr{T}_{0}> t\}  \sim \frac{V(0)}{\sigma_{22}+\hat\nu_\mathtt{i}(|u_2|^2)}\cdot t^{-1};
        \eeqnn

   \item[(4)] If $b_2=\sigma_{22}=0$ and $\hat\nu_\mathtt{i}(\mathbb{R}\times [x,\infty))\sim Cx^{-1-\alpha}$ as $x\to\infty$ for some $\alpha\in(0,1)$, there exists a constant $C>0$ such that as $t\to\infty$,
        \beqnn
        \mathbf{P}\{\mathcal{T}_A> t\} \sim \mathbf{P}\{\mathscr{T}_{0}> t\}  \sim C \cdot V(0) \cdot t^{-1/\alpha}.
        \eeqnn
 \end{enumerate}

 \end{corollary}
  \proof The first regime follows directly from Lemma~\ref{Grey} and Corollary~\ref{Thm408}.
   For the second regime,  we have
   \beqnn
   \mathbf{P}\{\mathcal{T}_A> t\}\leq\mathbf{P}\{\mathscr{T}_{0}> t\}= 1-\exp\{-V(0) \cdot \overline{v}_t\}\leq   V(0) \cdot \overline{v}_t.
   \eeqnn
  From (\ref{eqn3.25}), we have $G_0(x)\geq b_2 x$ for any $x\geq 0$.
  From Gr\"onwall's inequality, we also have for any $t\geq 1$,
  \beqnn
  \overline{v}_t\leq \overline{v}_1- \int_1^t b_2\overline{v}_sds  \leq \overline{v}_1e^{-b_2(t-1)}.
  \eeqnn
  The last two regimes for $\mathbf{P}\{\mathscr{T}_{0}> t\}$ can be proved following the proof of Corollary~\ref{Thm210} with the fact that $G_0(x)\sim [\sigma_{22}+\hat\nu_\mathtt{i}(|u_2|^2)] x^2$ in regime (3) and $G_0(x)\sim C x^{\alpha+1}$ in regime (4)  as $x\to 0+$.
  We now prove the last two regimes for $\mathbf{P}\{\mathcal{T}_A> t\}$.
  From Corollary~\ref{Thm407} and \ref{Thm408}, we have as $t\to\infty$,
  \beqnn
  \mathbf{P}\{\mathcal{T}_A> t\}\ar=\ar 1-\exp\big\{  - \phi^A_t( G^{A,-1}_0(\hat\nu_\mathtt{i}(A)),0)\cdot V(0) \big\} \sim   V(0)\cdot \phi^A_t( G^{A,-1}_0(\hat\nu_\mathtt{i}(A)),0).
  \eeqnn
  From (\ref{eqn4.44}), we also have
  \beqnn
  \phi^A_t( G^{A,-1}_0(\hat\nu_\mathtt{i}(A)),0)\ar=\ar G^{A,-1}_0(\hat\nu_\mathtt{i}(A)) - \int_0^t   G_0(\phi^A_s( G^{A,-1}_0(\hat\nu_\mathtt{i}(A)),0))ds.
  \eeqnn
  From this and the proof of Corollary~\ref{Thm210}, we also can prove that the last two regimes for $\mathbf{P}\{\mathcal{T}_A> t\}$.
  \qed


 \subsection{The sub-model $\{(P_a(t),V_a(t)):t\geq 0\}$ }

 We now study the distribution of induced jumps in the system $\{(P_a(t),V_a(t)):t\geq 0\}$ assuming that $a_2 > 0$. The analysis is much simpler than the preceding one because now the volatility process is recurrent. For any $T>0$, let
 \beqnn
 \mathcal{J}_A^a(T):= \#\big\{ t\in[0,T]: (\Delta P_a(t),\Delta V_a(t))\in A\big\}= \int_0^T \int_A\int_0^{V_a(s-)} N_0(ds,du,dx).
 \eeqnn
 Taking expectation on both sides of above equation,  we see that the expected number of jumps is given by
 \beqnn
 \mathbf{E}\big[\mathcal{J}_A^a(T)\big] \ar=\ar  \hat\nu_\mathtt{i}(A)\int_0^T \mathbf{E}[V_a(s)]ds.
 \eeqnn
 Taking expectation on both sides of (\ref{eqn4.20}), we also have
 \beqnn
 \mathbf{E}[V_a(t)]= \int_0^t (a_2-b_2 \mathbf{E}[V_a(s)])ds =\frac{a_2}{b_2} (1-e^{-b_2t})
 \quad\mbox{and}\quad
 \mathbf{E}\big[\mathcal{J}_A^a(T)\big]=   \hat\nu_\mathtt{i}(A)\cdot \frac{a_2}{b_2}\cdot \big[T-(1-e^{-b_2 T})/b_2\big].
 \eeqnn

 \begin{lemma}
 For any $T\in[0,\infty)$, we have
 \beqnn
 \mathbf{E}\big[\exp\{-\lambda\mathcal{J}_A^a(T)\}\big]= \exp\Big\{ - \int_0^T a_2 \psi_s^a(\lambda)ds\Big\},
 \eeqnn
  where $s\mapsto\psi_s^a(\lambda)$ is the unique solution to the following Riccati equation:
  \beqnn
  \psi_t^a(\lambda)\ar=\ar \hat\nu_\mathtt{i}(A)\cdot(1-e^{-\lambda})\cdot t - \int_0^t G_0(\psi_s^a(\lambda))ds.
  \eeqnn
 \end{lemma}
 \proof From the exponential formula of Poisson random measure; see \cite[p.8]{Bertoin1996}, we have
 \beqnn
 \mathbf{E}\big[\exp\{-\lambda\mathcal{J}_A^a(T)\}\big]\ar=\ar \mathbf{E}\Big[\exp\Big\{- \hat\nu_\mathtt{i}(A)\cdot(1-e^{-\lambda})\cdot\int_0^T V_a(s)ds\Big\}\Big].
 \eeqnn
 Since $\{(\int_0^t V_a(s)ds,V_a(t)):t\geq 0\}$ is affine,
 the result follows from \cite[Theorem 2.7]{DuffieFilipovicSchachermayer2003}.
 \qed

 \begin{remark}\label{concluding-rem}
 Let us compare exogenously and endogenously triggered jump cascades assuming that the volatility process strictly mean-reverts
 {and the arrival rate of exogenous socks  $\lambda_\mathtt{e}:=\nu_\mathtt{e}(\mathbb{U})$ is finite.}
 By Corollary \ref{cor-cluster} exogenously triggered clusters with jumps in a set $A$ arrive at rate $\lambda_\mathtt{e} \hat{\nu}_\mathtt{i}(A)$, and an external volatility shock of size $V(0)$ triggers $\frac{V(0)}{b_2} (1-e^{-b_2t})$ jumps by time $t>0$ on average. By contrast, endogenously triggered jump cascades arrive at a rate $a_2 \hat\nu_\mathtt{i}(A)$ and each cascade comprises $\frac{1}{b_2}(1-e^{-b_2t})$ jumps by time $t>0$ on average. This shows that both cascades have the same ``duration distribution'' but that {exogenously} triggered cascades usually comprise more jumps immediately and hence cluster more heavily.  This effect is illustrated by Figure 3.
\end{remark}

 \begin{appendix}
 
\renewcommand{\theequation}{\Alph{section}.\arabic{equation}}
\section{A cluster representation for $\{(P_a(t),V_a(t)):t\geq 0\}$}
	\setcounter{equation}{0}

This appendix proves Theorem \ref{thm-cluster}. The proof uses arguments given in Li \cite{Li2019} where the result is established for the volatility process. We assume throughout that $\sigma_{22} > 0$ and start with the following simple but useful lemma.

 \begin{lemma}\label{Thm404}
 For any two  finite measures $(u_2\wedge 1)\nu_1(du)$ and $(u_2\wedge 1)\nu_2(du)$ on $\mathbb{U}\setminus \{0\}$, we have $\nu_1(du)=\nu_2(du)$ if and only if for any $z=(z_1,z_2)\in  \mathbb{U}_*$,
 \beqlb\label{eqn4.12}
 \int_{\mathbb{U}}(e^{\langle z,u\rangle}-e^{z_1u_1})\nu_1(du)= \int_{\mathbb{U}}(e^{\langle z,u\rangle}-e^{z_1u_1})\nu_2(du).
 \eeqlb
 \end{lemma}
  \proof We first extend $\nu_1(du)$ and $\nu_2(du)$ to $\mathbb{U}$ with $\nu_1(\{0\})=\nu_2(\{0\})=0$. From (\ref{eqn4.12}) with $z$ replaced by $z+(0,1)$, we  have
 \beqlb\label{eqn4.13}
 \int_{\mathbb{U}}(e^{\langle z,u\rangle-u_2}-e^{z_1u_1})\nu_1(du)= \int_{\mathbb{U}}(e^{\langle z,u\rangle-u_2}-e^{z_1u_1})\nu_2(du). 
 \eeqlb
 Taking the difference between (\ref{eqn4.12}) and (\ref{eqn4.13}), we have
 \beqnn
  \int_{\mathbb{U}}e^{\langle z,u\rangle}(1-e^{-u_2})\nu_1(du)= \int_{\mathbb{U}}e^{\langle z,u\rangle}(1-e^{-u_2})\nu_2(du).
 \eeqnn
 By assumption $(1-e^{-u_2})\nu_1(du)$ and $(1-e^{-u_2})\nu_2(du)$ are finite measure on $\mathbb{U}$.
 By the one-to-one correspondence between measures and their characteristic function, $(1-e^{-u_2})\nu_1(du)=(1-e^{-u_2})\nu_2(du)$ and $\nu_1(du)=\nu_2(du)$.
  \qed

 For any $t>0$ and $u'\in\mathbb{U}$, the probability measure $Q_{0,t}(u',du)$ introduced in (\ref{eqn4.02}) is  infinitely divisible, i.e., for any $n\geq 1$,
 \beqlb\label{eqn4.08}
 \int_\mathbb{U} e^{\langle z,u\rangle}Q_{0,t}(u',du)=  \Big(\exp\big\{z_1\frac{u'_1}{n}+\psi^G_t(z)\frac{u'_2}{n}\big\}\Big)^n= \int_\mathbb{U} e^{\langle z,u\rangle}Q^{(*n)}_{0,t}(u'/n,du).
 \eeqlb
Using the the L\'evy-Khintchine formula for infinite divisible distributions (see \cite[Theorem 1]{Bertoin1996}), the representation (\ref{eqn3.24}) for the characteristic exponent $\{\psi^G_t(z): z\in  \mathbb{U}_*\}$ and Theorem 3.13 in \cite{Li2019} along with the assumption that $\sigma_{22}>0$, we obtain that
 \beqlb\label{eqn4.09}
 \psi^G_t(z)= b_1(t)z_1 \rangle + \sigma_{11}(t) |z_1|^2 +\int_{\mathbb{U}\setminus \{0\}} (e^{\langle z,u\rangle}-1- z_1u_1)\eta_t(du),
 \eeqlb
 where $b_1(t) \in \mathbb R,$ 
 $\sigma_{11}(t)\geq 0$ and $(|u_1|\wedge |u_1|^2+ |u_2|\wedge 1)\eta_t(du)$ is a finite measure on $\mathbb{U}\setminus \{0\}$.

 \begin{lemma}\label{Thm405}
 The family of $\sigma$-finite measures $\{\eta_t(du):t\geq 0 \}$ is an entrance law for $(Q_{0,t}^\circ)_{t\geq 0}$, i.e.
 $$\eta_{s+t}(du)= \eta_t Q_{0,s}^\circ(du), \qquad s,t\geq 0.$$
 \end{lemma}
 \proof In view of Lemma~\ref{Thm404}, it suffices to prove that for any $s,t\geq 0$
 \beqlb\label{eqn4.14}
 \int_{\mathbb{U}}(e^{\langle z,u\rangle}-e^{z_1u_1})\eta_{s+t}(du)= \int_{\mathbb{U}}(e^{\langle z,u\rangle}-e^{z_1u_1})\eta_tQ^{\circ}_{0,s}(du).
 \eeqlb
 From (\ref{eqn4.09}),
 \beqlb\label{eqn4.15}
 \int_{\mathbb{U}}(e^{\langle z,u\rangle}-e^{z_1u_1})\eta_{t}(du)=\int_{\mathbb{U}}(e^{\langle z,u\rangle}-e^{z_1u_1})\eta_{t}(du)=\psi^G_{t}(z_1,0)-\psi^G_{t}(z_1,z_2).
 \eeqlb
 Moreover, from the definition of $Q^{\circ}_{0,s}(u',du)$ and (\ref{eqn4.02}), we have
 \beqnn
 \int_{\mathbb{U}}(e^{\langle z,u\rangle}-e^{z_1u_1})Q^{\circ}_{0,s}(u',du)
 \ar=\ar\int_{\mathbb{U}}(e^{\langle z,u\rangle}-e^{z_1u_1})Q_{0,s}(u',du)\cr
 \ar=\ar  \exp\big\{z_1u'_1+\psi^G_s(z)u'_2\big\} -\exp\big\{z_1u'_1+\psi^G_s(z_1,0)u'_2\big\}.
 \eeqnn
 Taking this into the term on the right side of (\ref{eqn4.14}), we obtain from (\ref{eqn4.15}) that
 \beqnn
 \int_{\mathbb{U}}(e^{\langle z,u\rangle}-e^{z_1u_1})\eta_tQ^{\circ}_{0,s}(du)
 \ar=\ar \int_{\mathbb{U}}\eta_t(du') \int_{\mathbb{U}}(e^{\langle z,u\rangle}-e^{z_1u_1})Q^{\circ}_{0,s}(u',du)\cr
 \ar=\ar\int_{\mathbb{U}}[e^{z_1u'_1+\psi^G_s(z)u'_2} -e^{z_1u'_1+\psi^G_s(z_1,0)u'_2}]\eta_t(du')\cr
 \ar=\ar  \psi^G_t(z_1,\psi^G_s(z_1,0))-\psi^G_t(z_1,\psi^G_s(z)).
 \eeqnn
 From the semigroup property of $(\psi^G_t)_{t\geq 0}$, we also have $\psi^G_t(z_1,\psi^G_s(z_1,0))= \psi^G_{s+t}(z_1,0)$ and $\psi^G_t(z_1,\psi^G_s(z)) =\psi^G_{s+t}(z)$.
Along with (\ref{eqn4.15}) this yields the desired result.
 \qed

In view of the preceding lemma,  we can define an $\sigma$-finite measure $\mathbf{Q}(d\omega)$ on $\mathbf{D}([0,\infty),\mathbb{U})$ as follows: for any $0<t_1<t_2<\cdots<t_n$ and $u^{(1)},\cdots, u^{(n)} \in \mathbb{R}\times(0,\infty)$,
 \begin{align}\label{eqn4.31}
   & \mathbf{Q}(\omega(t_1)\in du^{(1)}, \omega(t_2)\in du^{(2)},\cdots \omega(t_n)\in du^{(n)}) \nonumber \\
:= & \eta_{t_1}(du^{(1)})Q_{0,t_2-t_1}^\circ(u^{(1)},du^{(2)})\cdots Q_{0,t_n-t_{n-1}}^\circ(u^{(n-1)},du^{(n)}).
 \end{align}

The following lemma provides the analogue of equation (3.17) in \cite{Li2019}.

 \begin{lemma}\label{Thm406}
 For any $t>0$, we have $\frac{1}{u'_2}Q_{0,t}((0,u'_2),du)\to \eta_t(du)$ as $u'_2\to 0+$.
 \end{lemma}
 \proof From (\ref{eqn4.15}) we have for any $z:=(z_1,z_2)\in \mathbb{U}_*$,
 \beqnn
 \lim_{u'_2\to 0+}\int_{\mathbb{U}}(e^{\langle z,u\rangle}-e^{z_1u_1})\frac{1}{u'_2} Q_{0,t}((0,u'_2),du)
 \ar=\ar\lim_{u'_2\to 0+} \frac{1}{u'_2} \big(e^{u'_2 \psi^G_t(z)} -e^{u'_2 \psi^G_t(z_1,0)}\big) \cr
 \ar =\ar \psi^G_t(z)-\psi^G_t(z_1,0)= \int_{\mathbb{U}}(e^{\langle z,u\rangle}-e^{z_1u_1})\eta_{t}(du).
 \eeqnn
 The same arguments as in the proof of Lemma~\ref{Thm404} yield
 \beqnn
 \lim_{u'_2\to 0+}\int_{\mathbb{U}}e^{\langle z,u\rangle}(1-e^{-u_2})\frac{1}{u'_2} Q_{0,t}((0,u'_2),du) = \int_{\mathbb{U}}e^{\langle z,u\rangle}(1-e^{-u_2})\eta_{t}(du),
 \eeqnn
 and hence the desired result.
 \qed

By Lemma~\ref{Thm406}, we have formally that
\begin{eqnarray*}
& &  \mathbf{Q}(\omega(t_1)\in du^{(1)}, \omega(t_2)\in du^{(2)},\cdots \omega(t_n)\in du^{(n)}) \nonumber \\
&=& \lim_{u_2 \to 0} \frac{1}{u_2}  Q_{0,t_1}^\circ((0,u_2); du^{(1)}) Q_{0,t_2-t_1}^\circ(u^{(1)},du^{(2)})\cdots Q_{0,t_n-t_{n-1}}^\circ(u^{(n-1)},du^{(n)}),
 \end{eqnarray*}
which shows that  $\mathbf{Q}(d\omega)$ is supported on $\mathbf{D}_0([0,\infty),\mathbb{U})$. We refer to the proof of Theorem 6.1 in \cite{Li2019} for further details.

\mbox{ }

\noindent \textsc{Proof of Theorem \ref{thm-cluster}.} It suffices to prove that $(\hat{P}_a(t),\hat{V}_a(t))$ is a Markov process with transition semigroup $(Q_{a,t})_{t\geq 0}$ on $\mathbb{U}$. For this, it is enough to prove that for any $0\leq t_1\leq r\leq  t_2$, any $z=(z_1,z_2)\in \mathbb{U}_*$ and every $\mathscr{E}_{t_1}$-measurable $\mathbb{C}$-valued random variable $X_{t_1}$,
 \beqlb\label{eqn4.17}
 \mathbf{E}[X_{t_1}\cdot e^{z_1P_a(t_2)+z_2V_a(t_2)} ]= \mathbf{E}\Big[ X_{t_1}\cdot \exp\Big\{z_1P_a(r)+\psi^G_{t_2-r}(z)V_a(r)+\int_0^{t_2-r} [a_1z_1+ a_2 \psi^G_s(z)] ds\Big\}\Big].
 \eeqlb
 From the definition of $(\mathscr{E}_t)_{t\geq 0}$, we just need to prove this statement with
 \beqnn
 X_{t_1}= \exp\Big\{ \int_0^{t_1} \int_{\mathbf{D}_0([0,\infty),\mathbb{U})} h(\omega)N_a(ds,d\omega)\Big\},
 \eeqnn
 where $h(\omega): \mathbf{D}([0,\infty),\mathbb{U}) \mapsto \mathbb{C}_- $.
 From this and (\ref{eqn4.18}),
 \beqlb\label{eqn4.32}
 \lefteqn{\mathbf{E}\Big[ \exp\Big\{  \int_0^{t_1}  \int_{\mathbf{D}_0([0,\infty),\mathbb{U})} h(\omega)N_a(ds,d\omega)+ z_1P_a(t_2)+z_2V_a(t_2)  \Big\} \Big]}\ar\ar\cr
 \ar=\ar \mathbf{E}\Big[ \exp\Big\{ a_1t_2z_1+ \int_0^{t_2} \int_{\mathbf{D}_0([0,\infty),\mathbb{U})} \big[h(\omega)\mathbf{1}_{\{s\leq t_1 \}}+\langle z,\omega(t_2-s)\rangle \big]N_a(ds,d\omega)     \Big\} \Big].
 \eeqlb
 By the exponential formula for Poisson random measures, the last term equals
 \beqlb\label{eqn4.30}
 \lefteqn{\exp\Big\{ a_1t_2z_1+  \int_0^{t_2} a_2 ds\int_{\mathbf{D}_0([0,\infty),\mathbb{U})} \big[\exp\{h(\omega)\mathbf{1}_{\{s\leq t_1 \}}+\langle z,\omega(t_2-s)\rangle\}-1\big]\mathbf{Q}(d\omega)     \Big\} }\ar\ar\cr
 \ar=\ar  \exp\Big\{ a_1rz_1+\int_0^r a_2 ds\int_{\mathbf{D}_0([0,\infty),\mathbb{U})} \big[\exp\{h(\omega)\mathbf{1}_{\{s\leq t_1 \}}+\langle z,\omega(t_2-s)\rangle\}-1\big]\mathbf{Q}(d\omega)\Big\} \cr
 \ar\ar \times  \exp\Big\{ a_1(t_2-r)z_1+ \int_{r}^{t_2} a_2 ds\int_{\mathbf{D}_0([0,\infty),\mathbb{U})} \big(e^{\langle z,\omega(t_2-s)\rangle}-1\big)\mathbf{Q}(d\omega)\Big\}.
 \eeqlb
 From the definition of $\mathbf{Q}(d\omega)$,  we have for any $s\leq r$,
 \beqnn
 \lefteqn{\int_{\mathbf{D}_0([0,\infty),\mathbb{U})} \exp\{h(\omega)\mathbf{1}_{\{s\leq t_1 \}}\}\big(e^{\langle z,\omega(t_2-s)\rangle}-1\big)\mathbf{Q}(d\omega)}\ar\ar\cr
 \ar=\ar \int_{\mathbf{D}_0([0,\infty),\mathbb{U})} \exp\{h(\omega)\mathbf{1}_{\{s\leq t_1 \}}\}\big[\exp\{  z_1\omega_1(r-s)+ \psi^G_{t_2-r} (z) \omega_2(r-s) \}-1\big]\mathbf{Q}(d\omega)
 \eeqnn
 and
 \beqnn
 \lefteqn{\int_{\mathbf{D}_0([0,\infty),\mathbb{U})} \big[\exp\{h(\omega)\mathbf{1}_{\{s\leq t_1 \}}+\langle z,\omega(t_2-s)\rangle\}-1\big]\mathbf{Q}(d\omega)}\ar\ar\cr
 \ar=\ar \int_{\mathbf{D}_0([0,\infty),\mathbb{U})} \big[\exp\{h(\omega)\mathbf{1}_{\{s\leq t_1 \}}\}-1\big]\mathbf{Q}(d\omega)
 + \int_{\mathbf{D}_0([0,\infty),\mathbb{U})} \exp\{h(\omega)\mathbf{1}_{\{s\leq t_1 \}}\}\big(e^{\langle z,\omega(t_2-s)\rangle}-1\big)\mathbf{Q}(d\omega)\cr
 \ar=\ar  \int_{\mathbf{D}_0([0,\infty),\mathbb{U})}  \big[\exp\{ h(\omega)\mathbf{1}_{\{s\leq t_1 \}}+ z_1\omega_1(r-s)+ \psi^G_{t_2-r} (z) \omega_2(r-s) \}-1\big]\mathbf{Q}(d\omega).
 \eeqnn
 Taking this back into the first term on the right side of the last equality in (\ref{eqn4.30}), this term equals
 \beqnn
 \exp\Big\{ a_1rz_1+\int_0^r a_2 ds\int_{\mathbf{D}_0([0,\infty),\mathbb{U})}  \big[\exp\{ h(\omega)\mathbf{1}_{\{s\leq t_1 \}}+ z_1\omega_1(r-s)+ \psi^G_{t_2-r} (z) \omega_2(r-s) \}-1\big]\mathbf{Q}(d\omega)\Big\} ,
 \eeqnn
 which equals to
 \beqnn
  \mathbf{E}\Big[ \exp\Big\{  \int_0^{t_1}  \int_{\mathbf{D}_0([0,\infty),\mathbb{U})} h(\omega)N_a(ds,d\omega)+ z_1P_a(r)+\psi^G_{t_2-r} (z)V_a(r)  \Big\} \Big].
 \eeqnn
 Moreover, from (\ref{eqn4.31}) and Lemma~\ref{Thm406} we have
 \beqnn
  \int_{\mathbf{D}([0,\infty),\mathbb{U})}\big(e^{\langle z,\omega(t_2-s)\rangle}-1\big) \mathbf{Q}(d\omega)
 \ar=\ar  \int_{\mathbb{U}} \big(e^{\langle z,u\rangle}-1\big)\eta_{t_2-s}(du) \cr
 \ar=\ar \lim_{u'_2\to 0+} \int_{\mathbb{U}} \big(e^{\langle z,u\rangle}-1\big)\frac{1}{u'_2}Q_{0,t_2-s}((0,u'_2),du) \cr
 \ar=\ar \lim_{u'_2\to 0+}\frac{1}{u'_2} ( e^{u'_2\psi^G_{t_2-s}(z)} -1)= \psi^G_{t_2-s}(z).
 \eeqnn
 Taking these back into (\ref{eqn4.32}), we have
 \beqnn
 \lefteqn{\mathbf{E}\Big[ \exp\Big\{  \int_0^{t_1}  \int_{\mathbf{D}([0,\infty),\mathbb{U})} h(\omega)N_a(ds,d\omega)+ z_1P_a(t_2)+z_2V_a(t_2)  \Big\} \Big]}\ar\ar\cr
 \ar=\ar  \mathbf{E}\Big[ \exp\Big\{  \int_0^{t_1}  \int_{\mathbf{D}([0,\infty),\mathbb{U})} h(\omega)N_a(ds,d\omega)+ z_1P_a(r)+\psi^G_{t_2-r}(z)V_a(r) + \int_{r}^{t_2} \big(a_1z_2+ a_2 \psi^G_{t_2-s}(z)\big) ds \Big\}\Big].
 \eeqnn
 Here we have got the desired result (\ref{eqn4.17}).
 \qed

 \end{appendix}

 \bibliographystyle{siam}

 \bibliography{ReferenceforVolatility}

\end{document}